\documentclass[twocolumn,
amsmath,
amssymb,
aps,superscriptaddress,a4paper,
]{revtex4-1}

\usepackage{graphicx}
\usepackage{mathtools}
\usepackage{amsmath}

\usepackage{xcolor}
\definecolor{Cerulean}{rgb}{0.,0.59,0.835}
\definecolor{RubineRed}{rgb}{0.61,0.07,0.12}
\usepackage[colorlinks,breaklinks,citecolor=blue,linkcolor=RubineRed,urlcolor=Cerulean]{hyperref}

\begin{document}

\title{\hspace{3cm } 
\newline
Attraction from frustration in ladder systems}
\author{Ivan Morera}
\affiliation{Departament de F{\'i}sica Qu{\`a}ntica i Astrof{\'i}sica, Facultat de F{\'i}sica, Universitat de Barcelona, E-08028 Barcelona, Spain}
\affiliation{Institut de Ci{\`e}ncies del Cosmos, Universitat de Barcelona, ICCUB, Mart{\'i} i Franqu{\`e}s 1, E-08028 Barcelona, Spain}
\affiliation{Department of Physics, Harvard University, Cambridge, Massachusetts 02138, USA}
\email[Corresponding author email: ]{imorera@icc.ub.edu}

\author{Annabelle Bohrdt}
\affiliation{Department of Physics, Harvard University, Cambridge, Massachusetts 02138, USA}
\affiliation{Department of Physics and Institute for Advanced Study, Technical University of Munich, 85748 Garching, Germany}
\affiliation{Munich Center for Quantum Science and Technology (MCQST), Schellingstr. 4, D-80799 M\"unchen, Germany}
\address{ITAMP, Harvard-Smithsonian Center for Astrophysics, Cambridge, MA 02138, USA}

\author{Wen Wei Ho}
\affiliation{Department of Physics, Harvard University, Cambridge, Massachusetts 02138, USA}
\affiliation{MIT-Harvard Center for Ultracold Atoms, Cambridge, MA, USA}
\affiliation{Department of Physics, Stanford University, Stanford, CA 94305, USA}

\author{Eugene Demler}
\affiliation{Department of Physics, Harvard University, Cambridge, Massachusetts 02138, USA}
\affiliation{Institute for Theoretical Physics, ETH Zurich, 8093 Zurich, Switzerland}
\date{\today}

\begin{abstract}
We analyze the formation of multi-particle bound states in ladders with frustrated kinetic energy in two component bosonic and two component fermionic systems. We focus on the regime of light doping relative to insulating states at half-filling, spin polarization close to 100 percent, and strong repulsive interactions. A special 
feature of these systems is that the binding energy scales with single particle tunneling $t$ rather than exchange interactions, since effective attraction arises from alleviating kinetic frustration. For two component Fermi systems on a zigzag ladder we find a bound state between a hole and a flipped spin (magnon) with a binding energy that can be as large as 0.6$t$. We demonstrate that magnon-hole attraction leads to formation of clusters comprised of several holes and magnons and expound on antiferromagentic correlations for the transverse spin components inside the clusters.  
We identify several many-body states that result from self-organization of multi-particle bound states, including a Luttinger liquid of  hole - magnon pairs and a density wave state of two hole - three magnon composites. 
We establish a symmetry between the spectra of Bose and Fermi systems and use it to establish the existence of antibound states in two component Bose mixtures with SU(2) symmetric repulsion on a zigzag ladder. We also consider Bose and Fermi systems on a square ladder with flux and demonstrate that both systems support bound states. We discuss experimental signatures of multi-particle bound states in both equilibrium and dynamical experiments. We point out intriguing connections between these systems and the quark bag model in QCD.

\end{abstract}

\maketitle

\section{Introduction}
\label{Sec1}

\subsection{Motivation}

Quantum frustrated systems hold the prospect of multifarious exotic many-body states, including spin liquids and unconventional superconductors (see  Ref.~\cite{lacroix2010introduction} and references therein). In this paper we consider two component mixtures of fermionic or bosonic particles on lattices with frustrated kinetic energy in the regime of  light doping relative to insulating states at half-filling, spin polarization close to 100\%, and repulsive SU(2) symmetric interactions. We demonstrate that such systems exhibit a common tendency to form multi-particle bound states with effective attraction between particles arising from kinetic frustration. In many cases we find that the ground state of such systems is best understood as  resulting from self-organization of multi-particle clusters. The systems under consideration can be realized using ultracold Bose and Fermi atoms in optical lattices with the currently available experimental techniques~\cite{Bakr2009,Sherson2010,PhysRevLett.114.213002,PhysRevLett.114.193001,PhysRevLett.115.263001,Haller2015,PhysRevA.92.063406,Greif953,Brown1385,Weitenberg2011,Schauss,PhysRevLett.94.086803,RevModPhys.83.1523,Atala2014,Fukuhara2013,PhysRevLett.108.225304,PhysRevLett.110.185301,PhysRevLett.111.185301,PhysRevLett.111.185302,Jotzu2014,PhysRevLett.112.043001,Goldman2016,Ane1602685,Schauss}. They may also be relevant for understanding twisted bilayer graphene close to ferromagentic insulating states~\cite{Sharpe605}. 

\begin{figure}[t!]
	\centering
	\includegraphics[width=1\columnwidth]{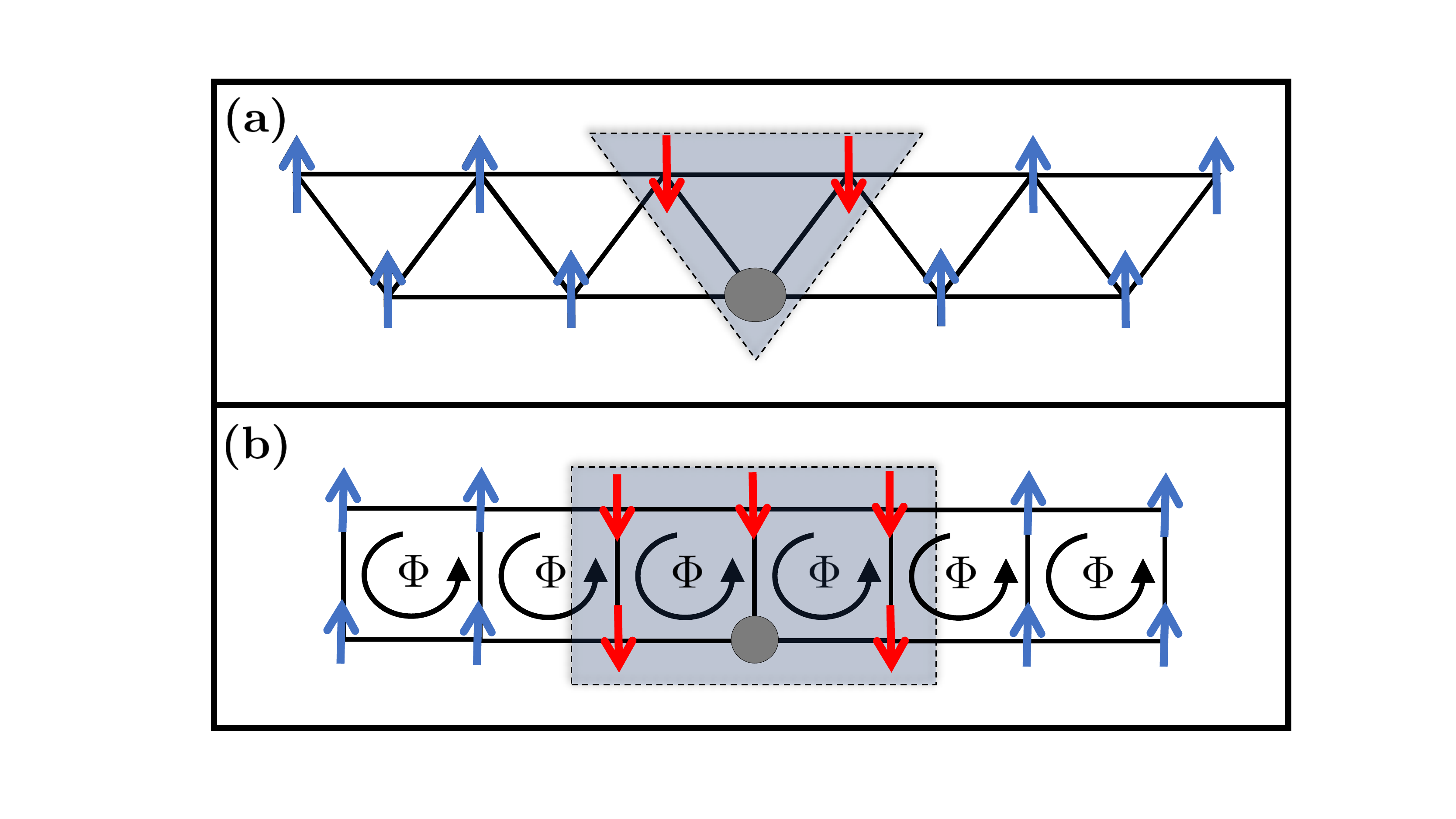}
	\caption{Schematic representation of multi-particle bound states of a single hole and one or several flipped spins (magnons) on top of a fully polarized insulating state. (a) corresponds to the non-bipartite zigzag ladder and (b) to the square ladder with a transverse magnetic flux $\Phi$.}
	\label{fig:HolMagRepr}
\end{figure}

We consider two specific lattice geometries: zigzag ladders and square ladders with flux (see Fig.~\ref{fig:HolMagRepr}), but we expect that our conclusions apply to broader classes of systems. 
Our starting ``vacuum'' is an insulating state with one fermion per site on a frustrated ladder (zigzag or regular ladder with a flux) and full spin polarization. When one of the spins is flipped, we call it a magnon. When a single fermion is removed altogether, we call it a hole. 
The starting point of our analysis is the observation of attraction between holes and magnons arising from frustration of the kinetic energy of holes (see discussion in Secs.~\ref{Sec2} and~\ref{Sec3}, and a related discussion for triangular lattices in~\cite{PhysRevB.97.140507}). We demonstrate that this attraction leads to the formation of multi-particle bound states, which we denote as nHmM, where integer $n$ and $m$ indicate the numbers of holes and magnons in the composite object (see Fig.~\ref{fig:HolMagRepr} for illustration), respectively.

The central objective of our paper is understanding the interplay of few- and many-body phenomena. Many-body systems are usually understood from the perspective of two-body correlations, such as the pairing amplitude in superconductors that arises due to formation of Cooper pairs. Traditional mean-field approaches rely on the order parameters defined to describe two particle correlations (superconductivity, magnetism, spin and charge density wave states). On the other hand, powerful theoretical methods have been developed for analyzing few-body states in vacuum, such as the celebrated Skorniakov-Ter-Martirosian (STM) equation~\cite{BRAATEN2006259,RevModPhys.89.035006}. However, the analysis of multi-particle composites within a many-body system remains an open problem in many areas of physics. For example, the primary objective of quantum chromodynamics (QCD) is developing accurate models of triplets of quarks binding into nucleons, which in turn combine to form nuclei. Intriguingly, we find many analogies between our system and the ``bag model'' of QCD~\cite{HASENFRATZ197875}. While in QCD gluons provide a ``bag'' that holds quarks together,  in the ladder systems discussed in this paper, magnons provide a ``bag'' which traps one or several holes, and in turn holes are holding the magnons together.

\subsection{Overview of results}
\begin{figure}[t!]
	\centering
	\includegraphics[width=1\columnwidth]{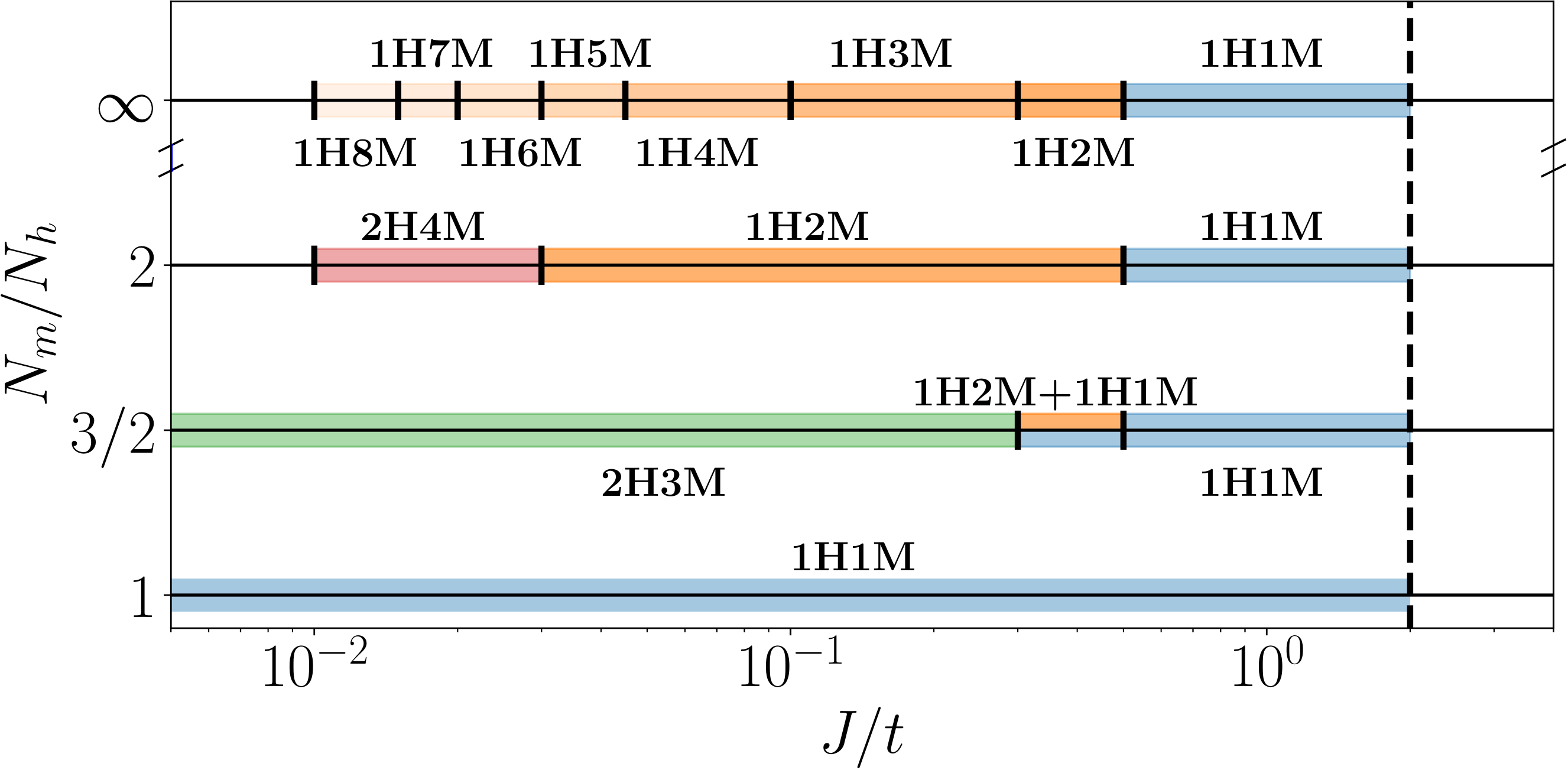}
	\caption{Phase diagram of the multi-particle bound states in the fermionic zigzag ladder for different $J/t$ and different ratios of the magnon to hole densities. The vertical dashed line represents the point $J/t\sim 2$ at which the hole-magnon bound states disappears. The $\infty$ line represents the case of a single hole with an arbitrary number of magnons.}
	\label{fig:PhasDiagr}
\end{figure}
Before presenting detailed microscopic calculations we summarize the main results of our work in a non-technical manner.

\begin{figure}[t!]
	\centering
	\includegraphics[width=\columnwidth]{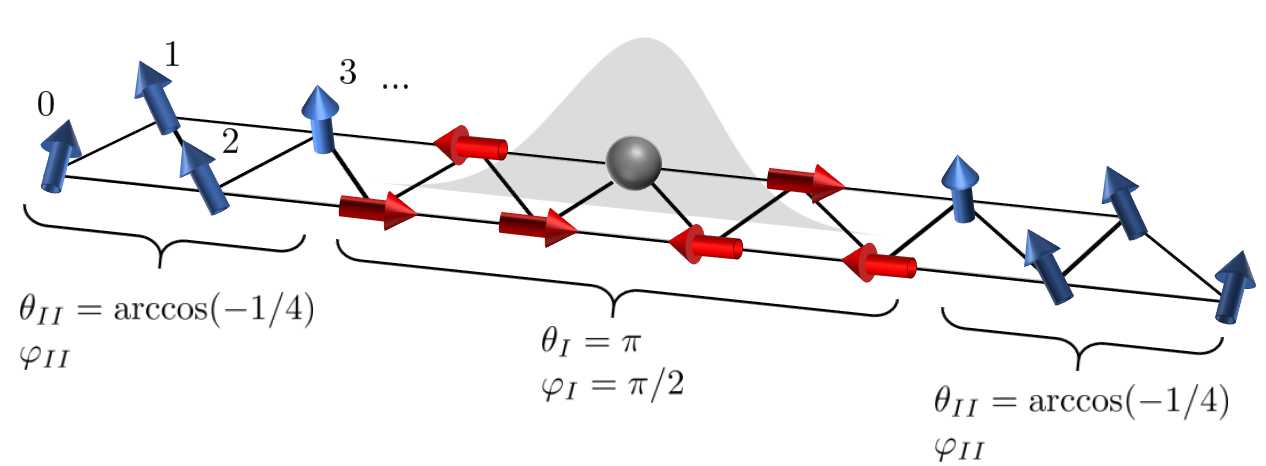}
	\caption{Schematic picture of the antiferromagentic spin bag surrounding a single hole. Inside the spinbag z-components of the spins are suppressed, while the xy components exhibit antiferromagnetic correlations between two sites of different legs. For clarity of presentation, in the figure we oriented transverse components to point along the y axis. Outside the spinbag, we find spiral winding of the xy components of the spins and z-magnetization approaching its asymptotic value. The two angles $\theta$ and $\varphi$ characterize the magnetic orders present in the system. They correspond to the angles between two adjacent spins in different legs in the xy plane and with respect to the z axis, respectively. We define two regions: Region (I) inside the spinbag and region (II) outside of it. In order to perform numerical calculations we map the ladder to a one-dimensional system in which the sites are indexed following the zigzag pattern specified by the numbers.} 
	\label{fig:AntiferroSB1}
\end{figure}

\begin{figure}[t!]
	\centering
	\includegraphics[width=1\columnwidth]{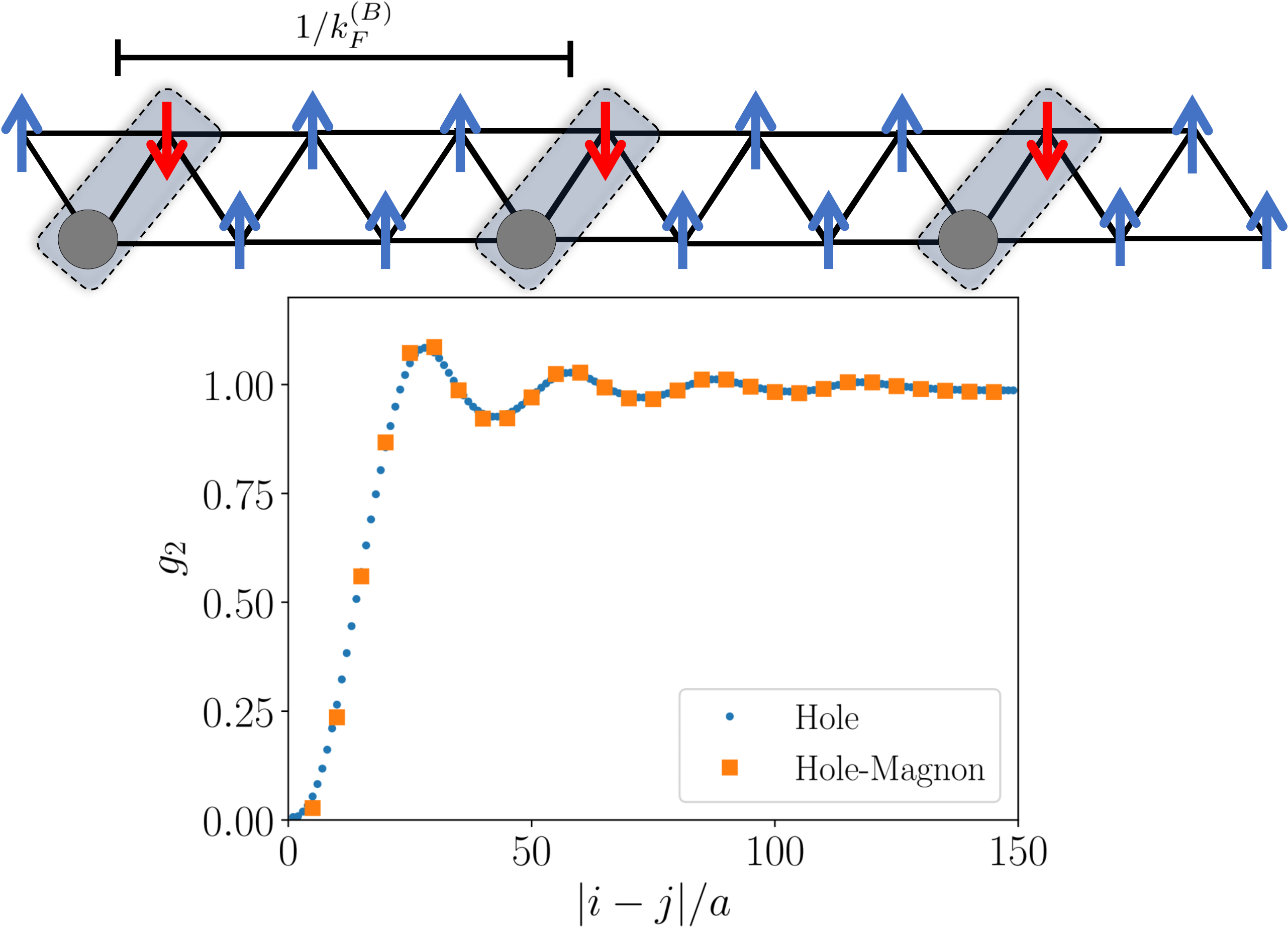}
	\caption{(a) Schematic representation of a self-organization of hole-magnon pairs in the fermionic zigzag ladder. Hole-magnon pairs form a Luttinger liquid with a characteristic Fermi momentum $k_F^{(B)}=\pi n_B$ given by the density of pairs $n_B$. (b) Pair correlation function of single holes and hole-magnon pairs in the fermionic zigzag ladder for $J/t=0$ and $N_h=N_m=10$.}
	\label{fig:PairCorrelation}
\end{figure}

{\it Fermions in a zigzag ladder}.
Kinetic frustration of holes in a zigzag ladder  gives rise to formation of several types of multi-particle magnon-hole bound states, such as shown in Figs.~\ref{fig:HolMagRepr} and~\ref{fig:PhasDiagr}. The nature of stable composite objects can be tuned by varying the ratio of interaction to single fermion hopping and the relative concentration of holes and magnons, see Sec.~\ref{Sec3}. We find large binding energies of composites, since attraction is driven by the kinetic energy of holes rather than the superexchange energy setting the amplitude of spin dynamics. For example, the binding energy of 1H1M (one hole and one magnon) can be as large as $0.6 t$, where $t$ is the fermion hopping. We also point out that the largest binding energy of the 1H1M pair is achieved in the limit of vanishing superexchange interactions, which corresponds to large on-site repulsion of two particles. Fig.~\ref{fig:PhasDiagr} shows several cuts through the phase diagram for simple ratios of the magnon and hole densities. 
We show that these bound states have a common tendency to exhibit antiferromagnetic correlations around the hole positions. In the case of a single hole, most of the magnons accumulate around the hole forming an antiferromagnetic region and the rest are pushed out of this region and form a spiral winding with a net $S^z$ magnetization as shown in Fig.~\ref{fig:AntiferroSB1}. 
When multiple composite objects are present in the system their statistics and interaction between them determine the nature of the ground state. 
In Fig.~\ref{fig:PairCorrelation} we report the pair correlation function of single holes and hole-magnon pairs. We observe that one closely follows the other one indicating that each hole in the system is paired with a magnon thus creating a fluid of hole-magnon pairs. In Sec.~\ref{Sec4a} we show that this fluid corresponds to a Luttinger liquid of fermionic 1H1M bound states, and in Sec.~\ref{Sec4b} we discuss the pair density wave state of 2H3M composites.

{\it Fermions and bosons in a square ladder with a flux}.
\newline
Quantum particles on a square ladder in the presence of static gauge field also experience frustration of kinetic energy as shown in Sec.~\ref{Sec6}. Optical lattices with synthetic gauge fields
have been realized experimentally for both fermions and bosons~\cite{Atala2014,PhysRevLett.111.185301,PhysRevLett.111.185302,PhysRevLett.108.225304,Jotzu2014,PhysRevLett.108.225303}, including square flux ladders~\cite{Atala2014,Fukuhara2013}. In Fig.~\ref{fig:BindingFluxLadder} we present results for the binding energy of 1H1M for different interaction strengths and different flux per plaquette. We find the largest binding energy of around $0.3t$ in the regime of vanishingly small superexchange energy. 

{\it Mott state of Bose atoms, antibound states}.
An important feature of hole-magnon systems for all lattice geometries is a  symmetry between the slightly doped Mott states of bosons and fermionic models. The low energy states of the hole-magnon Hamiltonian in the fermionic system can be mapped into the high energy states of the bosonic model, see Sec.~\ref{Sec4} for a detailed examination. This correspondence works in the regime of strong repulsion between particles, when we can exclude real ``doublons'' and describe both systems using $t$-$J$ type Hamiltonians.  The change of sign of the kinetic energy of holes in the two cases arises from the fact that a hole creation operator is an annihilation operator of the original particle. Thus the hole hopping term  has the opposite sign relative to the hopping of the original particle for fermions and the same sign for bosons (for details see discussion in Sec.~\ref{Sec2}). Superexchange interactions in the two systems also differ in sign. Bosonic systems have ferromagentic exchange interaction~\cite{PhysRevLett.91.090402} (we assume SU(2) symmetric repulsion between particles) while fermions have antiferromagentic superexchange interactions~\cite{PhysRev.79.350,Hofstetter2002}. The consequence of the ${\cal H}$ to $-{\cal H}$ mapping is that bound hole-magnon states of fermionic systems correspond to antibound states in the case of bosons. These antibound state can be understood as analogues of repulsively bound states observed in experiments in~\cite{Winkler2006} with one important differences that they are non-local, i.e. the bound ``elementary'' particles in our systems (i.e. magnons and holes) do not occupy the same site. It is worth noting that the symmetry between bound and antibound states applies not only to the basic 1H1M pairs but also to multi-particle clusters.

\begin{figure}[t!]
	\centering
	\includegraphics[width=1\columnwidth]{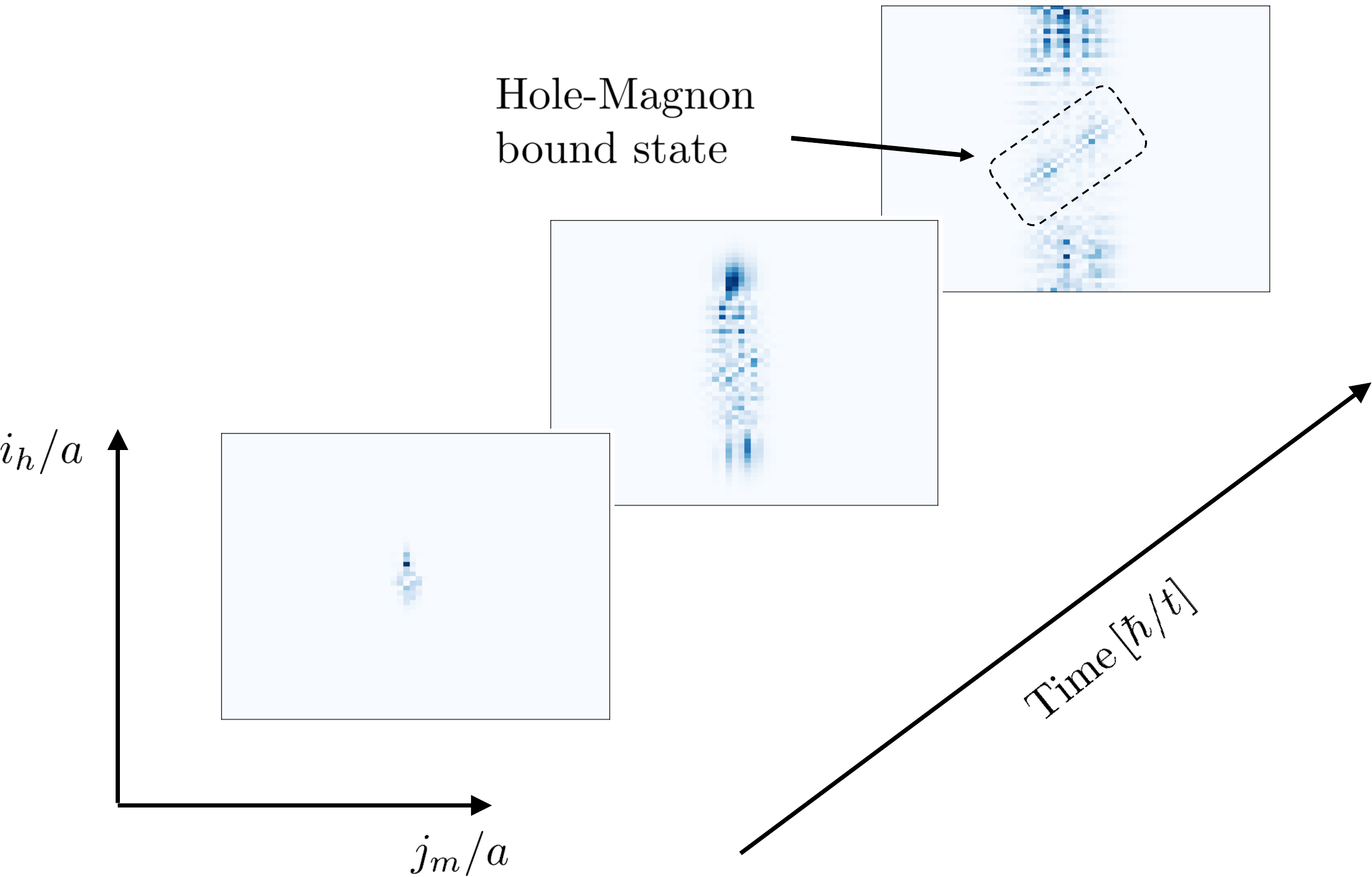}
	\caption{Dynamics of a hole-magnon pair in the zigzag ladder. Hole-magnon correlation function $\langle \hat{n}^h_i \hat{n}^m_j \rangle$ computed at different times for $J/t=0$. The initial state corresponds to putting a hole and a flipped spin in adjacent sites in the middle of the lattice. At long times a diagonal part can be discerned showing the correlated expansion of the hole and the magnon bound together. }
	\label{fig:DynamicCorr}
\end{figure}

{\it Dynamical probes of binding}.
An efficient method of exploring bound and antibound states of many-body systems is to use coherent quantum dynamics. To be concrete we consider the case of a 1H1M pair, however, this discussion can be generalized to arbitrary nHmM composites. The idea is as follows:  use external local potential and/or local optical spin control to prepare an initial state in which a hole and a magnon are localized on neighboring sites.   Switch off the localizing potential and let the system evolve under the many-body Hamiltonian. When there are no bound (antibound) states, the two particles will spread out essentially independently and the probability to find them next to each other should eventually decrease to zero (in a finite system it will saturate to a finite value set by inverse of the system size). On the other hand, when there is a bound (or antibound) state in the spectrum, we will find a finite probability that the magnon and the hole spread out together. The probability to find the two particles together after a long evolution time is given by the overlap between the bound state wavefunction and the initial state. In Fig.~\ref{fig:DynamicCorr} we show dynamics in the zigzag ladder starting from the initial state in which one hole and one magnon have been prepared on neighboring sites. We observe that the joint probability distribution function remains peaked where the two particles are close to each other. In Sec.~\ref{Sec5} we present a detailed analysis of the quantum dynamics of a hole-magnon pair in the fermionic zigzag ladder. The hole-magnon correlation at different times could be measured in current ultracold atom experiments with a quantum gas microscope~\cite{Bakr2009,Sherson2010,PhysRevLett.114.213002,PhysRevLett.114.193001,PhysRevLett.115.263001,Haller2015,PhysRevA.92.063406,Greif953,Brown1385,Weitenberg2011,Schauss}. 
A similar experimental technique has been used previously  in the Mott insulating regime of two component bosons to demonstrate the existence of two magnon bound states in the Heisenberg model in one-dimension~\cite{Fukuhara2013}.

\section{Microscopic model}
\label{Sec2}
\subsection{Kinetic frustration}
\begin{figure}[t!]
	\centering
	\includegraphics[width=1\columnwidth]{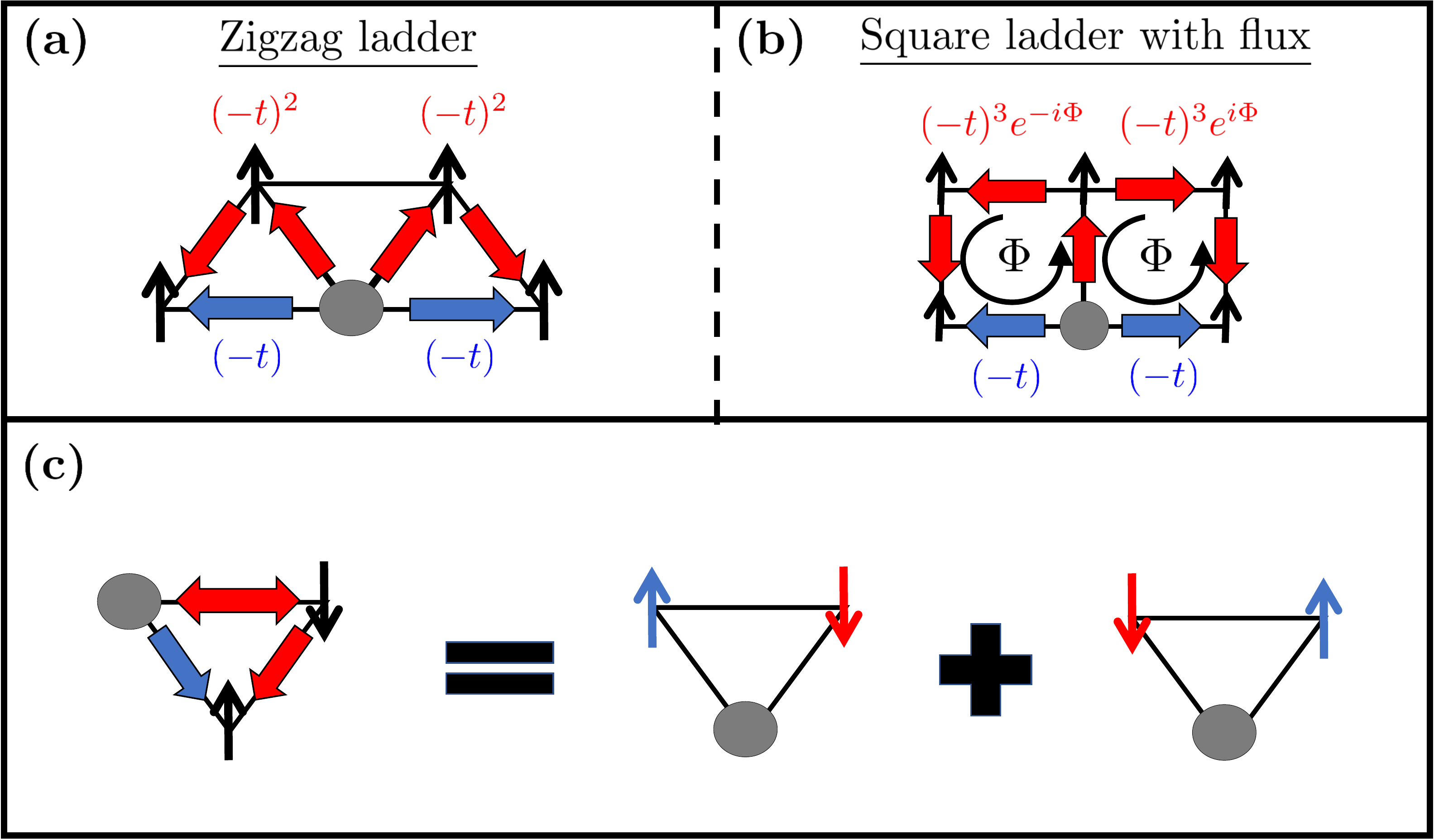}
	\caption{A single hole injected into a fully polarized insulating state exhibits kinetic frustration arising from destructive interference of different paths. This occurs for non-bipartite geometries such as the zigzag ladder (a) or for bipartite ones in the presence of magnetic flux (b). When a flipped spin is added into the system (c), the two different paths cannot interfere since they lead to different final states. These two final states correspond to the permutation of the two spins.}
	\label{fig:KineticFrustration}
\end{figure}

While the most familiar mechanism of bound state formation is attractive interaction between particles, kinetic frustration provides an alternative route.  
Kinetic frustration arises in lattice systems when different quantum paths of a propagating particle interfere destructively, see Fig.~\ref{fig:KineticFrustration} panel (a) and (b). In this situation the particle is not able to obtain the full kinetic energy corresponding to the number of nearest neighbor sites in a non-frustrated case. This frustration can be relieved by the presence of another distinguishable particle, see Fig.~\ref{fig:KineticFrustration} panel (c). When the original frustrated particle moves from the initial to the final point along the first trajectory, the second particle remains fixed at the same position. For the other path the ``frustrated'' particle moves from the same initial to the same final point, but its motion causes the second particle to move to a different site. For the two trajectories the second particle ends up in different positions, hence, these paths do not interfere and the ``frustrated'' particle can gain more kinetic energy. Moreover, since alleviation of the kinetic energy frustration is greater for shorter paths, it is energetically favorable for the two particles to stay close to each other in space.

The primary goal of this paper is to present several examples of  kinetic frustration giving rise to formation of multi-particle bound states in two component Bose and Fermi mixtures.

\subsection{$t$-$J$ models}
The $t-J$ model is given by,
\begin{equation}
\hat{H}_{t-J}=\hat{\mathcal{P}}\left[-t \sum_{\langle i,j\rangle,\sigma} \left(\hat{c}^{\dagger}_{i\sigma}\hat{c}_{j\sigma} +\text{h.c.} \right) \right] \hat{\mathcal{P}}+ J\sum_{\langle i,j\rangle}\mathbf{S}_i\mathbf{S}_j, 
\label{Eq:t-Jmodel}
\end{equation}
where $\hat{c}^{\dagger}_{i\sigma}$ creates a fermion or a boson at site $i$ with spin $\sigma = \uparrow,\downarrow$ and $\hat{\mathcal{P}}$ project onto the subspace of no double occupancies. The spin operators are defined by $S_i^{\gamma}=\sum_{\alpha,\beta} \hat{c}^{\dagger}_{i\alpha} \sigma_{\alpha,\beta}^{\gamma} \hat{c}_{i\beta}$ where $\sigma^{\gamma}$ are the Pauli matrices with $\gamma=x,y,z$.
The sum $\sum_{\langle i,j\rangle}$ is taken over first neighbors. 

The $t-J$ model contains two energy scales: the quantum tunneling (hopping) of particles $t$ and the superexchange $J$. The hopping $t$ gives rise to the kinetic energy of holes, and the superexchange term describes propagation and interaction of magnons. 
The $t-J$ model~\eqref{Eq:t-Jmodel} is an effective low energy model for the Hubbard model at very strong Hubbard interaction $U$ with $J= \pm 4t^2/U$, where $+$ ($-$) corresponds to the fermionic (bosonic) case~\cite{PhysRevLett.91.090402,Trotzky295,Brown540,Nichols383,Jepsen2020}. 

In order to study kinetic frustration in the $t-J$ model~\eqref{Eq:t-Jmodel} we have to specify the underlying lattice. 
Kinetic frustration can be easily achieved for fermions in lattice geometries where the number of legs enclosed in a minimal closed loop is odd. Therefore ladders are promising candidates where this condition can be satisfied. The minimal lattice fulfilling that condition is a zigzag ladder, in which a unit cell corresponds to two stacked triangles, see Fig.~\ref{fig:HolMagRepr} (a). Another promising platform for realizing a one-dimensional model with kinetic frustration is flux ladder systems, which have been recently realized in experiments with ultracold atoms~\cite{Atala2014,Tai2017}. In this system a static synthetic gauge field is used to provide the analogue of the Aharonov-Bohm flux of the ordinary magnetic field in condensed matter systems. A particularly attractive feature of such systems is the possibility of reaching high flux density per unit cell. In order to numerically simulate such systems we have performed exact diagonalization simulations for the cases of one hole and one magnon and density matrix renormalization group (DMRG) simulations for systems with more holes and magnons.

{\it Single particle dispersion for  the zigzag ladder}.
\newline
The dispersion relation for a single hole in a fully polarized zigzag ladder is given by $\epsilon_h(k)/t=\pm \{\cos(k)+\cos(2k)\}$, where $+$ ($-$) corresponds to the fermionic (bosonic) case. For fermions kinetic frustration manifests in the ground state of the hole since its ground state energy is given by $\epsilon_h(k_0)=-2.25 t$ and $k_0=\arccos(-1/4)$. This ground state energy is higher than the kinetic energy we would expect from counting the number of nearest neighbors $-4t$. For bosons kinetic frustration appears not in the ground state but for the highest energy single particle states. In this situation the highest energy obtained $2.25t$ is lower than the expected value of $4t$.
To understand this effect, one can change the sign of hopping $t \rightarrow -t$, so that the highest energy states of the positive $t$ model correspond to the lowest energy states of the negative $t$ model. Then one observes that in the negative $t$ case there is a relative minus sign between the two trajectories around a single triangle (blue and red paths in Fig.~\ref{fig:KineticFrustration}), which implies a destructive interference and thus a reduction of the kinetic energy gain due to hopping. 
This will play a crucial role when studying bound states of holes and magnons for the fermionic and bosonic systems.

{\it Single particle dispersion for  the square ladder with a flux}
\newline
For a square flux ladder we consider propagation of a single hole on top of a fully polarized insulating state. In this situation the dispersion relation of the hole reads $\epsilon_{\pm}({k})/t=-2\cos(\phi/2)\cos(k)\pm \sqrt{1 + 4\sin(\phi/2)^2 \sin(k)^2}$ where $\phi$ corresponds to the magnetic flux per plaquette (see e.g.~\cite{Tai2017}). The minimum of the band ($-$) is at $k_m=0$ ($k_m=\pm\sin(\phi/2)\sqrt{1-\cos(\phi/2)/(4\sin(\phi/2)^4)}$) for $\phi<\phi_c$ ($\phi>\phi_c$) where $\phi_c\approx 0.43\times \pi$, respectively. In this case two different bands are present $(\pm)$ because the system can be thought of as two coupled 1D systems. It is easy to check that the kinetic energy is always higher than the non-frustrated situation $\epsilon_{\pm}({k})>-3t$ for all momenta when the magnetic flux is turned on. Moreover, the spectrum is symmetric with respect to a change of sign of $t$, shifting the momentum by $\pi$, and exchanging the two bands ($\pm$). This symmetry shows that in this case frustration does not arise from the sign of the bare $t$ (recall that hole hopping has different signs for the bosonic and fermionic systems) but originates from the magnetic flux $\phi$ and appears for both lowest and highest energy single particle states. As a consequence, the dispersion of a single hole is the same for bosonic and fermionic systems.

\section{Bound States in fermionic Zigzag  ladder}
\label{Sec3}
The kinetic energy of a hole gets modified when one of the possible paths contains a magnon. This induces an effective interaction between the hole and the magnon, which can be either repulsive or attractive depending on the lattice geometry and particle statistics. Surprisingly this has implications beyond the formation of bound hole-magnon pairs. In this section we demonstrate that as one varies $J/t$ and relative concentrations of holes and magnons, different types of composite objects are formed. We will focus on the case of a fermionic zigzag ladder and discuss other systems in subsequent sections.

\subsection{Overview of bound states in fermionic zigzag ladder}
We present the phase diagram of multi-particle bound states in the fermionic zigzag ladder in Fig.~\ref{fig:PhasDiagr}. We note that we perform calculations at fixed ratios of the hole to magnon densities, and with system sizes that are much larger than the number of holes and magnons. Thus the phase diagram shown in Fig.~\ref{fig:PhasDiagr} should be understood as corresponding to the limit of vanishingly small concentrations of holes and magnons, with tuning parameters being $J/t$ and the ratio between the densities of holes and magnons. Going to finite density of dopants may result in  stabilization of other phases, which we plan to address in future work. 

To understand the results presented in Fig.~\ref{fig:PhasDiagr} it is useful to consider cuts through the phase diagram at fixed $N_m/N_h$ as we vary $J/t$. We find that generally large $J/t$  favors the decomposition of hole-magnon bound states into free holes and magnons. By contrast, for small $J/t$ we find stabilization of large composite objects. For the case $N_m/N_h=1$ we find that the bound hole-magnon state appears only when $J/t<2$. The origin of the suppression of hole-magnon binding for larger values of $J$ is discussed in the next subsection. We do not find formation of larger composites, which suggests that hole-magnon pairs repel each other for equal and small densities of holes and magnons. 

For $N_m/N_h=3/2$ we observe that at $J/t\sim 0.5$ a trimer bound states 1H2M is formed. Between $J/t\sim 0.3$ and $J/t\sim 0.5$ we find that hole-magnon pairs and trimers coexist, which suggests repulsive interaction between trimers and pairs. For smaller values $J/t<0.3$ a pentamer 2H3M composed of two holes and three magnons is formed. This indicates that the attraction between holes and magnons induces an effective attraction between two holes. This is the smallest bound state that we have found which includes a pair of holes. For $N_m/N_h=2$ we find a similar scenario but the hole-magnon trimer appears for a wider range of $J/t$ and an hexamer 2H4M, which includes a pair of holes, appears for a narrow range of $J/t$. In Sec.~\ref{Sec4} we discuss the many-body phases associated with the pentamer and the hole-magnon pair. 

Finally we study a single hole immersed in a fluid of magnons: this is the $N_m/N_h=\infty$ line in Fig.~\ref{fig:PhasDiagr}. By lowering the superexchange interaction $J/t$  we observe that the number of magnons bound to the hole increases. Magnons bound to the hole form a finite size cloud around the hole, a phenomenon that we will refer to as formation of a ``spinbag''. Within the spinbag region we observe strong antiferromagnetic correlations of the XY spin components between the two legs of the ladder as shown in Fig.~\ref{fig:AntiferroSB1}. 
This feature is a counterpart of the ferromagnetic spinbag (Nagaoka polaron) found in bipartite lattices~\cite{NAGAEV199239,alexandrov2007polarons,auerbach1994interacting,PhysRevB.64.024411}. Furthermore, similar to the Nagaoka polaron in the limit $J/t\rightarrow 0$ the bag size becomes of the order of the system size, which in our case indicates that the full system will exhibit antiferromagnetic correlations of the transverse spin components. Antiferromagnetic correlations surrounding the hole alleviate its kinetic frustration by making all possible paths distinguishable, thus lowering the energy. In this way, hole-magnon binding provides an explanation to the origin of antiferromagnetism in non-bipartite lattices~\cite{PhysRevLett.95.087202,PhysRevLett.112.187204}.

\subsection{Analysis of a 1H1M pair}
\begin{figure}[t!]
	\centering
	\includegraphics[width=1\columnwidth]{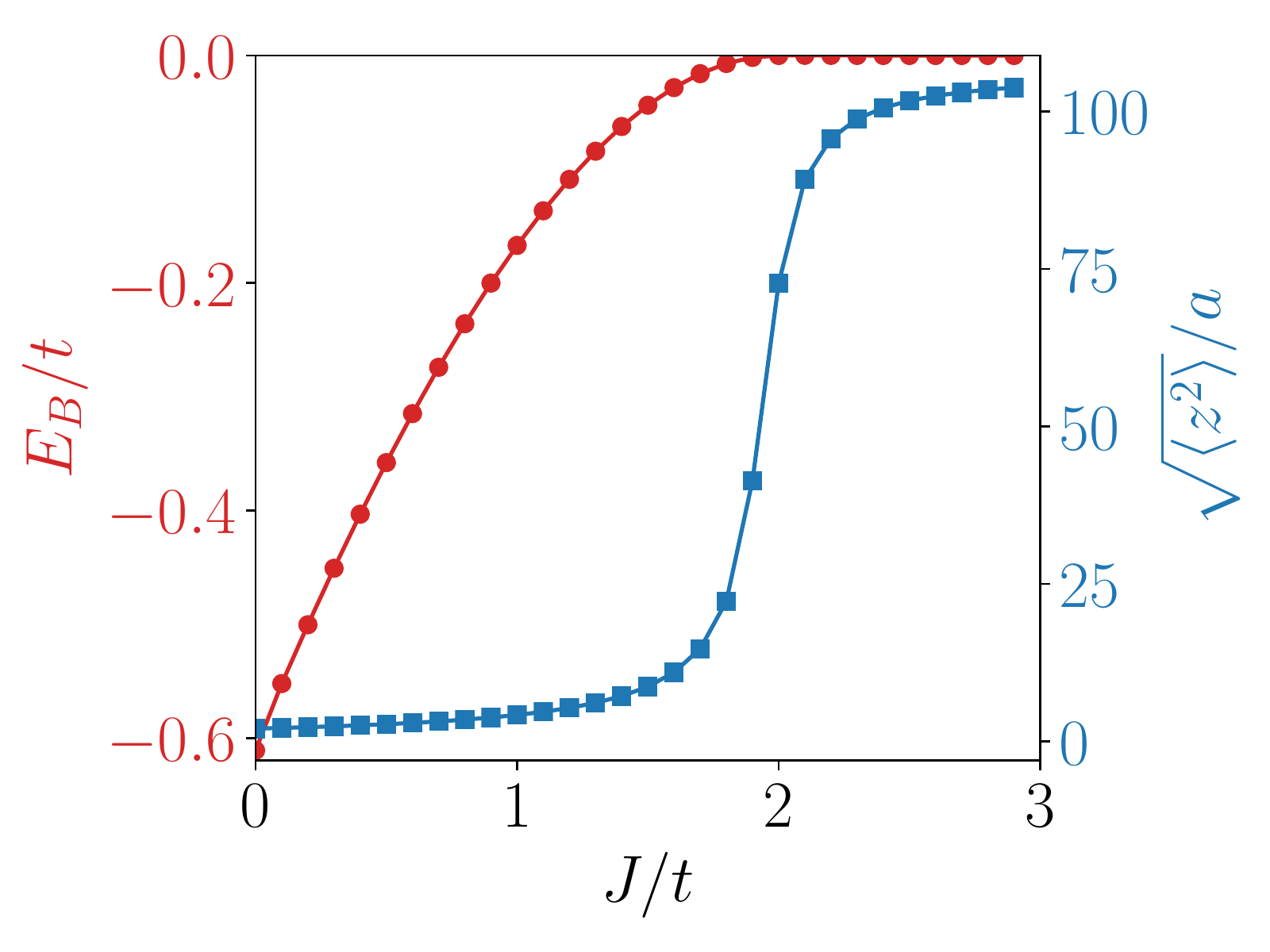} 
	\caption{Binding energy $E_B/t$ of the hole-magnon pair (red dots) on the fermionic zigzag ladder obtained in the limit of having an infinite ladder by performing a finite size scaling (see App.~\ref{AppendixA}) and the hole-magnon pair extension (blue squares) $\sqrt{\langle z^2 \rangle}/a$ for a ladder of $2\times200$ sites.}
	\label{fig:En_ext}
\end{figure}
In order to solve the hole-magnon problem of the fermionic $t-J$ model Eq.~\eqref{Eq:t-Jmodel} in a zigzag ladder we first map the ladder to a one-dimensional system in which the sites are indexed as shown in Fig.~\ref{fig:AntiferroSB1}. Then we pass from the original particle operators to operators of holes and magnons. In this way we reduce the problem to a one-dimensional two-body problem, see~\cite{Suppl} for a detailed examination.
Then the effective two-body problem can be solved and all the hole-magnon bound state properties can be obtained. We set the energy of the fully polarized state with one fermion on every site to be zero, i.e. we measure energies of all states relative to the fully polarized band insulating state.
The hole-magnon binding energy is defined as $E_B\equiv E_{1H1M}-E_{1H}-E_{1M}$ where $E_{1H1M}$, $E_{1H}$ and $E_{1M}$ are the groundstate energies for the hole-magnon, the single hole, and the single manon states respectively. We find the largest binding energy when $J/t=0$, see Fig.~\ref{fig:En_ext}. This demonstrates that kinetic frustration provides an effective attraction between a hole and a magnon. At the same time the superexchange interaction $J$ provides an effective repulsion between them. To see this effect one can consider the difference of superexchange energies of having a hole and a magnon at adjacent sites relative to the case of being at larger distances, $E_{J,\text{adjacent}}-E_{J,\text{separated}}=J/2$. This shows that in the fermionic system with $J>0$ having a hole and a magnon on adjacent sites increases the exchange energy. At $J/t=2$ the repulsion exactly cancels the attraction and the hole-magnon pair vanishes.
\begin{figure}[t!]
	\centering
	\includegraphics[width=1\columnwidth]{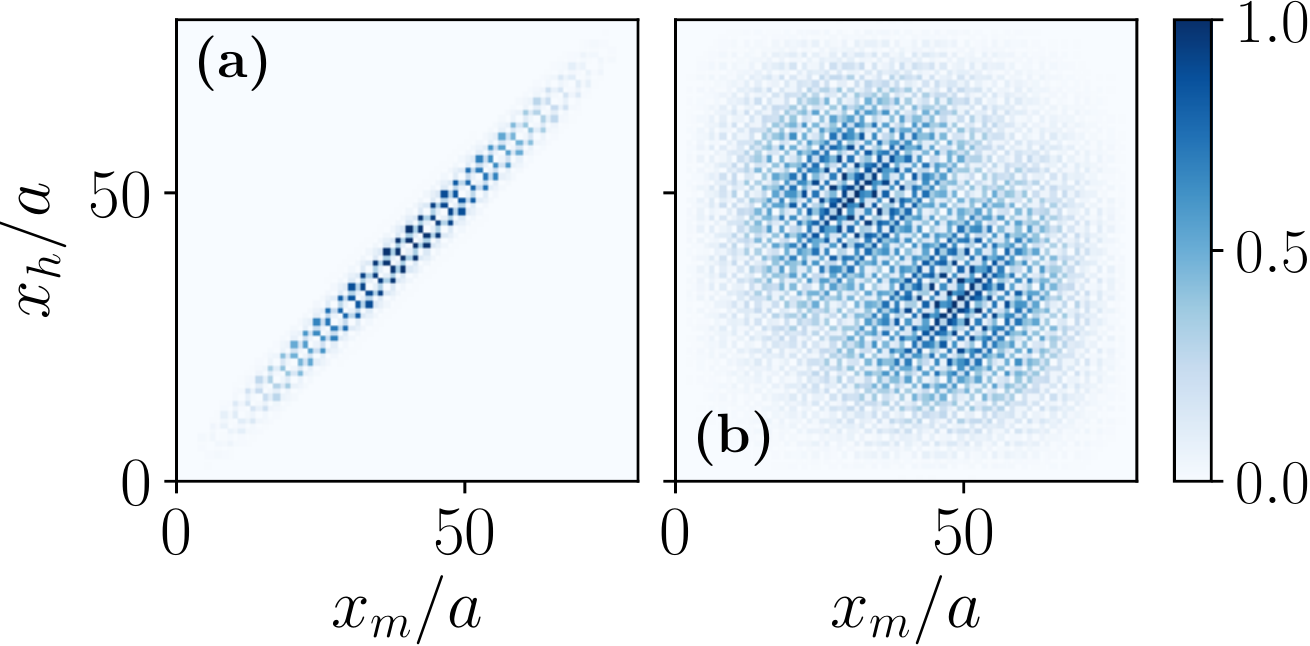} 
	\caption{Hole-magnon correlation function on the fermionic zigzag ladder for the case of $N_h=N_m=1$ and different values of the superexchange interaction $J/t=0.1,2.2$ for (a) and (b), respectively.}
	\label{fig:Corr1H1M}
\end{figure}
The hole-magnon bound state is characterized by having the two particles close to each other in real space. By construction they are hardcore particles because they cannot occupy the same site. By defining a relative distance between the hole and the flipped spin $z$ we compute the size of the bound state relative to the size of the system $\langle z^2 \rangle/\langle L^2\rangle$, see App.~\ref{AppendixA}. In the regime of strongest binding, $J/t\rightarrow 0$, the size of the bound state is much smaller than the system size and in fact, independent of it. With increasing $J$ and decreasing binding energy, the bound state expands (see Fig.~\ref{fig:En_ext}). At $J/t=2$ the binding energy goes to zero and the size of the bound state reaches the system size. Beyond that it becomes proportional to the system size, indicating a transition from a bound to a non-bound state (see App.~\ref{AppendixA} for a detailed finite size scaling analysis of the size of the hole-magnon composite object).
The hole-magnon correlation function $\langle n^h_i n^m_j \rangle$ can be used to see the real space structure of this bound state. For small values of $J/t$ we see that the positions of the magnon and the hole are staying close to each other, see Fig.~\ref{fig:Corr1H1M}. Moreover the correlation function has a peak on adjacent sites which also indicates a small spatial extent of the bound state. With increasing $J/t$ the distance between the hole and the magnon begins to increase and positions of the hole and the magnon become less correlated. These results indicate that analysis of the hole-magnon correlation function should provide direct experimental evidence of the bound state formation.

\subsection{Analysis of the 1H2M trimer}
\begin{figure}[t!]
	\centering
	\includegraphics[width=1\columnwidth]{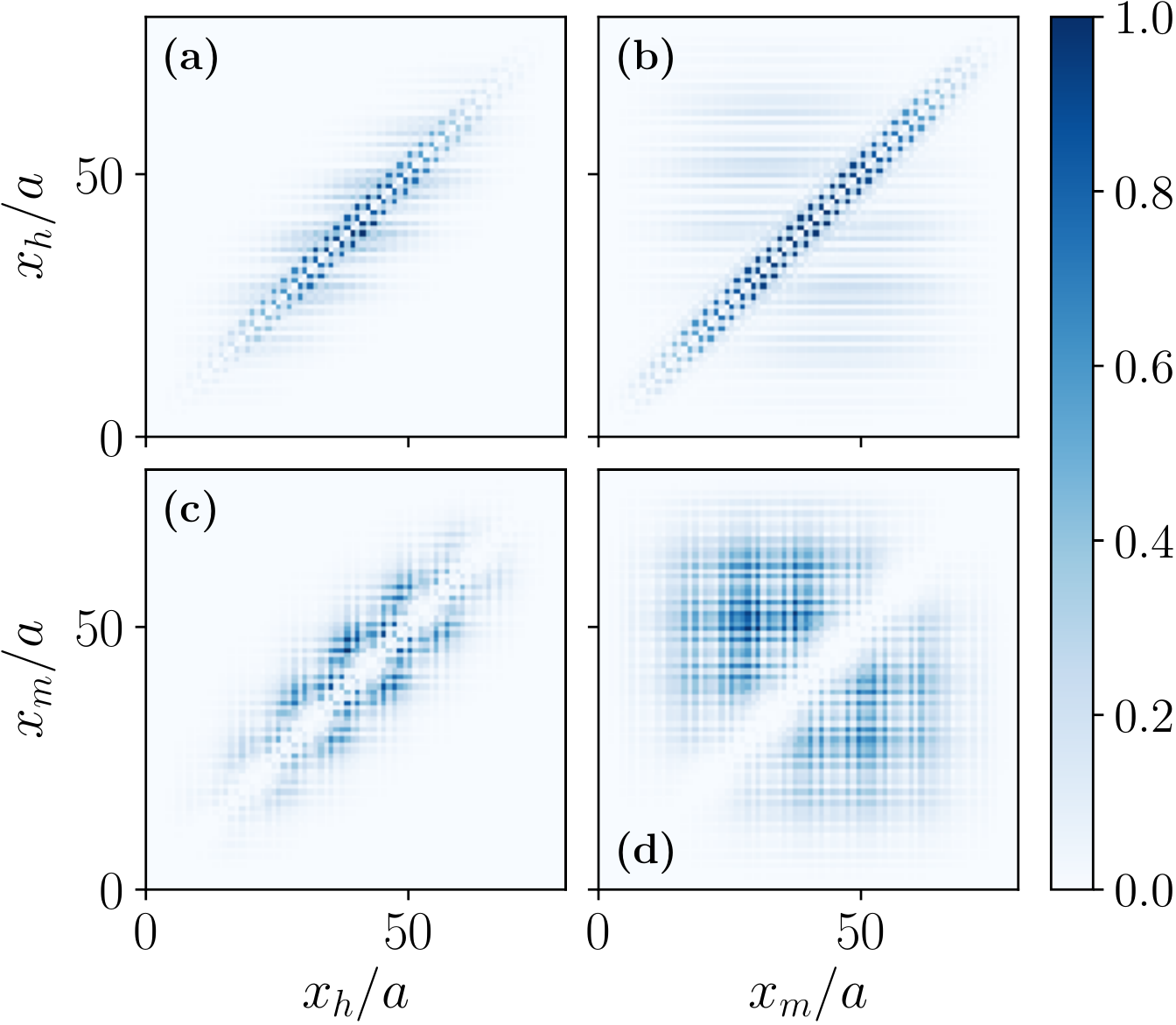} 
	\caption{Hole-magnon, (a) and (b), correlation function and magnon-magnon correlation function, (c) and (d), on the fermionic zigzag ladder for $N_h=1$, $N_m=2$ and two values of the superexchange interaction $J/t=0.2,0.6$.}
	\label{fig:Corr1H2M}
\end{figure}
In the previous subsection we discussed that superexchange interaction gives rise to an effective repulsion between a hole and a magnon, which competes with attraction arising from kinetic frustration. The relative strength of these opposing interactions determines the transition between trionic and pair binding on the $N_m/N_h=2$ line. For small values of $J/t$ the ground state of the system has 1H2M trimers. As $J/t$ increases, the system undergoes a transition in which trimers dissociate into 1H1M pairs and free magnons. Numerically we identify this transition by computing $E_B=E_{1H2M}-E_{1H1M}-E_{1M}$. When the binding energy $E_B$ is negative, a trimer state is stable. When $E_B$ is positive, trions become unstable to dissociating into hole-magnon pairs and free magnons. We observe that this transition occurs around $J/t\sim 0.5$. Correlation functions can also be used to detect the appearance of a trimer. Since these trimers contain two magnons, these bound states can be revealed by observing two magnon bunching in the correlation function $\langle n^m_i n^m_j\rangle$. On the other hand, for large values of $J/t$, when the system decomposes in a pair and a free magnon, the magnon-magnon correlation function shows that the two magnons are separated by a large distance comparable to the system size, see Fig.~\ref{fig:Corr1H2M}. Moreover the hole-magnon correlation function always shows that there is a magnon close to a hole (see the enhanced probability close to the diagonal $i_h=j_m$ line in Fig.~\ref{fig:Corr1H2M} upper row). This can be understood from the observation that for both trimers and 1H1M pairs, we should find a hole and a magnon close to each other.

\subsection{2H2M: Effective repulsive interactions}
\begin{figure}[t!]
	\centering
	\includegraphics[width=1\columnwidth]{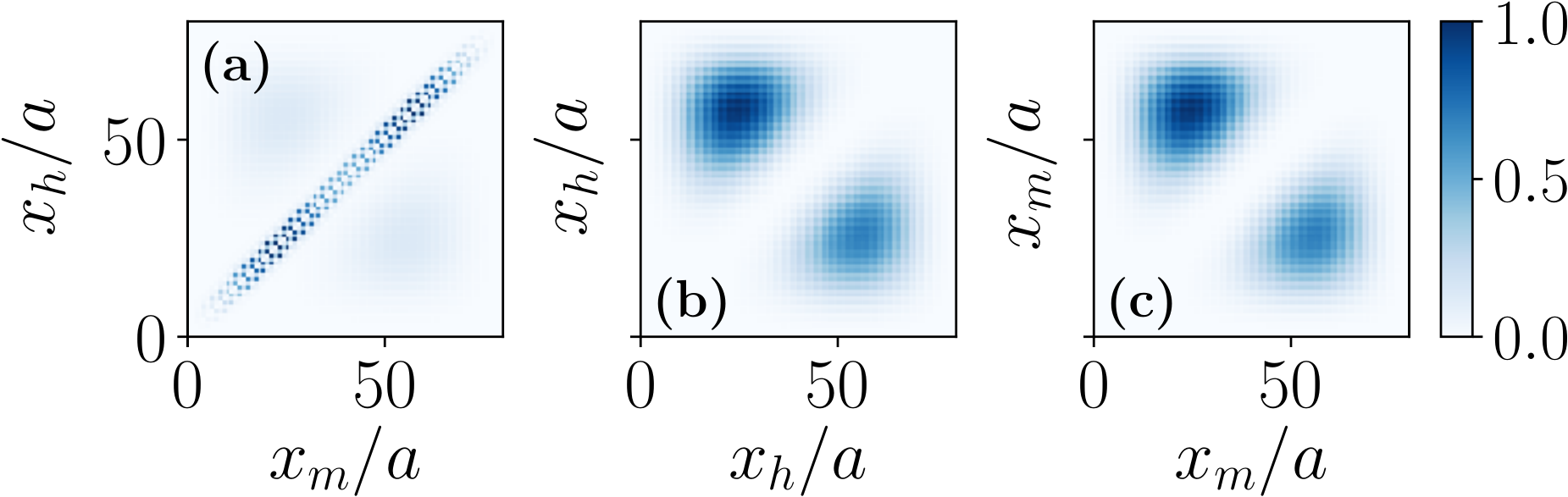} 
	\caption{(a) Hole-magnon, (b) hole-hole and (c) magnon-magnon correlation functions on the fermionic zigzag ladder for $N_h=2$, $N_m=2$ and $J/t=0.1$.}
	\label{fig:Corr2H2M}
\end{figure}
When multiple hole-magnon pairs are present in the system effective interactions between them appear. In particular, magnons can mediate effective interactions between holes. We analyze whether these effective interactions are enough to bind a pair of holes in zigzag ladders with nearly full spin polarization.
Surprisingly we do not find hole pairs for the cases of $2H1M$ and $2H2M$. In the case of two holes and two magnons we observe that the system decomposes into two hole-magnon pairs. The binding energy $E_{2H2M}-2E_{1H1M}$ approaches zero when we increase the system size for any value of $J/t$. This indicates that there is an effective repulsive force between hole-magnon pairs. By computing the hole-hole and magnon-magnon correlation we observe that the two holes and the two magnons are separated in space, see Fig.~\ref{fig:Corr2H2M}. On the other hand, the hole-magnon correlation still shows that a hole and a magnon sit close together. This indicates that the system forms two pairs which then stay away from each other.

\subsection{2H3M: Hole pairing}
\begin{figure}[t!]
	\centering
	\includegraphics[width=1\columnwidth]{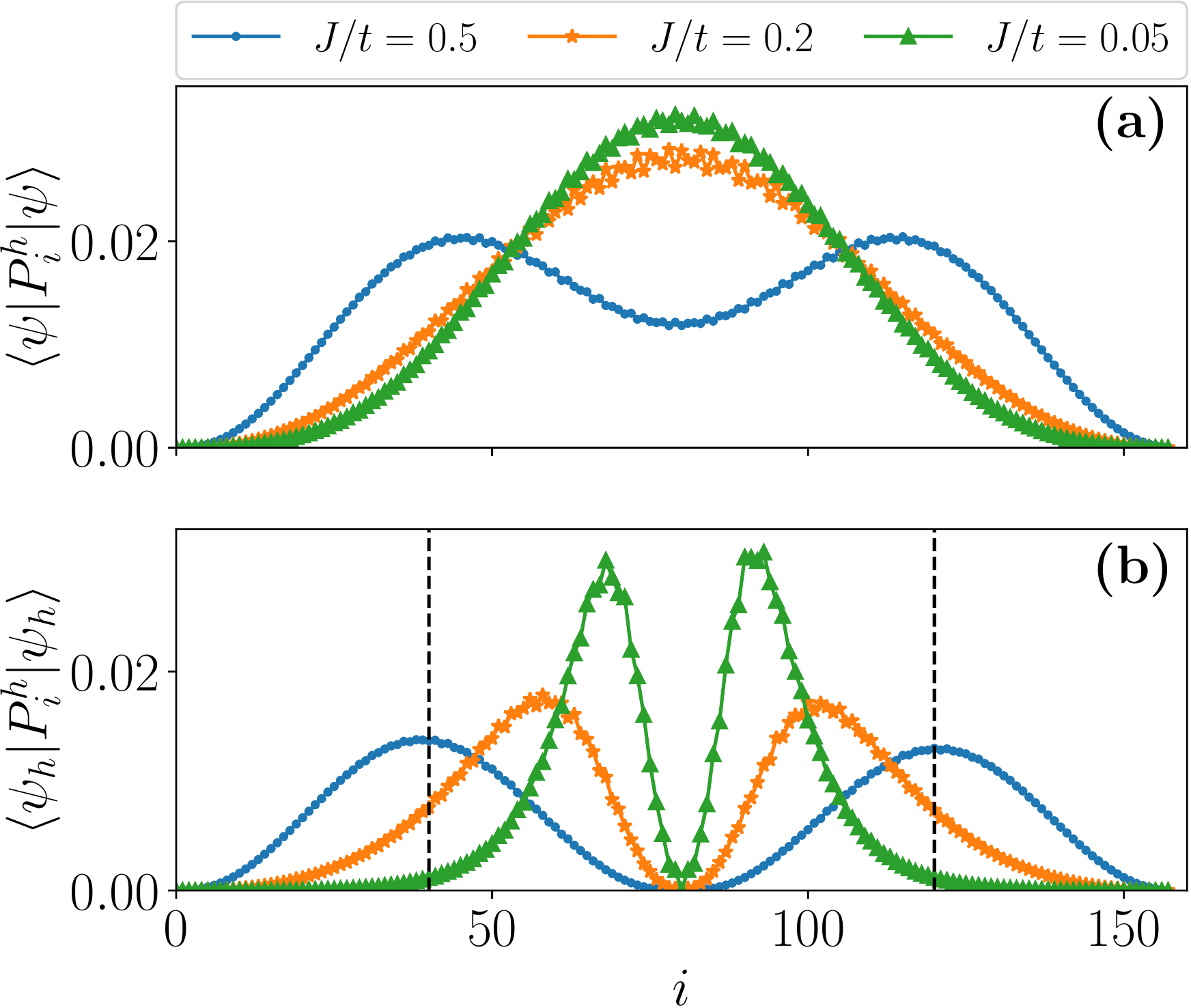}
	\caption{Hole density (a) for the case of $N_h=2$, $N_m=3$ and different values of the superexchange interaction on the fermionic zigzag ladder. Hole density of the projected state $|\psi_h\rangle$ (b) which has a fixed hole in the middle of the lattice, see main text. We removed the central point to improve visibility.}
	\label{fig:HolPair}
\end{figure}
In the zigzag ladder the smallest bound state containing a pair of holes is a pentamer formed by two holes and three magnons. We checked that the binding energy of the pentamer is negative with respect to all possible decompositions into smaller objects when $J/t<0.3$. We observe however that the binding energy with respect to the decomposition into a trimer and a pair is small. Therefore, although the pentamer is the groundstate of the system, already small temperatures will lead to the breaking of a pentamer into a 1H1M pair and a 1H2M trimer. In the remainder of this subsection we will focus on the properties of the ground state of the system. In order to detect the appearance of hole pairs we measure the hole density, see Fig.~\ref{fig:HolPair}. For large values of $J/t$ the hole density exhibits two humps indicating that the holes are separated by a large distance. With decreasing $J/t$ the two humps begin to approach to each other and finally fuse into a single one. After the fusion the two holes share the same region of space forming the pentamer together with three magnons. In order to quantify how far apart the two holes are, we fix the position of the first hole and compute the spatial distribution of the probability density of the second hole. Technically this is done by first projecting the groundstate wavefunction into the state which contains a hole in the center of the lattice $|\psi_h\rangle \equiv \hat{P}^h_L |\psi\rangle = \hat{c}_{L\uparrow}\hat{c}^{\dagger}_{L\uparrow}\hat{c}_{L\downarrow}\hat{c}^{\dagger}_{L\downarrow}|\psi\rangle$ and then computing the density of the second hole $\langle \psi_h | \hat{P}^h_i | \psi_h\rangle$~\cite{PhysRevB.55.6504}. We also normalize the state $\langle \psi_h|\psi_h\rangle=1$. For large values of $J/t$ this probability has maxima at the positions $L/4$ and $3L/4$, which is the expected result for two hardcore particles. When reducing $J/t$ the positions of the two maxima start to approach the center of the lattice. This shows that the two holes start approaching each other, which indicates effective attraction between them and formation of a bound state. 
As a function of $J/t$ the distance between the two maxima exhibits a minimum value of $2\times 10$ sites at $J/t\sim 0.05$, see Fig.~\ref{fig:HolPair}.
We conclude that the pentamer exists but it is a loosely bound object with a relatively large size.

\subsection{1HNM: Antiferromagnetic spinbag}
\begin{figure}[t!]
	\centering
	\includegraphics[width=1\columnwidth]{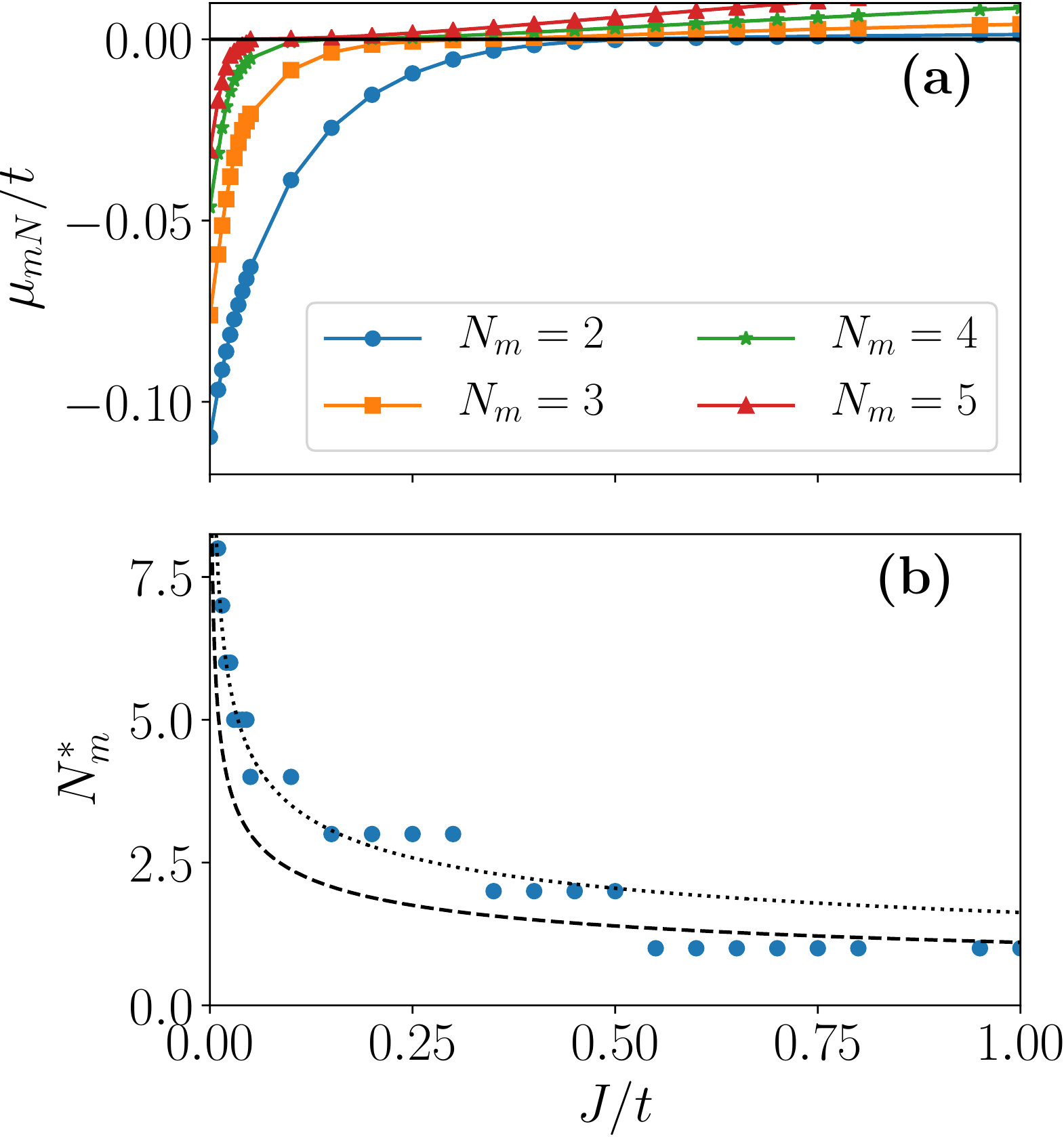} 
	\caption{Magnonic affinity energy $\mu_{mN}/t$ on the fermionic zigzag ladder (a) as a function of the superexchange interaction for the cases with $N_h=1$ and different numbers of magnons. Number of magnons forming the antiferromagnetic spinbag $N_m^*$ (b) as a function of the superexchange interaction $J/t$. Dashed line represents the analytical result for the antiferromagnetic spinbag $N_m^* \propto 1.10 (t/J)^{1/3}$, see App.~\ref{AppendixB}. Dotted line is the fit to numerical data which gives $N_m^* \propto 1.63 (t/J)^{1/3}$.}
	\label{fig:Spinbag_En}
\end{figure}
We now discuss the character of a magnetic polaron when a single hole makes a bound state with several magnons. With this goal in mind we studied systems that consist of a single hole and up to eight magnons. We define the binding energy $E_{B}^{(N)}\equiv E_{1HNM}-E_{1H}-NE_{1M}$ and then compute the difference $\mu_{mN} \equiv E_{B}^{(N)}-E_{B}^{(N-1)}$. This quantity plays the role of a magnonic affinity energy. We present its dependence on the superexchange interaction for different number of magnons in Fig.~\ref{fig:Spinbag_En} panel (a). The negative value of this quantity indicates that when the $N$-th magnon is added to the system, it is energetically favorable to attach it to the 1H(N-1)M bound state as an extra bound particle. Addition of magnons will continue to increase the size of the cluster bound to the hole as long as the magnon affinity remains negative. We refer to such clusters of bound magnons as a ``spinbag'' (see Fig.~\ref{fig:AntiferroSB1}). When the magnon affinity becomes positive, the excess magnons are pushed away from the bag. The point at which the magnonic affinity energy changes sign indicates the optimal number of magnons forming a spinbag $N_m^*$. For small values of $J/t$ the number of magnons inside the spinbag increases which also results in the expansion of its size, see Fig.~\ref{fig:Spinbag_En} panel (b). 
A simple scaling analysis can be used to understand the dependence of $N_m^*$ on the ratio $J/t$ (see \cite{Suppl} for a detailed analysis). For small values of $J/t$ the antiferromagnetic spinbag is frozen and it acts as an effective potential trapping the hole inside it. Therefore the kinetic energy of the hole decreases for larger spinbags $t/N_m^2$. On the other hand, creating the antiferromagnetic spinbag costs an extra magnetic energy $J N_m$. This competition between kinetic and magnetic energy gives the optimal number of magnons present in the spinbag $N_m^* \propto (t/J)^{1/3}$. Similar scalings have been found for the magnetic polaron in a two-dimensional square lattice~\cite{PhysRevLett.60.740,PhysRevB.64.024411,PhysRevX.8.011046}

\subsection{Phase diagram}
As we discussed before the fermionic $t-J$ model in a zigzag ladder exhibits a large zoo of bound states with different numbers of holes and magnons. We now summarize how we obtained transitions between different types of bound states as a function of $J/t$. Our first step was to compute binding energies defined previously for different numbers of the holes and magnons. We use this procedure to identify the lowest energy bound states for different values of $J/t$ and different ratios of the hole to magnon densities. Then we double the number of particles and compute the respective binding energy $E_{2N}-2E_N$. A negative binding energy denotes an effective attraction and particles will cluster and create a larger bound state. Instead a positive binding energy indicates an effective repulsive interaction and two separated bound states appear. This means that we have found the largest possible bound state for this ratio of number of magnons and holes at this value of $J/t$. Iterating this procedure for different values of $N_m/N_h$ and $J/t$ we obtain the phase diagram presented in Fig.~\ref{fig:PhasDiagr}.

\section{Many-body phases of fermionic zigzag ladders}
\label{Sec4}
\begin{figure}[t!]
	\centering
	\includegraphics[width=1\columnwidth]{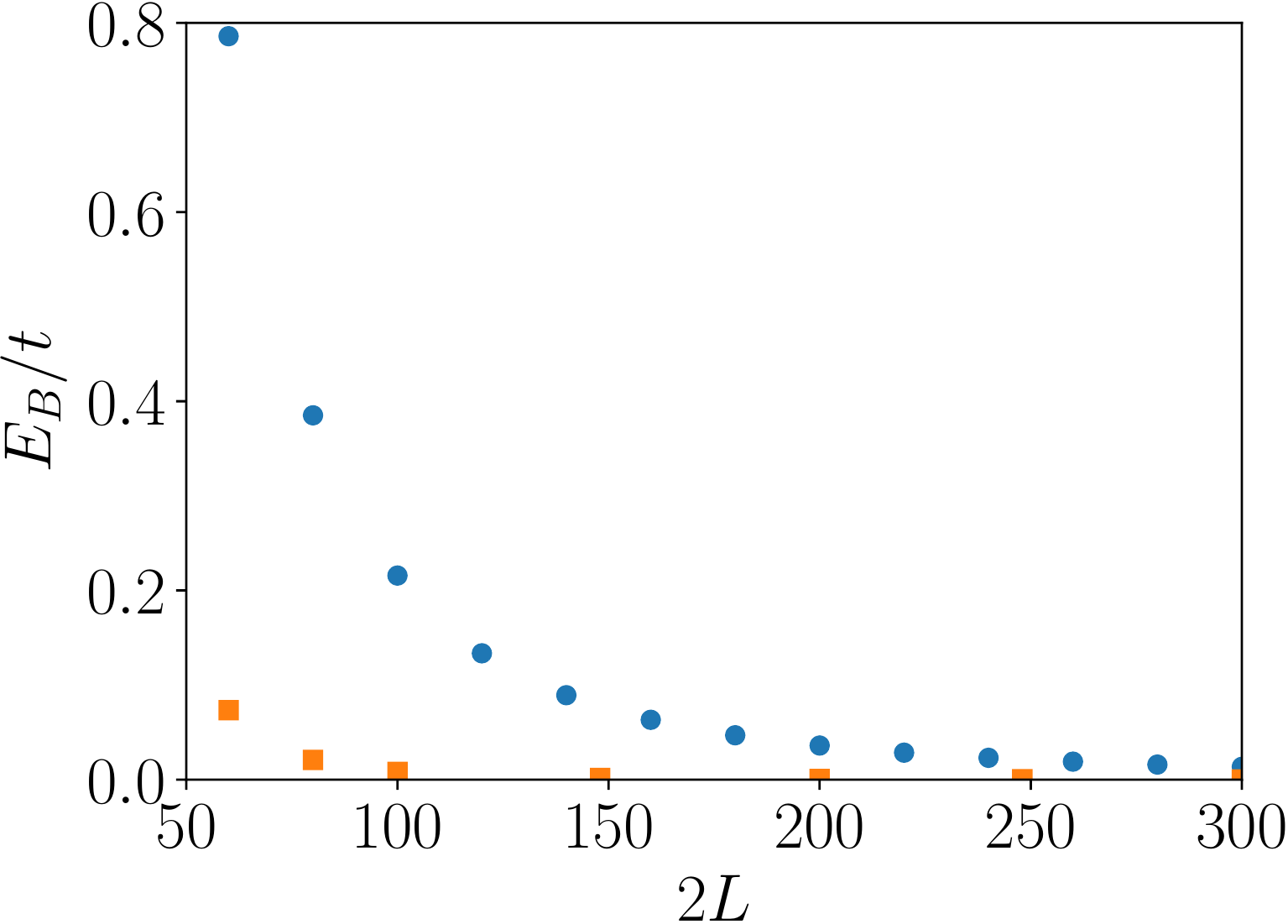} 
	\caption{Binding energies relative to the hole-magnon pair 1H1M $E_B=E_{10H10M}-10E_{1H1M}$ (blue dots) and the pentamer 2H3M $E_B=E_{6H9M}-3E_{2H3M}$ (orange squares) for the fermionic zigzag ladder as a function of the lattice length $2L$.}
	\label{fig:Energy_N}
\end{figure}
In this section we review many-body states that emerge in systems with many bound states. More formally we analyze zigzag ladders as we increase the number of holes and magnons while keeping the ratio $N_m/N_h$ fixed. After identifying ``optimal'' multi-particle bound states, we do not find further clustering. This suggest that optimal composites of holes and magnons exhibit repulsive interactions with each other. We begin by discussing the situation of small densities, when the interaction between different bound states is much smaller than the binding energy. In this regime the system can be understood as a dilute fluid of weakly interacting bound states, and the total energy of the system $E_{NB}$ should be equal to $N_BE_{1B}$, where $E_{1B}$ is the energy of a single bound state and $N_B$ is the number of bound states (see Fig.~\ref{fig:Energy_N}). We present results of these calculations for the cases of $10H10M$ and $6H9M$ where the energy of the system converges to $10 E_{1H1M}$ and $3 E_{2H3M}$, respectively. In the first case the energy converges slowly with the system size. In the second case the energy converges to the asymptotic value already for small sizes. This indicates strong repulsive interactions between pentamers which forces them to avoid spatial overlap.

\subsection{Luttinger liquid of composite fermions}
\label{Sec4a}
\begin{figure}[t!]
	\centering
	\includegraphics[width=1\columnwidth]{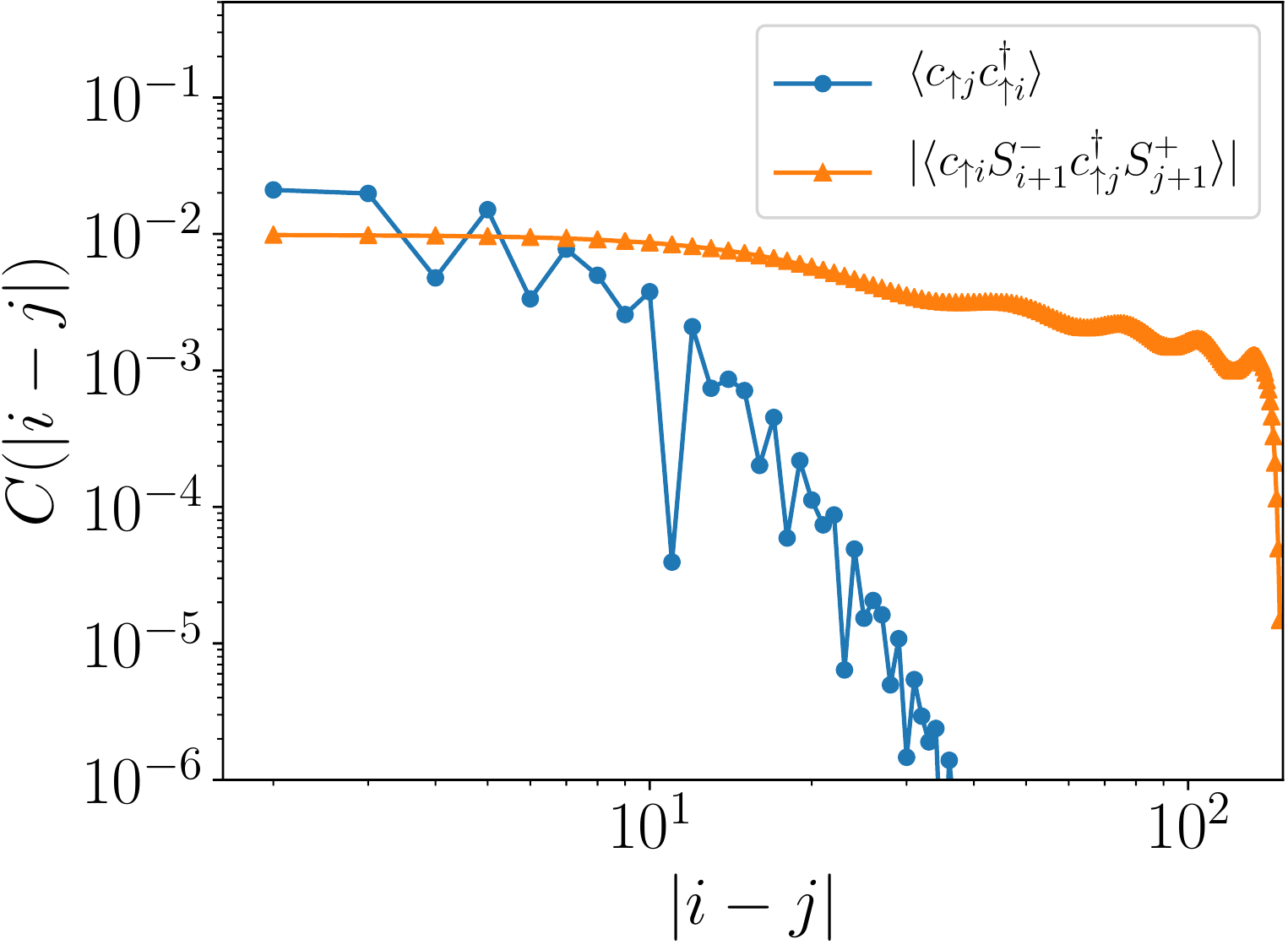} 
	\caption{Correlation functions of holes (dots) and hole-magnon pairs (upper triangles) on the fermionic zigzag ladder for $N_h=10$, $N_m=10$ and $J/t=0$. We note that operators $c_{\sigma}$ denote the original fermionic operators.}
	\label{fig:Correlation10H10M}
\end{figure}
Our next step is to establish the best physical picture of the quantum fluids of bound states. As our first example we analyze a system in which we increase the densities of holes and magnons but fix the ratio $N_m/N_h=1$. 1H1M pairs exhibit strong binding with the scale of the binding energy set by $t$ and effective repulsive interaction between pairs. This suggests that the many-body state can be understood as a Luttinger liquid of 1H1M pairs. In order to show the paired nature of the liquid we define the pair correlation function of hole-magnon pairs $g_2=\frac{\langle n_i^cn_j^c\rangle}{\langle n_i^c \rangle\langle n_j^c \rangle}$ for $i>j$ with $n_i^c=n_{i}^hn_{i+1}^m+n_{i}^mn_{i+1}^h$ being $n_i^h$ and $n_i^m$ the hole and magnon density on site $i$, respectively. 
This correlator shows that every hole in the system is accompanied by a magnon on an adjacent site, see~Fig.~\ref{fig:PairCorrelation}. This result strongly supports the hypothesis of the liquid of pairs, but we need to provide additional verification that the long range part of correlations is consistent with the Luttinger liquid phase.

The Luttinger liquid of bare holes and the Luttinger liquid of hole-magnon composites can be rigorously distinguished by identifying the operators that decay most slowly in space. To this end we compute two types of correlation functions: the correlation function of bare holes $\langle \hat{c}_{\uparrow i} \hat{c}^{\dagger}_{\uparrow j} \rangle$ and the correlation function of pairs $\langle \hat{c}_{\uparrow i} S^+_i \hat{c}^{\dagger}_{\uparrow j} S^-_j \rangle$ (see Fig.~\ref{fig:Correlation10H10M}, we also note that here operators $c_{\sigma}$ denote the original fermionic operators).
The hole correlation function has an exponential decay indicating that we cannot characterize the system as a Luttinger liquid of individual holes. On the other hand, the correlation function for pairs exhibits a much slower decay, which for large distances approaches a power-law decay with oscillations with a characteristic length scale $1/(\pi n_B)$ set by the pair density $n_B=N_h/(2L)=N_m/(2L)$ as predicted by Luttinger liquid theory~\cite{Giamarchi:743140}. This indicates that a paired Luttinger liquid appears when the number of magnons is equal to the number of holes and $J/t<2$.

\subsection{Pair density wave}
\label{Sec4b}
\begin{figure}[t!]
	\centering
	\includegraphics[width=1\columnwidth]{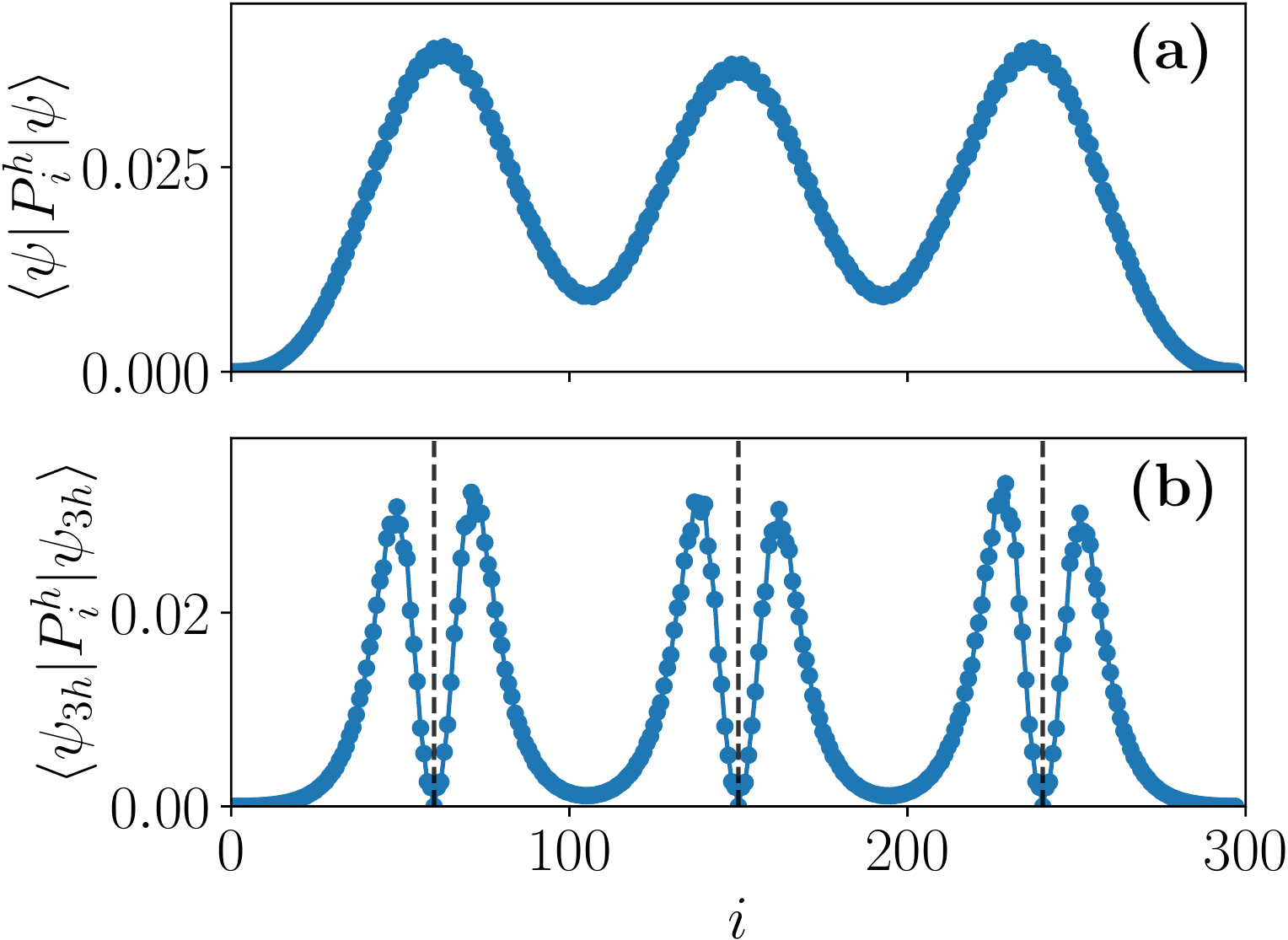} 
	\caption{Hole density on the fermionic zigzag ladder (a) for the case of $N_h=6$ and $N_m=9$ at $J/t=0.05$. Hole density of the projected wavefunction $|\psi_{3h}\rangle$ (b) which has three holes fixed at the positions specified by the dashed lines. }
	\label{fig:Density6H9M}
\end{figure}
In the earlier discussion of energetics we pointed out evidence of strong repulsive interactions between pentamers. The repulsive interactions between pentamers provides the opportunity to explore a many-body phase of pairs of holes. We will now discuss numerical evidence that pentamers form a crystal like phase. We call this state a pair density wave phase because every pentamer is a charge two object. The hole density presented in Fig.~\ref{fig:Density6H9M} provides a strong indication of the pair density wave. In a system with six holes we find three strong peaks in the density with very low density between the peaks. This  suggests that 2H3M pentamers  are strongly localized. Moreover by fixing the position of three holes and computing the hole density for the remaining three ones we observe that the latter are strongly localized close to the positions of the three fixed holes, see Fig.~\ref{fig:Density6H9M} panel (b). In the middle region between two bound pairs of holes the probability of finding another hole is negligible. Furthermore these densities seem very similar to the one found for a single pentamer in Fig.~\ref{fig:HolPair} panel (b). Before concluding this discussion we point out that to rigorously define a crystal phase in 1d in the thermodynamic limit requires spontaneous breaking of discrete translational symmetry. Breaking of translational symmetry is only possible for rational densities of pentamers. For irrational densities one can at most find quasi-long range order. Presence or absence of the long range crystal order is expected to depend strongly on the precise pentamer density. To identify true long range order in numerics calls for analyzing large system sizes. We postpone this investigation to subsequent publications.

\section{Bosons in zigzag ladder}
\label{Sec4}
\begin{figure}[t!]
	\centering
	\includegraphics[width=1\columnwidth]{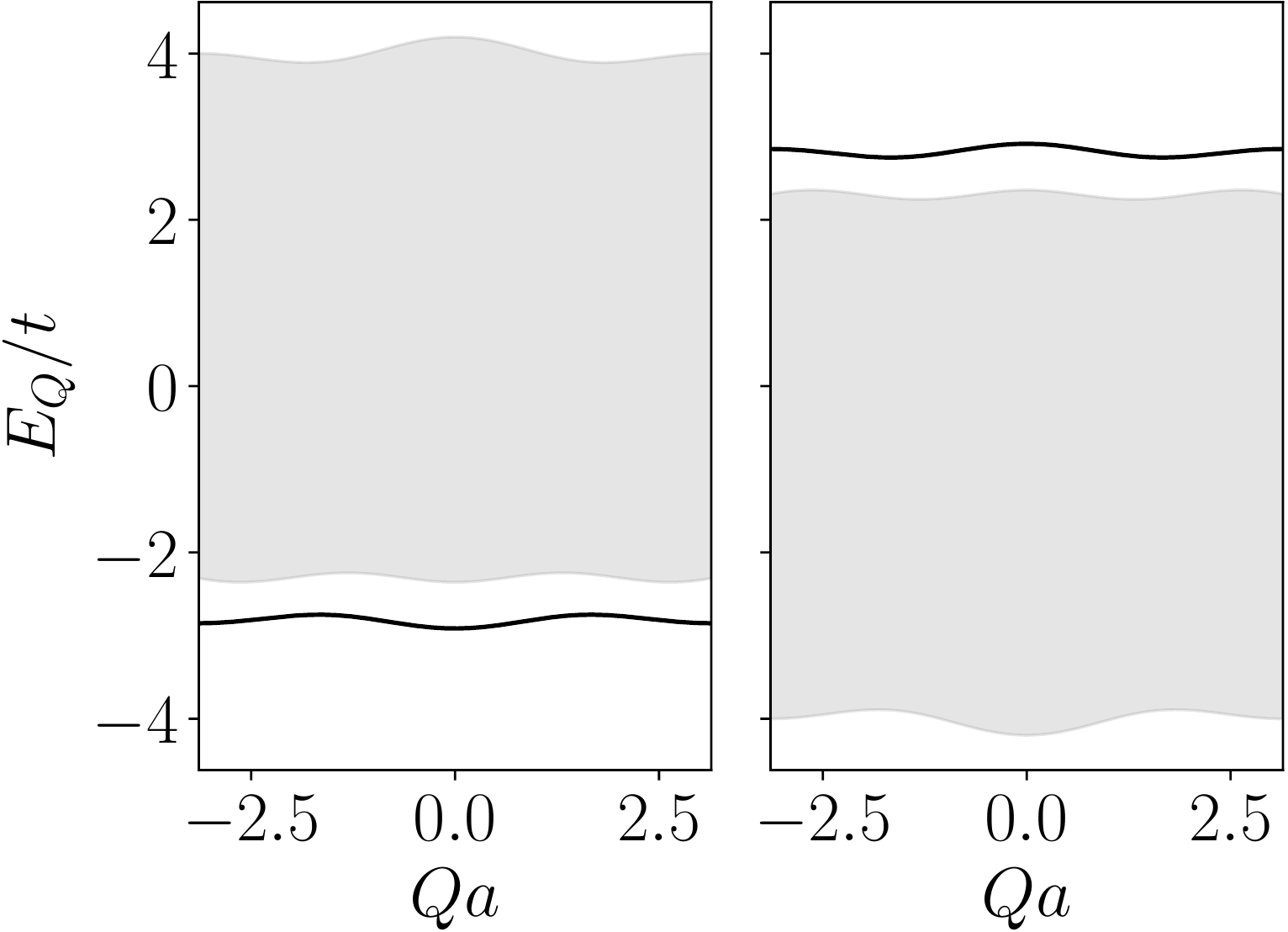} 
	\caption{Band structure of the hole-magnon problem on the zigzag ladder as a function of the total quasimomentum $Q$ for $J/t=0.1$. The grey area denotes the hole-magnon scattering continuum and the continuous line the bound state band. The left (right) panel corresponds to the fermionic (bosonic) case $t>0, J>0$ ($t<0, J<0$). The band structure is obtained by solving the effective two-body problem of a single hole and a magnon in the zigzag ladder, see App.~\ref{AppendixA}.}
	\label{fig:EnergyBand_BF}
\end{figure}
The fermionic $t-J$ model in a zigzag ladder presents the unique opportunity to explore the attraction between holes and magnons and thus attractive bound states coming from kinetic frustration. On the other hand, for the bosonic case repulsive hole-magnon bound states (antibound states) driven by kinetic frustration can be found. These states are also characterized by having the hole and magnon close in real space but their energy is higher than the scattering continuum, see Fig.~\ref{fig:EnergyBand_BF}. Thus these are not the groundstate of the system. As we will show there is a direct transformation relating the hole-magnon bound state in both models.

In order to obtain the relation between the fermionic and bosonic model it is convenient to express the Hamiltonian~\eqref{Eq:t-Jmodel} in terms of the hole and magnon operators rather than the original particles. Since we are working close to the fully polarized state we can introduce the transformations $t \, \hat{c}^{\dagger}_{\uparrow i}\hat{c}_{\uparrow j} \rightarrow \pm t\, \hat{h}^{\dagger}_j\hat{h}_i$ and $t \, \hat{c}^{\dagger}_{\downarrow i}\hat{c}_{\downarrow j} \rightarrow \pm t\, \hat{h}^{\dagger}_j\hat{h}_i S^+_i S^-_j$. The positive and negative signs correspond to the bosonic and fermionic case, respectively. These signs come from the commutation and anticommutation relations satisfied by bosons and fermions respectively. Moreover the superexchange coupling also changes sign when going from bosons to fermions $J=\pm 4t^2/U$ being positive for fermions and negative for bosons~\cite{PhysRevLett.91.090402,PhysRev.79.350,Hofstetter2002}. This establishes a connection between the fermionic and bosonic $t-J$ models $\hat{H}_{t-J}^F \Longleftrightarrow -\hat{H}_{t-J}^B$. If a hole-magnon bound state appears below the scattering continuum for the fermionic case it also appears above the continuum for the bosonic case, see Fig.~\ref{fig:EnergyBand_BF}. This sets a connection between attractive (fermionic) and repulsive (bosonic) bound states for $t-J$ models. 

\section{Experimental probes with dynamics}
\label{Sec5}
\begin{figure}[t!]
	\centering
	\includegraphics[width=1\columnwidth]{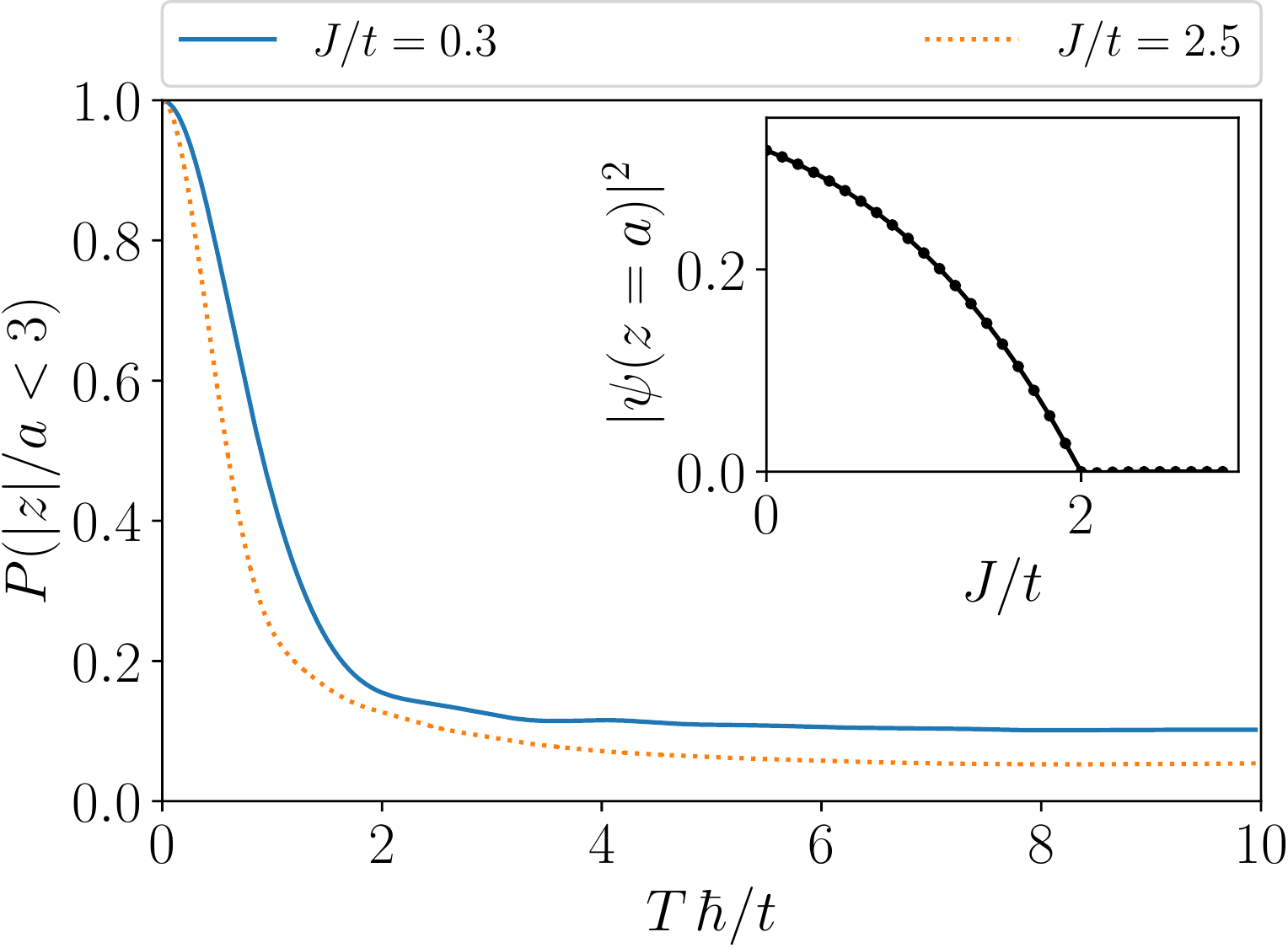} 
	\caption{Main panel: Probability of finding a hole and a flipped spin at a relative distance less than three lattice sites $P(|z|/a<3)$ as a function of time for a fermionic zigzag lattice of $2\times 50$ sites for two values of $J/t$. Inset panel: Probability of finding the hole and magnon at a relative distance $z=a$ as a function of $J/t$ obtained in the limit of having an infinite ladder by performing a finite size scaling, see App.~\ref{AppendixA}.}
	\label{fig:DynProb}
\end{figure}
So far we discussed hole-magnon and other composites at zero temperature. We expect that at temperature smaller than the binding energies our conclusions will remain accurate. However, when temperature exceeds the binding energy, we expect composites to dissociate.  Thus one of the practical challenges for experimental observation of the bound states that we discussed in previous sections is the requirement of cooling systems to temperatures below the binding energies. This difficulty is exacerbated by the requirement of having strongly polarized Fermi mixtures for which Pauli principle cuts off contact interactions and suppresses thermalization. An alternative is to perform dynamical experiments. These are based on starting a real time evolution of a fully polarized state which has a hole and a magnon localized on adjacent sites in the middle of the lattice. 
This state has an overlap with the hole-magnon bound state (around $0.3$). In the inset of Fig.~\ref{fig:DynProb} we present the probability of finding the hole and the magnon at nearest sites as a function of the superexchange coupling $J/t$. A finite overlap of the initial state with the bound state ensures that during the coherent evolution there is a finite probability for the hole and the magnon to stay close to each other as shown in Fig.~\ref{fig:DynProb}. 
In order to avoid finite size effects we have performed dynamical simulations up to the point where the hole or magnon touches the ends of the system. These finite time and finite size simulations make the probability to always saturate to a finite value even if no bound state is present. To dynamically discern the appearance of a hole-magnon bound state we require to have saturation values much larger than one over the system size. For small values of $J/t$ we observe saturation values that can be ten times larger than one over the system size. Hence such ``quantum walk'' like dynamics~\cite{Preiss1229} starting from the adjacent hole-magnon configuration can reveal the existence of the bound state. The dynamical observation of hole-magnon bound states could be adressed in current ultracold atom experiments with a quantum gas microscope~\cite{Bakr2009,Sherson2010,PhysRevLett.114.213002,PhysRevLett.114.193001,PhysRevLett.115.263001,Haller2015,PhysRevA.92.063406,Greif953,Brown1385,Weitenberg2011,Schauss}.

It is interesting to point out the similarity of the fermionic and bosonic systems under this time evolution. 
Since the two Hamiltonians are related by a minus sign $\hat{H}_{t-J}^F \Longleftrightarrow -\hat{H}_{t-J}^B$ but the procedure satisfies dynamical symmetry analogous to the one discussed in Ref.~\cite{Schneider2012} we expect the time evolution to be identical in both cases. Therefore far-from-equilibrium experiments can be used either to detect attractive (fermionc) or repulsive (bosonic) bound states.

\section{Fermions and bosons in flux ladders}
\label{Sec6}
\begin{figure}[t!]
	\centering
	\includegraphics[width=1\columnwidth]{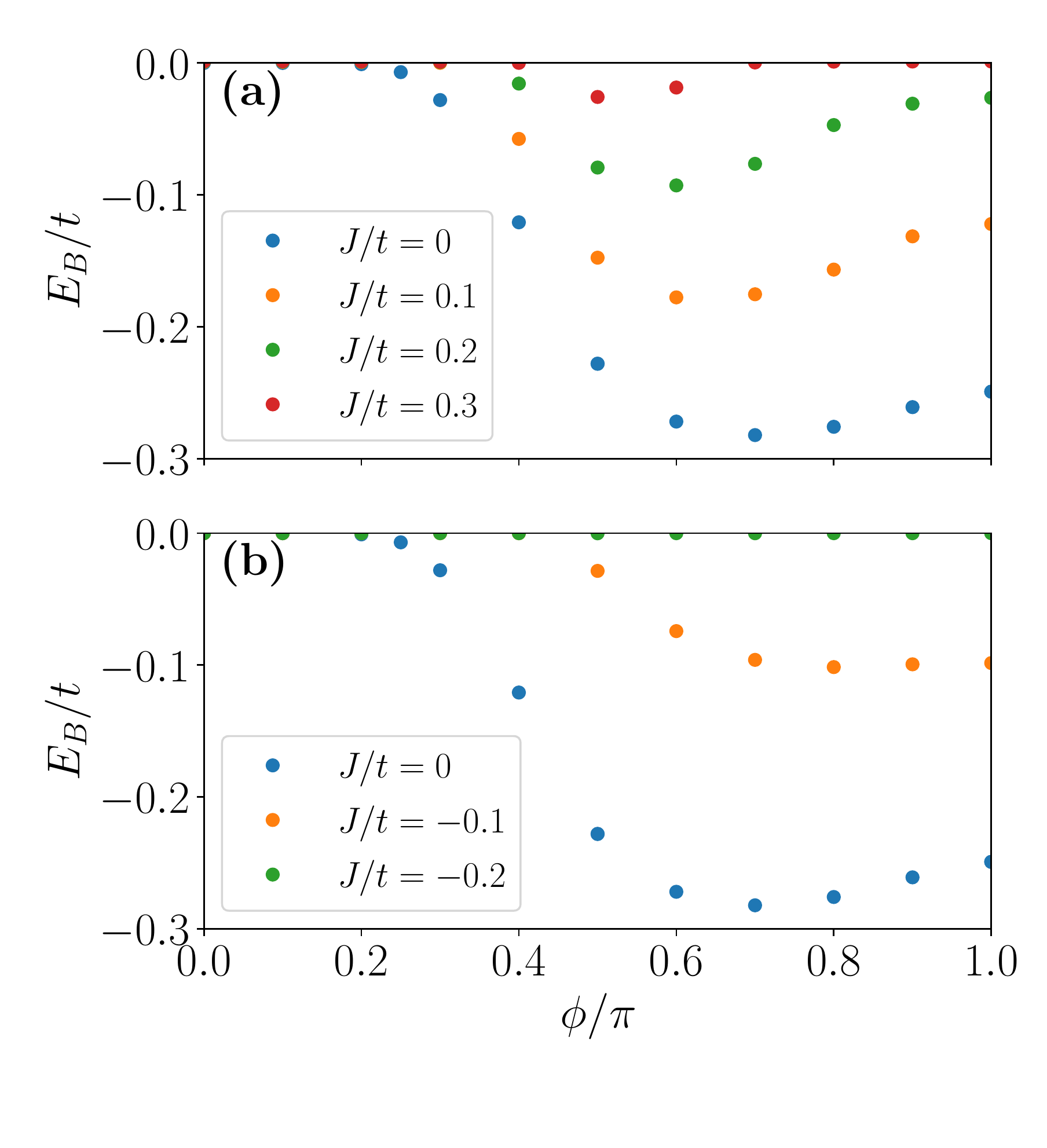} 
	\caption{Binding energy of the hole-magnon for a square ladder as a function of the transverse magnetic flux for different values of the superexchange interaction. The fermionic case corresponds to (a) and the bosonic one to (b).}
	\label{fig:BindingFluxLadder}
\end{figure}
A single hole propagating in a ferromagnetic background in a non-bipartite lattice always experiences kinetic frustration. However in a square lattice the kinetic energy of a single hole can also become frustrated if different paths contribute with different relative phases. Therefore we explore the appearance of hole-magnon bound states in a square ladder with magnetic flux.

Hole-magnon bound states are found and they exhibit very similar properties to the ones obtained for the zigzag ladder. Near the hole the spins tilt the XY plane and antiferromagnetic correlations are found, see Fig.~\ref{fig:HolMagRepr}. In order to quantify the parameter space where hole-magnon bound states can be observed we compute the respective binding energy $E_B\equiv E_{1H1M}-E_{1H}-E_{1M}$, see Fig.~\ref{fig:BindingFluxLadder}. The bound state appears for a wide range of values of the magnetic flux but the range of $J/t$ is small compared with the zigzag ladder. 

The single hole spectrum for the regular ladder with flux is symmetric under a change of sign of the single particle hopping. Thus starting with a spin polarized band insulator for fermions or a Mott insulator of bosons with $n=1$, we expect to find identical single particle spectra of individual holes. This proves that attractive bound states are present for both bosonic and fermionic systems. This is the main difference between bipartite and non-bipartite lattices. In Fig.~\ref{fig:BindingFluxLadder}. we present the binding energies for the fermionic panel (a) ($J/t>0$) and the bosonic panel (b) ($J/t<0$) situation. Since we still have the property that fermionic and bosonic models written in terms of hole and magnon operators differ by the sign, we infer that in both systems attractive and repulsive bound states are present. The wavefunction that describes an attractive/repulsive bound state for fermions also describes a repulsive/attractive bound state for bosons.

\section{Summary and Outlook}
The main result of our work is a demonstration of the effective attractive interaction arising  from kinetic frustration of distinguishable particles. These interactions lead to formation of several types of multi-particle complexes, which can then result in several types of interesting many-body phases. In this paper we focused on two component Bose and Fermi mixtures and two types of frustrated lattices: zigzag ladders and square ladders with a flux. We studied the regimes close to fully polarized insulating states, when these systems can be naturally described in terms of doped holes and magnons, i.e. flipped spins. We made concrete predictions for the phase diagrams of these models including identifying the most stable multi-particle bound states and understanding many-body states that emerge from the interaction of such composites. We expect that qualitatively, the results presented in our paper should be valid to general models with  kinetic frustration.

We used analysis of binding energies and correlation functions to discern the lowest energy multi-paritcle bound states for different values of the interaction strength and several representative density ratios. Examination of the correlation functions also allowed us to scrutinize the internal structure of these multi-particle bound states. We show that the system has a tendency to create an antiferromagnetic background around the position of the hole (see discussion in App.~\ref{AppendixB}). This antiferromagnetic background acts as an effective potential to the hole which confines it in this region. This object can be understood as a multi-particle bound state of many magnons and a single hole and we term it an antiferromagnetic spinbag. When adding a second hole to the system an effective attraction between the two holes can be found. We show that the minimal configuration exhibiting a pair of two holes is a pentamer composed of two holes and three magnons. 

We identified a generic dynamic symmetry between the fermionic and bosonic Hubbard models close to spin polarized insulating states. We demonstrated that expressing these models in terms of the hole and magnon operators  gives rise to effective Hamiltonians which differ  only in the overall sign. This symmetry maps the low energy states of the fermionic model to the high energy states of the bosonic model. Therefore when a bound state is present in the fermionic system we should find an antibound state for the bosonic system and vice versa: a bound state for bosonic system implies an antibound state for the fermionic system.

We discussed dynamical experiments which can be used to probe formation of multiparticle bound states. In particular, we considered a protocol for a fermionic system in a zigzag ladder in which a hole and a flipped spin are initialized on neighboring sites. After the hole and the magnon are released, we find that their coherent dynamics can be decomposed into two contributions: in the first one the hole and the magnon expand essentially independently of each other, and in the second one they move together. The latter part can be understood by observing that the initial configuration has a large overlap with the 1H1M bound state, which results in the hole-magnon pair expanding as a whole. Because of the dynamic symmetry between the fermionic and bosonic models, this procedure can be applied either for detecting a bound state for fermions or an antibound state for bosons. 
Moreover this probe could be generalized to systems with more holes or magnons in order to detect larger multi-particle bound or antibound states.

Our work extends beyond analysis of individual multi-particle bound states. We discuss several many-body phases that can be understood from the perspective of self organization of the multi-particle bound states. We argue that higher order correlation functions provide direct signatures of the emergent many-body states of composite objects and review two concrete examples. Our first example is a Luttinger liquid of 1 hole - 1 magnon pairs. We show that in this case composite operators for the hole-magnon pairs exhibit much slower spatial decay than operators of the original particles. The second example we discuss is a crystal of pentamers, in which every pentamer is comprised of two holes and three magnons. We refer to this phase as a pair density wave state since a pentamer is a charge two object. 

Our work on multi-particle bound state formation due to kinetic frustration can be extended in several directions. Powerful experimental tools developed for cold atomic ensembles in optical lattices allow to satisfy two criteria at the same time: relevance to current technologies and  richness of theoretically expected phenomena. A promising direction is to analyze ladders in which tunneling/interactions within the chains and between them are different. Applying potential gradients also allows to realize mixed dimensional systems, in which along certain directions there is exchange interaction but no single particle tunneling~\cite{10.21468/SciPostPhys.5.6.057}. In the case of bosonic systems one can also consider Hubbard models in which interactions are not SU(2) symmetric. This should translate into the anisotropy of exchange interactions~\cite{Jepsen2020,jepsen2021transverse}.  We discussed in Sec.~\ref{Sec3} that the Ising part of the interaction contributes to the repulsive interaction between a magnon and a hole. Thus by making the z-axis part of the  exchange interaction stronger, we expect to find suppression of magnon-hole binding, which should in turn affect stability of all multi-particle composites. A more detailed analysis of multi-particle formation and self-organization  in systems with magnetic flux is also an interesting future directions.  With increasing  magnetic flux we expect to find stronger frustration and stronger binding. This should make multi-particle systems, for which we find binding energies to be small, easier to observe in experiments.

Our results indicate that hole pairing is strongly suppressed for ladders relative to the 2D case.  A possible explanation is that the hole pairs in two dimensions form at finite angular momentum and in ladders they are suppressed because they cannot fully expand in the perpendicular direction. From the results of Ref.~\cite{PhysRevB.97.140507} one expects that 2D systems exhibit hole pairing arising from formation of 2 holes - 1 magnon trimers.
An interesting question then is to extend our work to a larger number of legs and study the transition from the quasi-1D geometry to the full 2D one.  

The antiferromagnetic spinbags that we find provide an intriguing analogy to the quark bag model used to explain the asymptotic freedom of QCD. In our system the antiferromagnetic background provides an effective static potential to the holes. While holes are fast degrees of freedom, they become trapped inside these antiferromagnetic regions. A possible future perspective would be to exploit this analogy and study the backreaction of the bag to the hole movement. We expect that this will allow to find excitations of composite particles in our system that are analogous to excitations of hadrons considered in the context of the bag model \cite{HASENFRATZ197875}. In particular we anticipate vibrational modes, that can be visualized as a hole ``rattling'' inside the AF bag, as well as surface modes, that correspond to local displacements of the bag with respect to its equilibrium position around the hole. The latter family of excitations should be particularly interesting in the 2D setting. Finally, this analogy could be used to explore other directions with the goal of providing new insight into interesting open questions regarding the nature of confinement. One particularly intriguing question is to use an experimental platform of frustrated quantum systems to realize an analogue of the transition between the hadronic and paired phases, expected in QCD at higher density. Many questions remain poorly understood regarding this phase transition, including the possibility of some intermediate exotic phases. In our frustrated system a similar situation could be explored in the ``pentamer phase'' when the density of holes to magnons is fixed at 2/3 but individual densities are being increased. As the density of pentamers increases, they should start to overlap, and at some critical density holes may become ``liberated'' from the bags. This will corresponds to breaking of pentamers, however, we expect that residual interaction between holes will still result in their pairing. Understanding this transition, including exploring the possibility of other phases, is an interesting open question.

\acknowledgments{We acknowledge helpful discussions with P. Dolgirev, M. Greiner, F. Grusdt, Eun-Ah Kim and P. Schauss. I. Morera thanks financial support from the
MINECO (Spain) Grants No. FIS2017- 87534-P and
FIS2017-84114-C2-1-P and the European Union
Regional Development Fund within the ERDF
Operational Program of Catalunya (project QUASICAT/
QuantumCat). A. Bohrdt is supported by the NSF through a grant for the Institute for Theoretical Atomic, Molecular, and Optical Physics at Harvard University and the Smithsonian Astrophysical Observatory. W. W. Ho is supported in part by the Stanford Institute of Theoretical Physics (SITP). E. Demler thanks Harvard-MIT CUA, the ARO grant number W911NF-20-1-0163, the NSF grant OAC-1934714, the NSF EAGER-QAC-QCH award number 2037687. DMRG computations have been performed using TeNPy~\cite{10.21468/SciPostPhysLectNotes.5}. }

\bibliographystyle{apsrev4-1}
\bibliography{paperbib}

\begin{thebibliography}{61}%
\makeatletter
\providecommand \@ifxundefined [1]{%
 \@ifx{#1\undefined}
}%
\providecommand \@ifnum [1]{%
 \ifnum #1\expandafter \@firstoftwo
 \else \expandafter \@secondoftwo
 \fi
}%
\providecommand \@ifx [1]{%
 \ifx #1\expandafter \@firstoftwo
 \else \expandafter \@secondoftwo
 \fi
}%
\providecommand \natexlab [1]{#1}%
\providecommand \enquote  [1]{``#1''}%
\providecommand \bibnamefont  [1]{#1}%
\providecommand \bibfnamefont [1]{#1}%
\providecommand \citenamefont [1]{#1}%
\providecommand \href@noop [0]{\@secondoftwo}%
\providecommand \href [0]{\begingroup \@sanitize@url \@href}%
\providecommand \@href[1]{\@@startlink{#1}\@@href}%
\providecommand \@@href[1]{\endgroup#1\@@endlink}%
\providecommand \@sanitize@url [0]{\catcode `\\12\catcode `\$12\catcode
  `\&12\catcode `\#12\catcode `\^12\catcode `\_12\catcode `\%12\relax}%
\providecommand \@@startlink[1]{}%
\providecommand \@@endlink[0]{}%
\providecommand \url  [0]{\begingroup\@sanitize@url \@url }%
\providecommand \@url [1]{\endgroup\@href {#1}{\urlprefix }}%
\providecommand \urlprefix  [0]{URL }%
\providecommand \Eprint [0]{\href }%
\providecommand \doibase [0]{http://dx.doi.org/}%
\providecommand \selectlanguage [0]{\@gobble}%
\providecommand \bibinfo  [0]{\@secondoftwo}%
\providecommand \bibfield  [0]{\@secondoftwo}%
\providecommand \translation [1]{[#1]}%
\providecommand \BibitemOpen [0]{}%
\providecommand \bibitemStop [0]{}%
\providecommand \bibitemNoStop [0]{.\EOS\space}%
\providecommand \EOS [0]{\spacefactor3000\relax}%
\providecommand \BibitemShut  [1]{\csname bibitem#1\endcsname}%
\let\auto@bib@innerbib\@empty
%</preamble>
\bibitem [{\citenamefont {Lacroix}(2010)}]{lacroix2010introduction}%
  \BibitemOpen
  \bibfield  {author} {\bibinfo {author} {\bibfnamefont {C.}~\bibnamefont
  {Lacroix}},\ }\href@noop {} {\emph {\bibinfo {title} {Introduction to
  frustrated magnetism : materials, experiments, theory}}}\ (\bibinfo
  {publisher} {Springer},\ \bibinfo {address} {Berlin London},\ \bibinfo {year}
  {2010})\BibitemShut {NoStop}%
\bibitem [{\citenamefont {Bakr}\ \emph {et~al.}(2009)\citenamefont {Bakr},
  \citenamefont {Gillen}, \citenamefont {Peng}, \citenamefont {F{\"o}lling},\
  and\ \citenamefont {Greiner}}]{Bakr2009}%
  \BibitemOpen
  \bibfield  {author} {\bibinfo {author} {\bibfnamefont {W.~S.}\ \bibnamefont
  {Bakr}}, \bibinfo {author} {\bibfnamefont {J.~I.}\ \bibnamefont {Gillen}},
  \bibinfo {author} {\bibfnamefont {A.}~\bibnamefont {Peng}}, \bibinfo {author}
  {\bibfnamefont {S.}~\bibnamefont {F{\"o}lling}}, \ and\ \bibinfo {author}
  {\bibfnamefont {M.}~\bibnamefont {Greiner}},\ }\href {\doibase
  10.1038/nature08482} {\bibfield  {journal} {\bibinfo  {journal} {Nature}\
  }\textbf {\bibinfo {volume} {462}},\ \bibinfo {pages} {74} (\bibinfo {year}
  {2009})}\BibitemShut {NoStop}%
\bibitem [{\citenamefont {Sherson}\ \emph {et~al.}(2010)\citenamefont
  {Sherson}, \citenamefont {Weitenberg}, \citenamefont {Endres}, \citenamefont
  {Cheneau}, \citenamefont {Bloch},\ and\ \citenamefont {Kuhr}}]{Sherson2010}%
  \BibitemOpen
  \bibfield  {author} {\bibinfo {author} {\bibfnamefont {J.~F.}\ \bibnamefont
  {Sherson}}, \bibinfo {author} {\bibfnamefont {C.}~\bibnamefont {Weitenberg}},
  \bibinfo {author} {\bibfnamefont {M.}~\bibnamefont {Endres}}, \bibinfo
  {author} {\bibfnamefont {M.}~\bibnamefont {Cheneau}}, \bibinfo {author}
  {\bibfnamefont {I.}~\bibnamefont {Bloch}}, \ and\ \bibinfo {author}
  {\bibfnamefont {S.}~\bibnamefont {Kuhr}},\ }\href {\doibase
  10.1038/nature09378} {\bibfield  {journal} {\bibinfo  {journal} {Nature}\
  }\textbf {\bibinfo {volume} {467}},\ \bibinfo {pages} {68} (\bibinfo {year}
  {2010})}\BibitemShut {NoStop}%
\bibitem [{\citenamefont {Parsons}\ \emph {et~al.}(2015)\citenamefont
  {Parsons}, \citenamefont {Huber}, \citenamefont {Mazurenko}, \citenamefont
  {Chiu}, \citenamefont {Setiawan}, \citenamefont {Wooley-Brown}, \citenamefont
  {Blatt},\ and\ \citenamefont {Greiner}}]{PhysRevLett.114.213002}%
  \BibitemOpen
  \bibfield  {author} {\bibinfo {author} {\bibfnamefont {M.~F.}\ \bibnamefont
  {Parsons}}, \bibinfo {author} {\bibfnamefont {F.}~\bibnamefont {Huber}},
  \bibinfo {author} {\bibfnamefont {A.}~\bibnamefont {Mazurenko}}, \bibinfo
  {author} {\bibfnamefont {C.~S.}\ \bibnamefont {Chiu}}, \bibinfo {author}
  {\bibfnamefont {W.}~\bibnamefont {Setiawan}}, \bibinfo {author}
  {\bibfnamefont {K.}~\bibnamefont {Wooley-Brown}}, \bibinfo {author}
  {\bibfnamefont {S.}~\bibnamefont {Blatt}}, \ and\ \bibinfo {author}
  {\bibfnamefont {M.}~\bibnamefont {Greiner}},\ }\href {\doibase
  10.1103/PhysRevLett.114.213002} {\bibfield  {journal} {\bibinfo  {journal}
  {Phys. Rev. Lett.}\ }\textbf {\bibinfo {volume} {114}},\ \bibinfo {pages}
  {213002} (\bibinfo {year} {2015})}\BibitemShut {NoStop}%
\bibitem [{\citenamefont {Cheuk}\ \emph {et~al.}(2015)\citenamefont {Cheuk},
  \citenamefont {Nichols}, \citenamefont {Okan}, \citenamefont {Gersdorf},
  \citenamefont {Ramasesh}, \citenamefont {Bakr}, \citenamefont {Lompe},\ and\
  \citenamefont {Zwierlein}}]{PhysRevLett.114.193001}%
  \BibitemOpen
  \bibfield  {author} {\bibinfo {author} {\bibfnamefont {L.~W.}\ \bibnamefont
  {Cheuk}}, \bibinfo {author} {\bibfnamefont {M.~A.}\ \bibnamefont {Nichols}},
  \bibinfo {author} {\bibfnamefont {M.}~\bibnamefont {Okan}}, \bibinfo {author}
  {\bibfnamefont {T.}~\bibnamefont {Gersdorf}}, \bibinfo {author}
  {\bibfnamefont {V.~V.}\ \bibnamefont {Ramasesh}}, \bibinfo {author}
  {\bibfnamefont {W.~S.}\ \bibnamefont {Bakr}}, \bibinfo {author}
  {\bibfnamefont {T.}~\bibnamefont {Lompe}}, \ and\ \bibinfo {author}
  {\bibfnamefont {M.~W.}\ \bibnamefont {Zwierlein}},\ }\href {\doibase
  10.1103/PhysRevLett.114.193001} {\bibfield  {journal} {\bibinfo  {journal}
  {Phys. Rev. Lett.}\ }\textbf {\bibinfo {volume} {114}},\ \bibinfo {pages}
  {193001} (\bibinfo {year} {2015})}\BibitemShut {NoStop}%
\bibitem [{\citenamefont {Omran}\ \emph {et~al.}(2015)\citenamefont {Omran},
  \citenamefont {Boll}, \citenamefont {Hilker}, \citenamefont {Kleinlein},
  \citenamefont {Salomon}, \citenamefont {Bloch},\ and\ \citenamefont
  {Gross}}]{PhysRevLett.115.263001}%
  \BibitemOpen
  \bibfield  {author} {\bibinfo {author} {\bibfnamefont {A.}~\bibnamefont
  {Omran}}, \bibinfo {author} {\bibfnamefont {M.}~\bibnamefont {Boll}},
  \bibinfo {author} {\bibfnamefont {T.~A.}\ \bibnamefont {Hilker}}, \bibinfo
  {author} {\bibfnamefont {K.}~\bibnamefont {Kleinlein}}, \bibinfo {author}
  {\bibfnamefont {G.}~\bibnamefont {Salomon}}, \bibinfo {author} {\bibfnamefont
  {I.}~\bibnamefont {Bloch}}, \ and\ \bibinfo {author} {\bibfnamefont
  {C.}~\bibnamefont {Gross}},\ }\href {\doibase 10.1103/PhysRevLett.115.263001}
  {\bibfield  {journal} {\bibinfo  {journal} {Phys. Rev. Lett.}\ }\textbf
  {\bibinfo {volume} {115}},\ \bibinfo {pages} {263001} (\bibinfo {year}
  {2015})}\BibitemShut {NoStop}%
\bibitem [{\citenamefont {Haller}\ \emph {et~al.}(2015)\citenamefont {Haller},
  \citenamefont {Hudson}, \citenamefont {Kelly}, \citenamefont {Cotta},
  \citenamefont {Peaudecerf}, \citenamefont {Bruce},\ and\ \citenamefont
  {Kuhr}}]{Haller2015}%
  \BibitemOpen
  \bibfield  {author} {\bibinfo {author} {\bibfnamefont {E.}~\bibnamefont
  {Haller}}, \bibinfo {author} {\bibfnamefont {J.}~\bibnamefont {Hudson}},
  \bibinfo {author} {\bibfnamefont {A.}~\bibnamefont {Kelly}}, \bibinfo
  {author} {\bibfnamefont {D.~A.}\ \bibnamefont {Cotta}}, \bibinfo {author}
  {\bibfnamefont {B.}~\bibnamefont {Peaudecerf}}, \bibinfo {author}
  {\bibfnamefont {G.~D.}\ \bibnamefont {Bruce}}, \ and\ \bibinfo {author}
  {\bibfnamefont {S.}~\bibnamefont {Kuhr}},\ }\href {\doibase
  10.1038/nphys3403} {\bibfield  {journal} {\bibinfo  {journal} {Nature
  Physics}\ }\textbf {\bibinfo {volume} {11}},\ \bibinfo {pages} {738}
  (\bibinfo {year} {2015})}\BibitemShut {NoStop}%
\bibitem [{\citenamefont {Edge}\ \emph {et~al.}(2015)\citenamefont {Edge},
  \citenamefont {Anderson}, \citenamefont {Jervis}, \citenamefont {McKay},
  \citenamefont {Day}, \citenamefont {Trotzky},\ and\ \citenamefont
  {Thywissen}}]{PhysRevA.92.063406}%
  \BibitemOpen
  \bibfield  {author} {\bibinfo {author} {\bibfnamefont {G.~J.~A.}\
  \bibnamefont {Edge}}, \bibinfo {author} {\bibfnamefont {R.}~\bibnamefont
  {Anderson}}, \bibinfo {author} {\bibfnamefont {D.}~\bibnamefont {Jervis}},
  \bibinfo {author} {\bibfnamefont {D.~C.}\ \bibnamefont {McKay}}, \bibinfo
  {author} {\bibfnamefont {R.}~\bibnamefont {Day}}, \bibinfo {author}
  {\bibfnamefont {S.}~\bibnamefont {Trotzky}}, \ and\ \bibinfo {author}
  {\bibfnamefont {J.~H.}\ \bibnamefont {Thywissen}},\ }\href {\doibase
  10.1103/PhysRevA.92.063406} {\bibfield  {journal} {\bibinfo  {journal} {Phys.
  Rev. A}\ }\textbf {\bibinfo {volume} {92}},\ \bibinfo {pages} {063406}
  (\bibinfo {year} {2015})}\BibitemShut {NoStop}%
\bibitem [{\citenamefont {Greif}\ \emph {et~al.}(2016)\citenamefont {Greif},
  \citenamefont {Parsons}, \citenamefont {Mazurenko}, \citenamefont {Chiu},
  \citenamefont {Blatt}, \citenamefont {Huber}, \citenamefont {Ji},\ and\
  \citenamefont {Greiner}}]{Greif953}%
  \BibitemOpen
  \bibfield  {author} {\bibinfo {author} {\bibfnamefont {D.}~\bibnamefont
  {Greif}}, \bibinfo {author} {\bibfnamefont {M.~F.}\ \bibnamefont {Parsons}},
  \bibinfo {author} {\bibfnamefont {A.}~\bibnamefont {Mazurenko}}, \bibinfo
  {author} {\bibfnamefont {C.~S.}\ \bibnamefont {Chiu}}, \bibinfo {author}
  {\bibfnamefont {S.}~\bibnamefont {Blatt}}, \bibinfo {author} {\bibfnamefont
  {F.}~\bibnamefont {Huber}}, \bibinfo {author} {\bibfnamefont
  {G.}~\bibnamefont {Ji}}, \ and\ \bibinfo {author} {\bibfnamefont
  {M.}~\bibnamefont {Greiner}},\ }\href {\doibase 10.1126/science.aad9041}
  {\bibfield  {journal} {\bibinfo  {journal} {Science}\ }\textbf {\bibinfo
  {volume} {351}},\ \bibinfo {pages} {953} (\bibinfo {year}
  {2016})}\BibitemShut {NoStop}%
\bibitem [{\citenamefont {Brown}\ \emph {et~al.}(2017)\citenamefont {Brown},
  \citenamefont {Mitra}, \citenamefont {Guardado-Sanchez}, \citenamefont
  {Schau{\ss}}, \citenamefont {Kondov}, \citenamefont {Khatami}, \citenamefont
  {Paiva}, \citenamefont {Trivedi}, \citenamefont {Huse},\ and\ \citenamefont
  {Bakr}}]{Brown1385}%
  \BibitemOpen
  \bibfield  {author} {\bibinfo {author} {\bibfnamefont {P.~T.}\ \bibnamefont
  {Brown}}, \bibinfo {author} {\bibfnamefont {D.}~\bibnamefont {Mitra}},
  \bibinfo {author} {\bibfnamefont {E.}~\bibnamefont {Guardado-Sanchez}},
  \bibinfo {author} {\bibfnamefont {P.}~\bibnamefont {Schau{\ss}}}, \bibinfo
  {author} {\bibfnamefont {S.~S.}\ \bibnamefont {Kondov}}, \bibinfo {author}
  {\bibfnamefont {E.}~\bibnamefont {Khatami}}, \bibinfo {author} {\bibfnamefont
  {T.}~\bibnamefont {Paiva}}, \bibinfo {author} {\bibfnamefont
  {N.}~\bibnamefont {Trivedi}}, \bibinfo {author} {\bibfnamefont {D.~A.}\
  \bibnamefont {Huse}}, \ and\ \bibinfo {author} {\bibfnamefont {W.~S.}\
  \bibnamefont {Bakr}},\ }\href {\doibase 10.1126/science.aam7838} {\bibfield
  {journal} {\bibinfo  {journal} {Science}\ }\textbf {\bibinfo {volume}
  {357}},\ \bibinfo {pages} {1385} (\bibinfo {year} {2017})}\BibitemShut
  {NoStop}%
\bibitem [{\citenamefont {Weitenberg}\ \emph {et~al.}(2011)\citenamefont
  {Weitenberg}, \citenamefont {Endres}, \citenamefont {Sherson}, \citenamefont
  {Cheneau}, \citenamefont {Schau{\ss}}, \citenamefont {Fukuhara},
  \citenamefont {Bloch},\ and\ \citenamefont {Kuhr}}]{Weitenberg2011}%
  \BibitemOpen
  \bibfield  {author} {\bibinfo {author} {\bibfnamefont {C.}~\bibnamefont
  {Weitenberg}}, \bibinfo {author} {\bibfnamefont {M.}~\bibnamefont {Endres}},
  \bibinfo {author} {\bibfnamefont {J.~F.}\ \bibnamefont {Sherson}}, \bibinfo
  {author} {\bibfnamefont {M.}~\bibnamefont {Cheneau}}, \bibinfo {author}
  {\bibfnamefont {P.}~\bibnamefont {Schau{\ss}}}, \bibinfo {author}
  {\bibfnamefont {T.}~\bibnamefont {Fukuhara}}, \bibinfo {author}
  {\bibfnamefont {I.}~\bibnamefont {Bloch}}, \ and\ \bibinfo {author}
  {\bibfnamefont {S.}~\bibnamefont {Kuhr}},\ }\href {\doibase
  10.1038/nature09827} {\bibfield  {journal} {\bibinfo  {journal} {Nature}\
  }\textbf {\bibinfo {volume} {471}},\ \bibinfo {pages} {319} (\bibinfo {year}
  {2011})}\BibitemShut {NoStop}%
\bibitem [{\citenamefont {Yang}\ \emph {et~al.}(2021)\citenamefont {Yang},
  \citenamefont {Liu}, \citenamefont {Mongkolkiattichai},\ and\ \citenamefont
  {Schauss}}]{Schauss}%
  \BibitemOpen
  \bibfield  {author} {\bibinfo {author} {\bibfnamefont {J.}~\bibnamefont
  {Yang}}, \bibinfo {author} {\bibfnamefont {L.}~\bibnamefont {Liu}}, \bibinfo
  {author} {\bibfnamefont {J.}~\bibnamefont {Mongkolkiattichai}}, \ and\
  \bibinfo {author} {\bibfnamefont {P.}~\bibnamefont {Schauss}},\ }\href@noop
  {} {\  (\bibinfo {year} {2021})},\ \Eprint
  {http://arxiv.org/abs/arXiv:2102.11862} {arXiv:2102.11862} \BibitemShut
  {NoStop}%
\bibitem [{\citenamefont {S\o{}rensen}\ \emph {et~al.}(2005)\citenamefont
  {S\o{}rensen}, \citenamefont {Demler},\ and\ \citenamefont
  {Lukin}}]{PhysRevLett.94.086803}%
  \BibitemOpen
  \bibfield  {author} {\bibinfo {author} {\bibfnamefont {A.~S.}\ \bibnamefont
  {S\o{}rensen}}, \bibinfo {author} {\bibfnamefont {E.}~\bibnamefont {Demler}},
  \ and\ \bibinfo {author} {\bibfnamefont {M.~D.}\ \bibnamefont {Lukin}},\
  }\href {\doibase 10.1103/PhysRevLett.94.086803} {\bibfield  {journal}
  {\bibinfo  {journal} {Phys. Rev. Lett.}\ }\textbf {\bibinfo {volume} {94}},\
  \bibinfo {pages} {086803} (\bibinfo {year} {2005})}\BibitemShut {NoStop}%
\bibitem [{\citenamefont {Dalibard}\ \emph {et~al.}(2011)\citenamefont
  {Dalibard}, \citenamefont {Gerbier}, \citenamefont
  {Juzeli\ifmmode~\bar{u}\else \={u}\fi{}nas},\ and\ \citenamefont
  {\"Ohberg}}]{RevModPhys.83.1523}%
  \BibitemOpen
  \bibfield  {author} {\bibinfo {author} {\bibfnamefont {J.}~\bibnamefont
  {Dalibard}}, \bibinfo {author} {\bibfnamefont {F.}~\bibnamefont {Gerbier}},
  \bibinfo {author} {\bibfnamefont {G.}~\bibnamefont
  {Juzeli\ifmmode~\bar{u}\else \={u}\fi{}nas}}, \ and\ \bibinfo {author}
  {\bibfnamefont {P.}~\bibnamefont {\"Ohberg}},\ }\href {\doibase
  10.1103/RevModPhys.83.1523} {\bibfield  {journal} {\bibinfo  {journal} {Rev.
  Mod. Phys.}\ }\textbf {\bibinfo {volume} {83}},\ \bibinfo {pages} {1523}
  (\bibinfo {year} {2011})}\BibitemShut {NoStop}%
\bibitem [{\citenamefont {Atala}\ \emph {et~al.}(2014)\citenamefont {Atala},
  \citenamefont {Aidelsburger}, \citenamefont {Lohse}, \citenamefont
  {Barreiro}, \citenamefont {Paredes},\ and\ \citenamefont
  {Bloch}}]{Atala2014}%
  \BibitemOpen
  \bibfield  {author} {\bibinfo {author} {\bibfnamefont {M.}~\bibnamefont
  {Atala}}, \bibinfo {author} {\bibfnamefont {M.}~\bibnamefont {Aidelsburger}},
  \bibinfo {author} {\bibfnamefont {M.}~\bibnamefont {Lohse}}, \bibinfo
  {author} {\bibfnamefont {J.~T.}\ \bibnamefont {Barreiro}}, \bibinfo {author}
  {\bibfnamefont {B.}~\bibnamefont {Paredes}}, \ and\ \bibinfo {author}
  {\bibfnamefont {I.}~\bibnamefont {Bloch}},\ }\href {\doibase
  10.1038/nphys2998} {\bibfield  {journal} {\bibinfo  {journal} {Nature
  Physics}\ }\textbf {\bibinfo {volume} {10}},\ \bibinfo {pages} {588}
  (\bibinfo {year} {2014})}\BibitemShut {NoStop}%
\bibitem [{\citenamefont {Fukuhara}\ \emph {et~al.}(2013)\citenamefont
  {Fukuhara}, \citenamefont {Schau{\ss}}, \citenamefont {Endres}, \citenamefont
  {Hild}, \citenamefont {Cheneau}, \citenamefont {Bloch},\ and\ \citenamefont
  {Gross}}]{Fukuhara2013}%
  \BibitemOpen
  \bibfield  {author} {\bibinfo {author} {\bibfnamefont {T.}~\bibnamefont
  {Fukuhara}}, \bibinfo {author} {\bibfnamefont {P.}~\bibnamefont
  {Schau{\ss}}}, \bibinfo {author} {\bibfnamefont {M.}~\bibnamefont {Endres}},
  \bibinfo {author} {\bibfnamefont {S.}~\bibnamefont {Hild}}, \bibinfo {author}
  {\bibfnamefont {M.}~\bibnamefont {Cheneau}}, \bibinfo {author} {\bibfnamefont
  {I.}~\bibnamefont {Bloch}}, \ and\ \bibinfo {author} {\bibfnamefont
  {C.}~\bibnamefont {Gross}},\ }\href {\doibase 10.1038/nature12541} {\bibfield
   {journal} {\bibinfo  {journal} {Nature}\ }\textbf {\bibinfo {volume}
  {502}},\ \bibinfo {pages} {76} (\bibinfo {year} {2013})}\BibitemShut
  {NoStop}%
\bibitem [{\citenamefont {Struck}\ \emph {et~al.}(2012)\citenamefont {Struck},
  \citenamefont {\"Olschl\"ager}, \citenamefont {Weinberg}, \citenamefont
  {Hauke}, \citenamefont {Simonet}, \citenamefont {Eckardt}, \citenamefont
  {Lewenstein}, \citenamefont {Sengstock},\ and\ \citenamefont
  {Windpassinger}}]{PhysRevLett.108.225304}%
  \BibitemOpen
  \bibfield  {author} {\bibinfo {author} {\bibfnamefont {J.}~\bibnamefont
  {Struck}}, \bibinfo {author} {\bibfnamefont {C.}~\bibnamefont
  {\"Olschl\"ager}}, \bibinfo {author} {\bibfnamefont {M.}~\bibnamefont
  {Weinberg}}, \bibinfo {author} {\bibfnamefont {P.}~\bibnamefont {Hauke}},
  \bibinfo {author} {\bibfnamefont {J.}~\bibnamefont {Simonet}}, \bibinfo
  {author} {\bibfnamefont {A.}~\bibnamefont {Eckardt}}, \bibinfo {author}
  {\bibfnamefont {M.}~\bibnamefont {Lewenstein}}, \bibinfo {author}
  {\bibfnamefont {K.}~\bibnamefont {Sengstock}}, \ and\ \bibinfo {author}
  {\bibfnamefont {P.}~\bibnamefont {Windpassinger}},\ }\href {\doibase
  10.1103/PhysRevLett.108.225304} {\bibfield  {journal} {\bibinfo  {journal}
  {Phys. Rev. Lett.}\ }\textbf {\bibinfo {volume} {108}},\ \bibinfo {pages}
  {225304} (\bibinfo {year} {2012})}\BibitemShut {NoStop}%
\bibitem [{\citenamefont {Cooper}\ and\ \citenamefont
  {Dalibard}(2013)}]{PhysRevLett.110.185301}%
  \BibitemOpen
  \bibfield  {author} {\bibinfo {author} {\bibfnamefont {N.~R.}\ \bibnamefont
  {Cooper}}\ and\ \bibinfo {author} {\bibfnamefont {J.}~\bibnamefont
  {Dalibard}},\ }\href {\doibase 10.1103/PhysRevLett.110.185301} {\bibfield
  {journal} {\bibinfo  {journal} {Phys. Rev. Lett.}\ }\textbf {\bibinfo
  {volume} {110}},\ \bibinfo {pages} {185301} (\bibinfo {year}
  {2013})}\BibitemShut {NoStop}%
\bibitem [{\citenamefont {Aidelsburger}\ \emph {et~al.}(2013)\citenamefont
  {Aidelsburger}, \citenamefont {Atala}, \citenamefont {Lohse}, \citenamefont
  {Barreiro}, \citenamefont {Paredes},\ and\ \citenamefont
  {Bloch}}]{PhysRevLett.111.185301}%
  \BibitemOpen
  \bibfield  {author} {\bibinfo {author} {\bibfnamefont {M.}~\bibnamefont
  {Aidelsburger}}, \bibinfo {author} {\bibfnamefont {M.}~\bibnamefont {Atala}},
  \bibinfo {author} {\bibfnamefont {M.}~\bibnamefont {Lohse}}, \bibinfo
  {author} {\bibfnamefont {J.~T.}\ \bibnamefont {Barreiro}}, \bibinfo {author}
  {\bibfnamefont {B.}~\bibnamefont {Paredes}}, \ and\ \bibinfo {author}
  {\bibfnamefont {I.}~\bibnamefont {Bloch}},\ }\href {\doibase
  10.1103/PhysRevLett.111.185301} {\bibfield  {journal} {\bibinfo  {journal}
  {Phys. Rev. Lett.}\ }\textbf {\bibinfo {volume} {111}},\ \bibinfo {pages}
  {185301} (\bibinfo {year} {2013})}\BibitemShut {NoStop}%
\bibitem [{\citenamefont {Miyake}\ \emph {et~al.}(2013)\citenamefont {Miyake},
  \citenamefont {Siviloglou}, \citenamefont {Kennedy}, \citenamefont {Burton},\
  and\ \citenamefont {Ketterle}}]{PhysRevLett.111.185302}%
  \BibitemOpen
  \bibfield  {author} {\bibinfo {author} {\bibfnamefont {H.}~\bibnamefont
  {Miyake}}, \bibinfo {author} {\bibfnamefont {G.~A.}\ \bibnamefont
  {Siviloglou}}, \bibinfo {author} {\bibfnamefont {C.~J.}\ \bibnamefont
  {Kennedy}}, \bibinfo {author} {\bibfnamefont {W.~C.}\ \bibnamefont {Burton}},
  \ and\ \bibinfo {author} {\bibfnamefont {W.}~\bibnamefont {Ketterle}},\
  }\href {\doibase 10.1103/PhysRevLett.111.185302} {\bibfield  {journal}
  {\bibinfo  {journal} {Phys. Rev. Lett.}\ }\textbf {\bibinfo {volume} {111}},\
  \bibinfo {pages} {185302} (\bibinfo {year} {2013})}\BibitemShut {NoStop}%
\bibitem [{\citenamefont {Jotzu}\ \emph {et~al.}(2014)\citenamefont {Jotzu},
  \citenamefont {Messer}, \citenamefont {Desbuquois}, \citenamefont {Lebrat},
  \citenamefont {Uehlinger}, \citenamefont {Greif},\ and\ \citenamefont
  {Esslinger}}]{Jotzu2014}%
  \BibitemOpen
  \bibfield  {author} {\bibinfo {author} {\bibfnamefont {G.}~\bibnamefont
  {Jotzu}}, \bibinfo {author} {\bibfnamefont {M.}~\bibnamefont {Messer}},
  \bibinfo {author} {\bibfnamefont {R.}~\bibnamefont {Desbuquois}}, \bibinfo
  {author} {\bibfnamefont {M.}~\bibnamefont {Lebrat}}, \bibinfo {author}
  {\bibfnamefont {T.}~\bibnamefont {Uehlinger}}, \bibinfo {author}
  {\bibfnamefont {D.}~\bibnamefont {Greif}}, \ and\ \bibinfo {author}
  {\bibfnamefont {T.}~\bibnamefont {Esslinger}},\ }\href {\doibase
  10.1038/nature13915} {\bibfield  {journal} {\bibinfo  {journal} {Nature}\
  }\textbf {\bibinfo {volume} {515}},\ \bibinfo {pages} {237} (\bibinfo {year}
  {2014})}\BibitemShut {NoStop}%
\bibitem [{\citenamefont {Celi}\ \emph {et~al.}(2014)\citenamefont {Celi},
  \citenamefont {Massignan}, \citenamefont {Ruseckas}, \citenamefont {Goldman},
  \citenamefont {Spielman}, \citenamefont {Juzeli\ifmmode~\bar{u}\else
  \={u}\fi{}nas},\ and\ \citenamefont {Lewenstein}}]{PhysRevLett.112.043001}%
  \BibitemOpen
  \bibfield  {author} {\bibinfo {author} {\bibfnamefont {A.}~\bibnamefont
  {Celi}}, \bibinfo {author} {\bibfnamefont {P.}~\bibnamefont {Massignan}},
  \bibinfo {author} {\bibfnamefont {J.}~\bibnamefont {Ruseckas}}, \bibinfo
  {author} {\bibfnamefont {N.}~\bibnamefont {Goldman}}, \bibinfo {author}
  {\bibfnamefont {I.~B.}\ \bibnamefont {Spielman}}, \bibinfo {author}
  {\bibfnamefont {G.}~\bibnamefont {Juzeli\ifmmode~\bar{u}\else
  \={u}\fi{}nas}}, \ and\ \bibinfo {author} {\bibfnamefont {M.}~\bibnamefont
  {Lewenstein}},\ }\href {\doibase 10.1103/PhysRevLett.112.043001} {\bibfield
  {journal} {\bibinfo  {journal} {Phys. Rev. Lett.}\ }\textbf {\bibinfo
  {volume} {112}},\ \bibinfo {pages} {043001} (\bibinfo {year}
  {2014})}\BibitemShut {NoStop}%
\bibitem [{\citenamefont {Goldman}\ \emph {et~al.}(2016)\citenamefont
  {Goldman}, \citenamefont {Budich},\ and\ \citenamefont
  {Zoller}}]{Goldman2016}%
  \BibitemOpen
  \bibfield  {author} {\bibinfo {author} {\bibfnamefont {N.}~\bibnamefont
  {Goldman}}, \bibinfo {author} {\bibfnamefont {J.~C.}\ \bibnamefont {Budich}},
  \ and\ \bibinfo {author} {\bibfnamefont {P.}~\bibnamefont {Zoller}},\ }\href
  {\doibase 10.1038/nphys3803} {\bibfield  {journal} {\bibinfo  {journal}
  {Nature Physics}\ }\textbf {\bibinfo {volume} {12}},\ \bibinfo {pages} {639}
  (\bibinfo {year} {2016})}\BibitemShut {NoStop}%
\bibitem [{\citenamefont {An}\ \emph {et~al.}(2017)\citenamefont {An},
  \citenamefont {Meier},\ and\ \citenamefont {Gadway}}]{Ane1602685}%
  \BibitemOpen
  \bibfield  {author} {\bibinfo {author} {\bibfnamefont {F.~A.}\ \bibnamefont
  {An}}, \bibinfo {author} {\bibfnamefont {E.~J.}\ \bibnamefont {Meier}}, \
  and\ \bibinfo {author} {\bibfnamefont {B.}~\bibnamefont {Gadway}},\ }\href
  {\doibase 10.1126/sciadv.1602685} {\bibfield  {journal} {\bibinfo  {journal}
  {Science Advances}\ }\textbf {\bibinfo {volume} {3}} (\bibinfo {year}
  {2017}),\ 10.1126/sciadv.1602685}\BibitemShut {NoStop}%
\bibitem [{\citenamefont {Sharpe}\ \emph {et~al.}(2019)\citenamefont {Sharpe},
  \citenamefont {Fox}, \citenamefont {Barnard}, \citenamefont {Finney},
  \citenamefont {Watanabe}, \citenamefont {Taniguchi}, \citenamefont
  {Kastner},\ and\ \citenamefont {Goldhaber-Gordon}}]{Sharpe605}%
  \BibitemOpen
  \bibfield  {author} {\bibinfo {author} {\bibfnamefont {A.~L.}\ \bibnamefont
  {Sharpe}}, \bibinfo {author} {\bibfnamefont {E.~J.}\ \bibnamefont {Fox}},
  \bibinfo {author} {\bibfnamefont {A.~W.}\ \bibnamefont {Barnard}}, \bibinfo
  {author} {\bibfnamefont {J.}~\bibnamefont {Finney}}, \bibinfo {author}
  {\bibfnamefont {K.}~\bibnamefont {Watanabe}}, \bibinfo {author}
  {\bibfnamefont {T.}~\bibnamefont {Taniguchi}}, \bibinfo {author}
  {\bibfnamefont {M.~A.}\ \bibnamefont {Kastner}}, \ and\ \bibinfo {author}
  {\bibfnamefont {D.}~\bibnamefont {Goldhaber-Gordon}},\ }\href {\doibase
  10.1126/science.aaw3780} {\bibfield  {journal} {\bibinfo  {journal}
  {Science}\ }\textbf {\bibinfo {volume} {365}},\ \bibinfo {pages} {605}
  (\bibinfo {year} {2019})}\BibitemShut {NoStop}%
\bibitem [{\citenamefont {Zhang}\ \emph {et~al.}(2018)\citenamefont {Zhang},
  \citenamefont {Zhu},\ and\ \citenamefont {Batista}}]{PhysRevB.97.140507}%
  \BibitemOpen
  \bibfield  {author} {\bibinfo {author} {\bibfnamefont {S.-S.}\ \bibnamefont
  {Zhang}}, \bibinfo {author} {\bibfnamefont {W.}~\bibnamefont {Zhu}}, \ and\
  \bibinfo {author} {\bibfnamefont {C.~D.}\ \bibnamefont {Batista}},\ }\href
  {\doibase 10.1103/PhysRevB.97.140507} {\bibfield  {journal} {\bibinfo
  {journal} {Phys. Rev. B}\ }\textbf {\bibinfo {volume} {97}},\ \bibinfo
  {pages} {140507} (\bibinfo {year} {2018})}\BibitemShut {NoStop}%
\bibitem [{\citenamefont {Braaten}\ and\ \citenamefont
  {Hammer}(2006)}]{BRAATEN2006259}%
  \BibitemOpen
  \bibfield  {author} {\bibinfo {author} {\bibfnamefont {E.}~\bibnamefont
  {Braaten}}\ and\ \bibinfo {author} {\bibfnamefont {H.-W.}\ \bibnamefont
  {Hammer}},\ }\href {\doibase https://doi.org/10.1016/j.physrep.2006.03.001}
  {\bibfield  {journal} {\bibinfo  {journal} {Physics Reports}\ }\textbf
  {\bibinfo {volume} {428}},\ \bibinfo {pages} {259} (\bibinfo {year}
  {2006})}\BibitemShut {NoStop}%
\bibitem [{\citenamefont {Greene}\ \emph {et~al.}(2017)\citenamefont {Greene},
  \citenamefont {Giannakeas},\ and\ \citenamefont
  {P\'erez-R\'{\i}os}}]{RevModPhys.89.035006}%
  \BibitemOpen
  \bibfield  {author} {\bibinfo {author} {\bibfnamefont {C.~H.}\ \bibnamefont
  {Greene}}, \bibinfo {author} {\bibfnamefont {P.}~\bibnamefont {Giannakeas}},
  \ and\ \bibinfo {author} {\bibfnamefont {J.}~\bibnamefont
  {P\'erez-R\'{\i}os}},\ }\href {\doibase 10.1103/RevModPhys.89.035006}
  {\bibfield  {journal} {\bibinfo  {journal} {Rev. Mod. Phys.}\ }\textbf
  {\bibinfo {volume} {89}},\ \bibinfo {pages} {035006} (\bibinfo {year}
  {2017})}\BibitemShut {NoStop}%
\bibitem [{\citenamefont {Hasenfratz}\ and\ \citenamefont
  {Kuti}(1978)}]{HASENFRATZ197875}%
  \BibitemOpen
  \bibfield  {author} {\bibinfo {author} {\bibfnamefont {P.}~\bibnamefont
  {Hasenfratz}}\ and\ \bibinfo {author} {\bibfnamefont {J.}~\bibnamefont
  {Kuti}},\ }\href {\doibase https://doi.org/10.1016/0370-1573(78)90076-5}
  {\bibfield  {journal} {\bibinfo  {journal} {Physics Reports}\ }\textbf
  {\bibinfo {volume} {40}},\ \bibinfo {pages} {75 } (\bibinfo {year}
  {1978})}\BibitemShut {NoStop}%
\bibitem [{\citenamefont {Jim\'enez-Garc\'{\i}a}\ \emph
  {et~al.}(2012)\citenamefont {Jim\'enez-Garc\'{\i}a}, \citenamefont {LeBlanc},
  \citenamefont {Williams}, \citenamefont {Beeler}, \citenamefont {Perry},\
  and\ \citenamefont {Spielman}}]{PhysRevLett.108.225303}%
  \BibitemOpen
  \bibfield  {author} {\bibinfo {author} {\bibfnamefont {K.}~\bibnamefont
  {Jim\'enez-Garc\'{\i}a}}, \bibinfo {author} {\bibfnamefont {L.~J.}\
  \bibnamefont {LeBlanc}}, \bibinfo {author} {\bibfnamefont {R.~A.}\
  \bibnamefont {Williams}}, \bibinfo {author} {\bibfnamefont {M.~C.}\
  \bibnamefont {Beeler}}, \bibinfo {author} {\bibfnamefont {A.~R.}\
  \bibnamefont {Perry}}, \ and\ \bibinfo {author} {\bibfnamefont {I.~B.}\
  \bibnamefont {Spielman}},\ }\href {\doibase 10.1103/PhysRevLett.108.225303}
  {\bibfield  {journal} {\bibinfo  {journal} {Phys. Rev. Lett.}\ }\textbf
  {\bibinfo {volume} {108}},\ \bibinfo {pages} {225303} (\bibinfo {year}
  {2012})}\BibitemShut {NoStop}%
\bibitem [{\citenamefont {Duan}\ \emph {et~al.}(2003)\citenamefont {Duan},
  \citenamefont {Demler},\ and\ \citenamefont {Lukin}}]{PhysRevLett.91.090402}%
  \BibitemOpen
  \bibfield  {author} {\bibinfo {author} {\bibfnamefont {L.-M.}\ \bibnamefont
  {Duan}}, \bibinfo {author} {\bibfnamefont {E.}~\bibnamefont {Demler}}, \ and\
  \bibinfo {author} {\bibfnamefont {M.~D.}\ \bibnamefont {Lukin}},\ }\href
  {\doibase 10.1103/PhysRevLett.91.090402} {\bibfield  {journal} {\bibinfo
  {journal} {Phys. Rev. Lett.}\ }\textbf {\bibinfo {volume} {91}},\ \bibinfo
  {pages} {090402} (\bibinfo {year} {2003})}\BibitemShut {NoStop}%
\bibitem [{\citenamefont {Anderson}(1950)}]{PhysRev.79.350}%
  \BibitemOpen
  \bibfield  {author} {\bibinfo {author} {\bibfnamefont {P.~W.}\ \bibnamefont
  {Anderson}},\ }\href {\doibase 10.1103/PhysRev.79.350} {\bibfield  {journal}
  {\bibinfo  {journal} {Phys. Rev.}\ }\textbf {\bibinfo {volume} {79}},\
  \bibinfo {pages} {350} (\bibinfo {year} {1950})}\BibitemShut {NoStop}%
\bibitem [{\citenamefont {Hofstetter}\ \emph {et~al.}(2002)\citenamefont
  {Hofstetter}, \citenamefont {Cirac}, \citenamefont {Zoller}, \citenamefont
  {Demler},\ and\ \citenamefont {Lukin}}]{Hofstetter2002}%
  \BibitemOpen
  \bibfield  {author} {\bibinfo {author} {\bibfnamefont {W.}~\bibnamefont
  {Hofstetter}}, \bibinfo {author} {\bibfnamefont {J.~I.}\ \bibnamefont
  {Cirac}}, \bibinfo {author} {\bibfnamefont {P.}~\bibnamefont {Zoller}},
  \bibinfo {author} {\bibfnamefont {E.}~\bibnamefont {Demler}}, \ and\ \bibinfo
  {author} {\bibfnamefont {M.~D.}\ \bibnamefont {Lukin}},\ }\href {\doibase
  10.1103/PhysRevLett.89.220407} {\bibfield  {journal} {\bibinfo  {journal}
  {Phys. Rev. Lett.}\ }\textbf {\bibinfo {volume} {89}},\ \bibinfo {pages}
  {220407} (\bibinfo {year} {2002})}\BibitemShut {NoStop}%
\bibitem [{\citenamefont {Winkler}\ \emph {et~al.}(2006)\citenamefont
  {Winkler}, \citenamefont {Thalhammer}, \citenamefont {Lang}, \citenamefont
  {Grimm}, \citenamefont {Hecker~Denschlag}, \citenamefont {Daley},
  \citenamefont {Kantian}, \citenamefont {B{\"u}chler},\ and\ \citenamefont
  {Zoller}}]{Winkler2006}%
  \BibitemOpen
  \bibfield  {author} {\bibinfo {author} {\bibfnamefont {K.}~\bibnamefont
  {Winkler}}, \bibinfo {author} {\bibfnamefont {G.}~\bibnamefont {Thalhammer}},
  \bibinfo {author} {\bibfnamefont {F.}~\bibnamefont {Lang}}, \bibinfo {author}
  {\bibfnamefont {R.}~\bibnamefont {Grimm}}, \bibinfo {author} {\bibfnamefont
  {J.}~\bibnamefont {Hecker~Denschlag}}, \bibinfo {author} {\bibfnamefont
  {A.~J.}\ \bibnamefont {Daley}}, \bibinfo {author} {\bibfnamefont
  {A.}~\bibnamefont {Kantian}}, \bibinfo {author} {\bibfnamefont {H.~P.}\
  \bibnamefont {B{\"u}chler}}, \ and\ \bibinfo {author} {\bibfnamefont
  {P.}~\bibnamefont {Zoller}},\ }\href {\doibase 10.1038/nature04918}
  {\bibfield  {journal} {\bibinfo  {journal} {Nature}\ }\textbf {\bibinfo
  {volume} {441}},\ \bibinfo {pages} {853} (\bibinfo {year}
  {2006})}\BibitemShut {NoStop}%
\bibitem [{\citenamefont {Trotzky}\ \emph {et~al.}(2008)\citenamefont
  {Trotzky}, \citenamefont {Cheinet}, \citenamefont {F{\"o}lling},
  \citenamefont {Feld}, \citenamefont {Schnorrberger}, \citenamefont {Rey},
  \citenamefont {Polkovnikov}, \citenamefont {Demler}, \citenamefont {Lukin},\
  and\ \citenamefont {Bloch}}]{Trotzky295}%
  \BibitemOpen
  \bibfield  {author} {\bibinfo {author} {\bibfnamefont {S.}~\bibnamefont
  {Trotzky}}, \bibinfo {author} {\bibfnamefont {P.}~\bibnamefont {Cheinet}},
  \bibinfo {author} {\bibfnamefont {S.}~\bibnamefont {F{\"o}lling}}, \bibinfo
  {author} {\bibfnamefont {M.}~\bibnamefont {Feld}}, \bibinfo {author}
  {\bibfnamefont {U.}~\bibnamefont {Schnorrberger}}, \bibinfo {author}
  {\bibfnamefont {A.~M.}\ \bibnamefont {Rey}}, \bibinfo {author} {\bibfnamefont
  {A.}~\bibnamefont {Polkovnikov}}, \bibinfo {author} {\bibfnamefont {E.~A.}\
  \bibnamefont {Demler}}, \bibinfo {author} {\bibfnamefont {M.~D.}\
  \bibnamefont {Lukin}}, \ and\ \bibinfo {author} {\bibfnamefont
  {I.}~\bibnamefont {Bloch}},\ }\href {\doibase 10.1126/science.1150841}
  {\bibfield  {journal} {\bibinfo  {journal} {Science}\ }\textbf {\bibinfo
  {volume} {319}},\ \bibinfo {pages} {295} (\bibinfo {year}
  {2008})}\BibitemShut {NoStop}%
\bibitem [{\citenamefont {Brown}\ \emph {et~al.}(2015)\citenamefont {Brown},
  \citenamefont {Wyllie}, \citenamefont {Koller}, \citenamefont {Goldschmidt},
  \citenamefont {Foss-Feig},\ and\ \citenamefont {Porto}}]{Brown540}%
  \BibitemOpen
  \bibfield  {author} {\bibinfo {author} {\bibfnamefont {R.~C.}\ \bibnamefont
  {Brown}}, \bibinfo {author} {\bibfnamefont {R.}~\bibnamefont {Wyllie}},
  \bibinfo {author} {\bibfnamefont {S.~B.}\ \bibnamefont {Koller}}, \bibinfo
  {author} {\bibfnamefont {E.~A.}\ \bibnamefont {Goldschmidt}}, \bibinfo
  {author} {\bibfnamefont {M.}~\bibnamefont {Foss-Feig}}, \ and\ \bibinfo
  {author} {\bibfnamefont {J.~V.}\ \bibnamefont {Porto}},\ }\href {\doibase
  10.1126/science.aaa1385} {\bibfield  {journal} {\bibinfo  {journal}
  {Science}\ }\textbf {\bibinfo {volume} {348}},\ \bibinfo {pages} {540}
  (\bibinfo {year} {2015})}\BibitemShut {NoStop}%
\bibitem [{\citenamefont {Nichols}\ \emph {et~al.}(2019)\citenamefont
  {Nichols}, \citenamefont {Cheuk}, \citenamefont {Okan}, \citenamefont
  {Hartke}, \citenamefont {Mendez}, \citenamefont {Senthil}, \citenamefont
  {Khatami}, \citenamefont {Zhang},\ and\ \citenamefont
  {Zwierlein}}]{Nichols383}%
  \BibitemOpen
  \bibfield  {author} {\bibinfo {author} {\bibfnamefont {M.~A.}\ \bibnamefont
  {Nichols}}, \bibinfo {author} {\bibfnamefont {L.~W.}\ \bibnamefont {Cheuk}},
  \bibinfo {author} {\bibfnamefont {M.}~\bibnamefont {Okan}}, \bibinfo {author}
  {\bibfnamefont {T.~R.}\ \bibnamefont {Hartke}}, \bibinfo {author}
  {\bibfnamefont {E.}~\bibnamefont {Mendez}}, \bibinfo {author} {\bibfnamefont
  {T.}~\bibnamefont {Senthil}}, \bibinfo {author} {\bibfnamefont
  {E.}~\bibnamefont {Khatami}}, \bibinfo {author} {\bibfnamefont
  {H.}~\bibnamefont {Zhang}}, \ and\ \bibinfo {author} {\bibfnamefont {M.~W.}\
  \bibnamefont {Zwierlein}},\ }\href {\doibase 10.1126/science.aat4387}
  {\bibfield  {journal} {\bibinfo  {journal} {Science}\ }\textbf {\bibinfo
  {volume} {363}},\ \bibinfo {pages} {383} (\bibinfo {year}
  {2019})}\BibitemShut {NoStop}%
\bibitem [{\citenamefont {Jepsen}\ \emph {et~al.}(2020)\citenamefont {Jepsen},
  \citenamefont {Amato-Grill}, \citenamefont {Dimitrova}, \citenamefont {Ho},
  \citenamefont {Demler},\ and\ \citenamefont {Ketterle}}]{Jepsen2020}%
  \BibitemOpen
  \bibfield  {author} {\bibinfo {author} {\bibfnamefont {P.~N.}\ \bibnamefont
  {Jepsen}}, \bibinfo {author} {\bibfnamefont {J.}~\bibnamefont {Amato-Grill}},
  \bibinfo {author} {\bibfnamefont {I.}~\bibnamefont {Dimitrova}}, \bibinfo
  {author} {\bibfnamefont {W.~W.}\ \bibnamefont {Ho}}, \bibinfo {author}
  {\bibfnamefont {E.}~\bibnamefont {Demler}}, \ and\ \bibinfo {author}
  {\bibfnamefont {W.}~\bibnamefont {Ketterle}},\ }\href {\doibase
  10.1038/s41586-020-3033-y} {\bibfield  {journal} {\bibinfo  {journal}
  {Nature}\ }\textbf {\bibinfo {volume} {588}},\ \bibinfo {pages} {403}
  (\bibinfo {year} {2020})}\BibitemShut {NoStop}%
\bibitem [{\citenamefont {Tai}\ \emph {et~al.}(2017)\citenamefont {Tai},
  \citenamefont {Lukin}, \citenamefont {Rispoli}, \citenamefont {Schittko},
  \citenamefont {Menke}, \citenamefont {Borgnia}, \citenamefont {Preiss},
  \citenamefont {Grusdt}, \citenamefont {Kaufman},\ and\ \citenamefont
  {Greiner}}]{Tai2017}%
  \BibitemOpen
  \bibfield  {author} {\bibinfo {author} {\bibfnamefont {M.~E.}\ \bibnamefont
  {Tai}}, \bibinfo {author} {\bibfnamefont {A.}~\bibnamefont {Lukin}}, \bibinfo
  {author} {\bibfnamefont {M.}~\bibnamefont {Rispoli}}, \bibinfo {author}
  {\bibfnamefont {R.}~\bibnamefont {Schittko}}, \bibinfo {author}
  {\bibfnamefont {T.}~\bibnamefont {Menke}}, \bibinfo {author} {\bibfnamefont
  {D.}~\bibnamefont {Borgnia}}, \bibinfo {author} {\bibfnamefont {P.~M.}\
  \bibnamefont {Preiss}}, \bibinfo {author} {\bibfnamefont {F.}~\bibnamefont
  {Grusdt}}, \bibinfo {author} {\bibfnamefont {A.~M.}\ \bibnamefont {Kaufman}},
  \ and\ \bibinfo {author} {\bibfnamefont {M.}~\bibnamefont {Greiner}},\ }\href
  {\doibase 10.1038/nature22811} {\bibfield  {journal} {\bibinfo  {journal}
  {Nature}\ }\textbf {\bibinfo {volume} {546}},\ \bibinfo {pages} {519}
  (\bibinfo {year} {2017})}\BibitemShut {NoStop}%
\bibitem [{\citenamefont {Nagaev}(1992)}]{NAGAEV199239}%
  \BibitemOpen
  \bibfield  {author} {\bibinfo {author} {\bibfnamefont {E.}~\bibnamefont
  {Nagaev}},\ }\href {\doibase https://doi.org/10.1016/0304-8853(92)90011-C}
  {\bibfield  {journal} {\bibinfo  {journal} {Journal of Magnetism and Magnetic
  Materials}\ }\textbf {\bibinfo {volume} {110}},\ \bibinfo {pages} {39}
  (\bibinfo {year} {1992})}\BibitemShut {NoStop}%
\bibitem [{\citenamefont {Alexandrov}(2007)}]{alexandrov2007polarons}%
  \BibitemOpen
  \bibfield  {author} {\bibinfo {author} {\bibfnamefont {A.~S.}\ \bibnamefont
  {Alexandrov}},\ }\href@noop {} {\emph {\bibinfo {title} {Polarons in advanced
  materials}}}\ (\bibinfo  {publisher} {Canopus Pub},\ \bibinfo {address}
  {Dordrecht},\ \bibinfo {year} {2007})\BibitemShut {NoStop}%
\bibitem [{\citenamefont {Auerbach}(1994)}]{auerbach1994interacting}%
  \BibitemOpen
  \bibfield  {author} {\bibinfo {author} {\bibfnamefont {A.}~\bibnamefont
  {Auerbach}},\ }\href@noop {} {\emph {\bibinfo {title} {Interacting Electrons
  and Quantum Magnetism}}}\ (\bibinfo  {publisher} {Springer New York},\
  \bibinfo {address} {New York, NY},\ \bibinfo {year} {1994})\BibitemShut
  {NoStop}%
\bibitem [{\citenamefont {White}\ and\ \citenamefont
  {Affleck}(2001)}]{PhysRevB.64.024411}%
  \BibitemOpen
  \bibfield  {author} {\bibinfo {author} {\bibfnamefont {S.~R.}\ \bibnamefont
  {White}}\ and\ \bibinfo {author} {\bibfnamefont {I.}~\bibnamefont
  {Affleck}},\ }\href {\doibase 10.1103/PhysRevB.64.024411} {\bibfield
  {journal} {\bibinfo  {journal} {Phys. Rev. B}\ }\textbf {\bibinfo {volume}
  {64}},\ \bibinfo {pages} {024411} (\bibinfo {year} {2001})}\BibitemShut
  {NoStop}%
\bibitem [{\citenamefont {Haerter}\ and\ \citenamefont
  {Shastry}(2005)}]{PhysRevLett.95.087202}%
  \BibitemOpen
  \bibfield  {author} {\bibinfo {author} {\bibfnamefont {J.~O.}\ \bibnamefont
  {Haerter}}\ and\ \bibinfo {author} {\bibfnamefont {B.~S.}\ \bibnamefont
  {Shastry}},\ }\href {\doibase 10.1103/PhysRevLett.95.087202} {\bibfield
  {journal} {\bibinfo  {journal} {Phys. Rev. Lett.}\ }\textbf {\bibinfo
  {volume} {95}},\ \bibinfo {pages} {087202} (\bibinfo {year}
  {2005})}\BibitemShut {NoStop}%
\bibitem [{\citenamefont {Sposetti}\ \emph {et~al.}(2014)\citenamefont
  {Sposetti}, \citenamefont {Bravo}, \citenamefont {Trumper}, \citenamefont
  {Gazza},\ and\ \citenamefont {Manuel}}]{PhysRevLett.112.187204}%
  \BibitemOpen
  \bibfield  {author} {\bibinfo {author} {\bibfnamefont {C.~N.}\ \bibnamefont
  {Sposetti}}, \bibinfo {author} {\bibfnamefont {B.}~\bibnamefont {Bravo}},
  \bibinfo {author} {\bibfnamefont {A.~E.}\ \bibnamefont {Trumper}}, \bibinfo
  {author} {\bibfnamefont {C.~J.}\ \bibnamefont {Gazza}}, \ and\ \bibinfo
  {author} {\bibfnamefont {L.~O.}\ \bibnamefont {Manuel}},\ }\href {\doibase
  10.1103/PhysRevLett.112.187204} {\bibfield  {journal} {\bibinfo  {journal}
  {Phys. Rev. Lett.}\ }\textbf {\bibinfo {volume} {112}},\ \bibinfo {pages}
  {187204} (\bibinfo {year} {2014})}\BibitemShut {NoStop}%
\bibitem [{Sup()}]{Suppl}%
  \BibitemOpen
  \href@noop {} {\ }\bibinfo {note} {See Supplementary Material}\BibitemShut
  {NoStop}%
\bibitem [{\citenamefont {White}\ and\ \citenamefont
  {Scalapino}(1997)}]{PhysRevB.55.6504}%
  \BibitemOpen
  \bibfield  {author} {\bibinfo {author} {\bibfnamefont {S.~R.}\ \bibnamefont
  {White}}\ and\ \bibinfo {author} {\bibfnamefont {D.~J.}\ \bibnamefont
  {Scalapino}},\ }\href {\doibase 10.1103/PhysRevB.55.6504} {\bibfield
  {journal} {\bibinfo  {journal} {Phys. Rev. B}\ }\textbf {\bibinfo {volume}
  {55}},\ \bibinfo {pages} {6504} (\bibinfo {year} {1997})}\BibitemShut
  {NoStop}%
\bibitem [{\citenamefont {Shraiman}\ and\ \citenamefont
  {Siggia}(1988)}]{PhysRevLett.60.740}%
  \BibitemOpen
  \bibfield  {author} {\bibinfo {author} {\bibfnamefont {B.~I.}\ \bibnamefont
  {Shraiman}}\ and\ \bibinfo {author} {\bibfnamefont {E.~D.}\ \bibnamefont
  {Siggia}},\ }\href {\doibase 10.1103/PhysRevLett.60.740} {\bibfield
  {journal} {\bibinfo  {journal} {Phys. Rev. Lett.}\ }\textbf {\bibinfo
  {volume} {60}},\ \bibinfo {pages} {740} (\bibinfo {year} {1988})}\BibitemShut
  {NoStop}%
\bibitem [{\citenamefont {Grusdt}\ \emph
  {et~al.}(2018{\natexlab{a}})\citenamefont {Grusdt}, \citenamefont
  {K\'anasz-Nagy}, \citenamefont {Bohrdt}, \citenamefont {Chiu}, \citenamefont
  {Ji}, \citenamefont {Greiner}, \citenamefont {Greif},\ and\ \citenamefont
  {Demler}}]{PhysRevX.8.011046}%
  \BibitemOpen
  \bibfield  {author} {\bibinfo {author} {\bibfnamefont {F.}~\bibnamefont
  {Grusdt}}, \bibinfo {author} {\bibfnamefont {M.}~\bibnamefont
  {K\'anasz-Nagy}}, \bibinfo {author} {\bibfnamefont {A.}~\bibnamefont
  {Bohrdt}}, \bibinfo {author} {\bibfnamefont {C.~S.}\ \bibnamefont {Chiu}},
  \bibinfo {author} {\bibfnamefont {G.}~\bibnamefont {Ji}}, \bibinfo {author}
  {\bibfnamefont {M.}~\bibnamefont {Greiner}}, \bibinfo {author} {\bibfnamefont
  {D.}~\bibnamefont {Greif}}, \ and\ \bibinfo {author} {\bibfnamefont
  {E.}~\bibnamefont {Demler}},\ }\href {\doibase 10.1103/PhysRevX.8.011046}
  {\bibfield  {journal} {\bibinfo  {journal} {Phys. Rev. X}\ }\textbf {\bibinfo
  {volume} {8}},\ \bibinfo {pages} {011046} (\bibinfo {year}
  {2018}{\natexlab{a}})}\BibitemShut {NoStop}%
\bibitem [{\citenamefont {Giamarchi}(2004)}]{Giamarchi:743140}%
  \BibitemOpen
  \bibfield  {author} {\bibinfo {author} {\bibfnamefont {T.}~\bibnamefont
  {Giamarchi}},\ }\href {\doibase 10.1093/acprof:oso/9780198525004.001.0001}
  {\emph {\bibinfo {title} {{Quantum physics in one dimension}}}},\
  International series of monographs on physics\ (\bibinfo  {publisher}
  {Clarendon Press},\ \bibinfo {address} {Oxford},\ \bibinfo {year}
  {2004})\BibitemShut {NoStop}%
\bibitem [{\citenamefont {Preiss}\ \emph {et~al.}(2015)\citenamefont {Preiss},
  \citenamefont {Ma}, \citenamefont {Tai}, \citenamefont {Lukin}, \citenamefont
  {Rispoli}, \citenamefont {Zupancic}, \citenamefont {Lahini}, \citenamefont
  {Islam},\ and\ \citenamefont {Greiner}}]{Preiss1229}%
  \BibitemOpen
  \bibfield  {author} {\bibinfo {author} {\bibfnamefont {P.~M.}\ \bibnamefont
  {Preiss}}, \bibinfo {author} {\bibfnamefont {R.}~\bibnamefont {Ma}}, \bibinfo
  {author} {\bibfnamefont {M.~E.}\ \bibnamefont {Tai}}, \bibinfo {author}
  {\bibfnamefont {A.}~\bibnamefont {Lukin}}, \bibinfo {author} {\bibfnamefont
  {M.}~\bibnamefont {Rispoli}}, \bibinfo {author} {\bibfnamefont
  {P.}~\bibnamefont {Zupancic}}, \bibinfo {author} {\bibfnamefont
  {Y.}~\bibnamefont {Lahini}}, \bibinfo {author} {\bibfnamefont
  {R.}~\bibnamefont {Islam}}, \ and\ \bibinfo {author} {\bibfnamefont
  {M.}~\bibnamefont {Greiner}},\ }\href {\doibase 10.1126/science.1260364}
  {\bibfield  {journal} {\bibinfo  {journal} {Science}\ }\textbf {\bibinfo
  {volume} {347}},\ \bibinfo {pages} {1229} (\bibinfo {year}
  {2015})}\BibitemShut {NoStop}%
\bibitem [{\citenamefont {Schneider}\ \emph {et~al.}(2012)\citenamefont
  {Schneider}, \citenamefont {Hackerm{\"u}ller}, \citenamefont {Ronzheimer},
  \citenamefont {Will}, \citenamefont {Braun}, \citenamefont {Best},
  \citenamefont {Bloch}, \citenamefont {Demler}, \citenamefont {Mandt},
  \citenamefont {Rasch},\ and\ \citenamefont {Rosch}}]{Schneider2012}%
  \BibitemOpen
  \bibfield  {author} {\bibinfo {author} {\bibfnamefont {U.}~\bibnamefont
  {Schneider}}, \bibinfo {author} {\bibfnamefont {L.}~\bibnamefont
  {Hackerm{\"u}ller}}, \bibinfo {author} {\bibfnamefont {J.~P.}\ \bibnamefont
  {Ronzheimer}}, \bibinfo {author} {\bibfnamefont {S.}~\bibnamefont {Will}},
  \bibinfo {author} {\bibfnamefont {S.}~\bibnamefont {Braun}}, \bibinfo
  {author} {\bibfnamefont {T.}~\bibnamefont {Best}}, \bibinfo {author}
  {\bibfnamefont {I.}~\bibnamefont {Bloch}}, \bibinfo {author} {\bibfnamefont
  {E.}~\bibnamefont {Demler}}, \bibinfo {author} {\bibfnamefont
  {S.}~\bibnamefont {Mandt}}, \bibinfo {author} {\bibfnamefont
  {D.}~\bibnamefont {Rasch}}, \ and\ \bibinfo {author} {\bibfnamefont
  {A.}~\bibnamefont {Rosch}},\ }\href {\doibase 10.1038/nphys2205} {\bibfield
  {journal} {\bibinfo  {journal} {Nature Physics}\ }\textbf {\bibinfo {volume}
  {8}},\ \bibinfo {pages} {213} (\bibinfo {year} {2012})}\BibitemShut {NoStop}%
\bibitem [{\citenamefont {Grusdt}\ \emph
  {et~al.}(2018{\natexlab{b}})\citenamefont {Grusdt}, \citenamefont {Zhu},
  \citenamefont {Shi},\ and\ \citenamefont
  {Demler}}]{10.21468/SciPostPhys.5.6.057}%
  \BibitemOpen
  \bibfield  {author} {\bibinfo {author} {\bibfnamefont {F.}~\bibnamefont
  {Grusdt}}, \bibinfo {author} {\bibfnamefont {Z.}~\bibnamefont {Zhu}},
  \bibinfo {author} {\bibfnamefont {T.}~\bibnamefont {Shi}}, \ and\ \bibinfo
  {author} {\bibfnamefont {E.}~\bibnamefont {Demler}},\ }\href {\doibase
  10.21468/SciPostPhys.5.6.057} {\bibfield  {journal} {\bibinfo  {journal}
  {SciPost Phys.}\ }\textbf {\bibinfo {volume} {5}},\ \bibinfo {pages} {57}
  (\bibinfo {year} {2018}{\natexlab{b}})}\BibitemShut {NoStop}%
\bibitem [{\citenamefont {Jepsen}\ \emph {et~al.}(2021)\citenamefont {Jepsen},
  \citenamefont {Ho}, \citenamefont {Amato-Grill}, \citenamefont {Dimitrova},
  \citenamefont {Demler},\ and\ \citenamefont
  {Ketterle}}]{jepsen2021transverse}%
  \BibitemOpen
  \bibfield  {author} {\bibinfo {author} {\bibfnamefont {P.~N.}\ \bibnamefont
  {Jepsen}}, \bibinfo {author} {\bibfnamefont {W.~W.}\ \bibnamefont {Ho}},
  \bibinfo {author} {\bibfnamefont {J.}~\bibnamefont {Amato-Grill}}, \bibinfo
  {author} {\bibfnamefont {I.}~\bibnamefont {Dimitrova}}, \bibinfo {author}
  {\bibfnamefont {E.}~\bibnamefont {Demler}}, \ and\ \bibinfo {author}
  {\bibfnamefont {W.}~\bibnamefont {Ketterle}},\ }\href@noop {} {} (\bibinfo
  {year} {2021}),\ \Eprint {http://arxiv.org/abs/2103.07866} {arXiv:2103.07866
  [cond-mat.quant-gas]} \BibitemShut {NoStop}%
\bibitem [{\citenamefont {Hauschild}\ and\ \citenamefont
  {Pollmann}(2018)}]{10.21468/SciPostPhysLectNotes.5}%
  \BibitemOpen
  \bibfield  {author} {\bibinfo {author} {\bibfnamefont {J.}~\bibnamefont
  {Hauschild}}\ and\ \bibinfo {author} {\bibfnamefont {F.}~\bibnamefont
  {Pollmann}},\ }\href {\doibase 10.21468/SciPostPhysLectNotes.5} {\bibfield
  {journal} {\bibinfo  {journal} {SciPost Phys. Lect. Notes}\ ,\ \bibinfo
  {pages} {5}} (\bibinfo {year} {2018})}\BibitemShut {NoStop}%
\bibitem [{\citenamefont {Ogata}\ and\ \citenamefont
  {Shiba}(1990)}]{PhysRevB.41.2326}%
  \BibitemOpen
  \bibfield  {author} {\bibinfo {author} {\bibfnamefont {M.}~\bibnamefont
  {Ogata}}\ and\ \bibinfo {author} {\bibfnamefont {H.}~\bibnamefont {Shiba}},\
  }\href {\doibase 10.1103/PhysRevB.41.2326} {\bibfield  {journal} {\bibinfo
  {journal} {Phys. Rev. B}\ }\textbf {\bibinfo {volume} {41}},\ \bibinfo
  {pages} {2326} (\bibinfo {year} {1990})}\BibitemShut {NoStop}%
\bibitem [{\citenamefont {Ren}\ and\ \citenamefont
  {Anderson}(1993)}]{PhysRevB.48.16662}%
  \BibitemOpen
  \bibfield  {author} {\bibinfo {author} {\bibfnamefont {Y.}~\bibnamefont
  {Ren}}\ and\ \bibinfo {author} {\bibfnamefont {P.~W.}\ \bibnamefont
  {Anderson}},\ }\href {\doibase 10.1103/PhysRevB.48.16662} {\bibfield
  {journal} {\bibinfo  {journal} {Phys. Rev. B}\ }\textbf {\bibinfo {volume}
  {48}},\ \bibinfo {pages} {16662} (\bibinfo {year} {1993})}\BibitemShut
  {NoStop}%
\bibitem [{\citenamefont {Zaanen}\ \emph {et~al.}(2001)\citenamefont {Zaanen},
  \citenamefont {Osman}, \citenamefont {Kruis}, \citenamefont {Nussinov},\ and\
  \citenamefont {Tworzydlo}}]{10.1080/13642810108208566}%
  \BibitemOpen
  \bibfield  {author} {\bibinfo {author} {\bibfnamefont {J.}~\bibnamefont
  {Zaanen}}, \bibinfo {author} {\bibfnamefont {O.~Y.}\ \bibnamefont {Osman}},
  \bibinfo {author} {\bibfnamefont {H.~V.}\ \bibnamefont {Kruis}}, \bibinfo
  {author} {\bibfnamefont {Z.}~\bibnamefont {Nussinov}}, \ and\ \bibinfo
  {author} {\bibfnamefont {J.}~\bibnamefont {Tworzydlo}},\ }\href {\doibase
  10.1080/13642810108208566} {\bibfield  {journal} {\bibinfo  {journal}
  {Philosophical Magazine B}\ }\textbf {\bibinfo {volume} {81}},\ \bibinfo
  {pages} {1485} (\bibinfo {year} {2001})}\BibitemShut {NoStop}%
\bibitem [{\citenamefont {Kruis}\ \emph {et~al.}(2004)\citenamefont {Kruis},
  \citenamefont {McCulloch}, \citenamefont {Nussinov},\ and\ \citenamefont
  {Zaanen}}]{PhysRevB.70.075109}%
  \BibitemOpen
  \bibfield  {author} {\bibinfo {author} {\bibfnamefont {H.~V.}\ \bibnamefont
  {Kruis}}, \bibinfo {author} {\bibfnamefont {I.~P.}\ \bibnamefont
  {McCulloch}}, \bibinfo {author} {\bibfnamefont {Z.}~\bibnamefont {Nussinov}},
  \ and\ \bibinfo {author} {\bibfnamefont {J.}~\bibnamefont {Zaanen}},\ }\href
  {\doibase 10.1103/PhysRevB.70.075109} {\bibfield  {journal} {\bibinfo
  {journal} {Phys. Rev. B}\ }\textbf {\bibinfo {volume} {70}},\ \bibinfo
  {pages} {075109} (\bibinfo {year} {2004})}\BibitemShut {NoStop}%
\bibitem [{\citenamefont {Hilker}\ \emph {et~al.}(2017)\citenamefont {Hilker},
  \citenamefont {Salomon}, \citenamefont {Grusdt}, \citenamefont {Omran},
  \citenamefont {Boll}, \citenamefont {Demler}, \citenamefont {Bloch},\ and\
  \citenamefont {Gross}}]{Hilker484}%
  \BibitemOpen
  \bibfield  {author} {\bibinfo {author} {\bibfnamefont {T.~A.}\ \bibnamefont
  {Hilker}}, \bibinfo {author} {\bibfnamefont {G.}~\bibnamefont {Salomon}},
  \bibinfo {author} {\bibfnamefont {F.}~\bibnamefont {Grusdt}}, \bibinfo
  {author} {\bibfnamefont {A.}~\bibnamefont {Omran}}, \bibinfo {author}
  {\bibfnamefont {M.}~\bibnamefont {Boll}}, \bibinfo {author} {\bibfnamefont
  {E.}~\bibnamefont {Demler}}, \bibinfo {author} {\bibfnamefont
  {I.}~\bibnamefont {Bloch}}, \ and\ \bibinfo {author} {\bibfnamefont
  {C.}~\bibnamefont {Gross}},\ }\href {\doibase 10.1126/science.aam8990}
  {\bibfield  {journal} {\bibinfo  {journal} {Science}\ }\textbf {\bibinfo
  {volume} {357}},\ \bibinfo {pages} {484} (\bibinfo {year}
  {2017})}\BibitemShut {NoStop}%
\bibitem [{\citenamefont {Bohrdt}\ \emph {et~al.}(2018)\citenamefont {Bohrdt},
  \citenamefont {Greif}, \citenamefont {Demler}, \citenamefont {Knap},\ and\
  \citenamefont {Grusdt}}]{PhysRevB.97.125117}%
  \BibitemOpen
  \bibfield  {author} {\bibinfo {author} {\bibfnamefont {A.}~\bibnamefont
  {Bohrdt}}, \bibinfo {author} {\bibfnamefont {D.}~\bibnamefont {Greif}},
  \bibinfo {author} {\bibfnamefont {E.}~\bibnamefont {Demler}}, \bibinfo
  {author} {\bibfnamefont {M.}~\bibnamefont {Knap}}, \ and\ \bibinfo {author}
  {\bibfnamefont {F.}~\bibnamefont {Grusdt}},\ }\href {\doibase
  10.1103/PhysRevB.97.125117} {\bibfield  {journal} {\bibinfo  {journal} {Phys.
  Rev. B}\ }\textbf {\bibinfo {volume} {97}},\ \bibinfo {pages} {125117}
  (\bibinfo {year} {2018})}\BibitemShut {NoStop}%
\end{thebibliography}%

\clearpage
\appendix
\section{Hole-magnon bound state}
\label{AppendixA}
\subsection{Two-body problem}
By doping a fully polarized insulator in a frustrated geometry with vacancies (i.e. holes) and flipped spins (i.e. magnons) effective interactions between them appear. In this section we consider an insulator polarized along the $z$ axis in a fermionic zigzag ladder and study the hole-magnon problem. Holes and magnons can rigorously be created by employing the transformations $\hat{c}_{\uparrow i}=\hat {h}^{\dagger}_i$ and $\hat{c}_{\downarrow i}=\hat{h}^{\dagger}_i S^+_i$. In this way the original $2L-N_h$ particles problem can be reduced to a problem of $N_h$ holes and $N_m$ magnons, being $2L$ the total number of sites in the ladder. 
Under these transformations the $t-J$ Hamiltonian becomes,
\begin{eqnarray}
\hat{H}_{tJ}& =& \pm t\sum_{i,\,e} \left( \hat{h}^{\dagger}_{i+e}\hat{h}_{i}+\text{h.c.} \right)+ J\sum_{i,\,e} \mathbf{S}_i \mathbf{S}_{i+e} \nonumber \\
&\pm &
t\sum_{i,\,e} \left( \hat{h}^{\dagger}_{i+e}\hat{h}_{i} S^-_i S^+_{i+e} + \text{h.c.}  \right),
\label{Eq:tJHolMag}
\end{eqnarray}
where $e=a,2a$ being $a$ the lattice spacing and the $+$ and $-$ signs correspond to the fermionic and bosonic case, respectively. They appear because of the anticonmutation and commutation relations satisfied by the hole operators in each case. 
Under these transformations the fully polarized insulating state $|\text{FP}\rangle$ becomes the vacuum of holes and magnons. Holes and magnons can be created by applying the respective creation operators. In particular the hole-magnon state is defined as $\hat{h}^{\dagger}_i S^-_j |\text{FP}\rangle \equiv |ij\rangle$. 
We can solve this two-particle problem by separating the center $R/a=(i+j)/2$ and relative $z/a=i-j$ motion using the set of states
\begin{equation}
|\psi\rangle = \sum_{ij} e^{i Q R} \psi_Q(z) |ij\rangle,
\end{equation}
where we introduce the total quasi-momentum $Q$ of the pair and the wavefunction in relative position space $\psi_Q(z)$. Since the interaction part of the Hamiltonian~\eqref{Eq:tJHolMag} does not couple states with different quasi-momentum $Q$ we can obtain the equation of motion for the relative part,
\begin{eqnarray}
&&E_Q \psi_Q(z) = \sum_{e=\pm a,\pm 2a} \left( \pm te^{iQe/2} + \frac{J}{2}e^{-iQe/2} \right)\psi_Q(z+e) \nonumber \\
&+& \sum_{e=\pm a,\pm 2a}\delta(z-e) \left( \pm t \psi_Q(-e) + \frac{J}{2} \psi_Q(e)  \right),
\label{Eq:EoM}
\end{eqnarray}
where the energy of the pair $E_Q$ depends parametrically on the total quasi-momentum. We measure the hole-magnon energy $E_Q$ with respect to the energy of the ferromagnetic background. The hole-magnon interaction contains an exchange term of strength $t$ and a bare nearest-neighbor interaction of strength $J/2$. For fermions (bosons) the nearest-neighbor interaction $J>0$ ($J<0$) has a repulsive (attractive) effect on the pair. Moreover, since parity is a good quantum number of our model the exchange term can be cast into a diagonal form $\psi_Q(-e)=\pm\psi_Q(e)$ by considering a symmetric or antisymmetric state with respect to the hole-magnon exchange. In this way, we notice that the exchange term has an attractive (repulsive) effect for the antisymmetric (symmetric) solution in the fermionic model. In the bosonic model we observe the opposite. Finally, since by definition we cannot have a hole and a magnon in the same position we impose a hard-core constrain in the solution $\psi_Q(0)=0$ by including a very strong on-site repulsion.

We numerically solve Eq.~\eqref{Eq:EoM} and we obtain a hole-magnon bound state for the fermionic model and an antibound state in the bosonic one, see Fig.~\ref{fig:EnergyBand_BF}. The bound and antibound states are antisymmetric with respect to the exchange of the hole and magnon positions. Thus we do not observe binding when the solution is symmetric. The appearance of an antisymmetric bound (antibound) state can be explained by the presence of the interexchange interaction which leads to an effective attraction (repulsion) between the hole and the magnon in the fermionic (bosonic) model.

\begin{figure}[t]
    \centering
    \includegraphics[width=1\columnwidth]{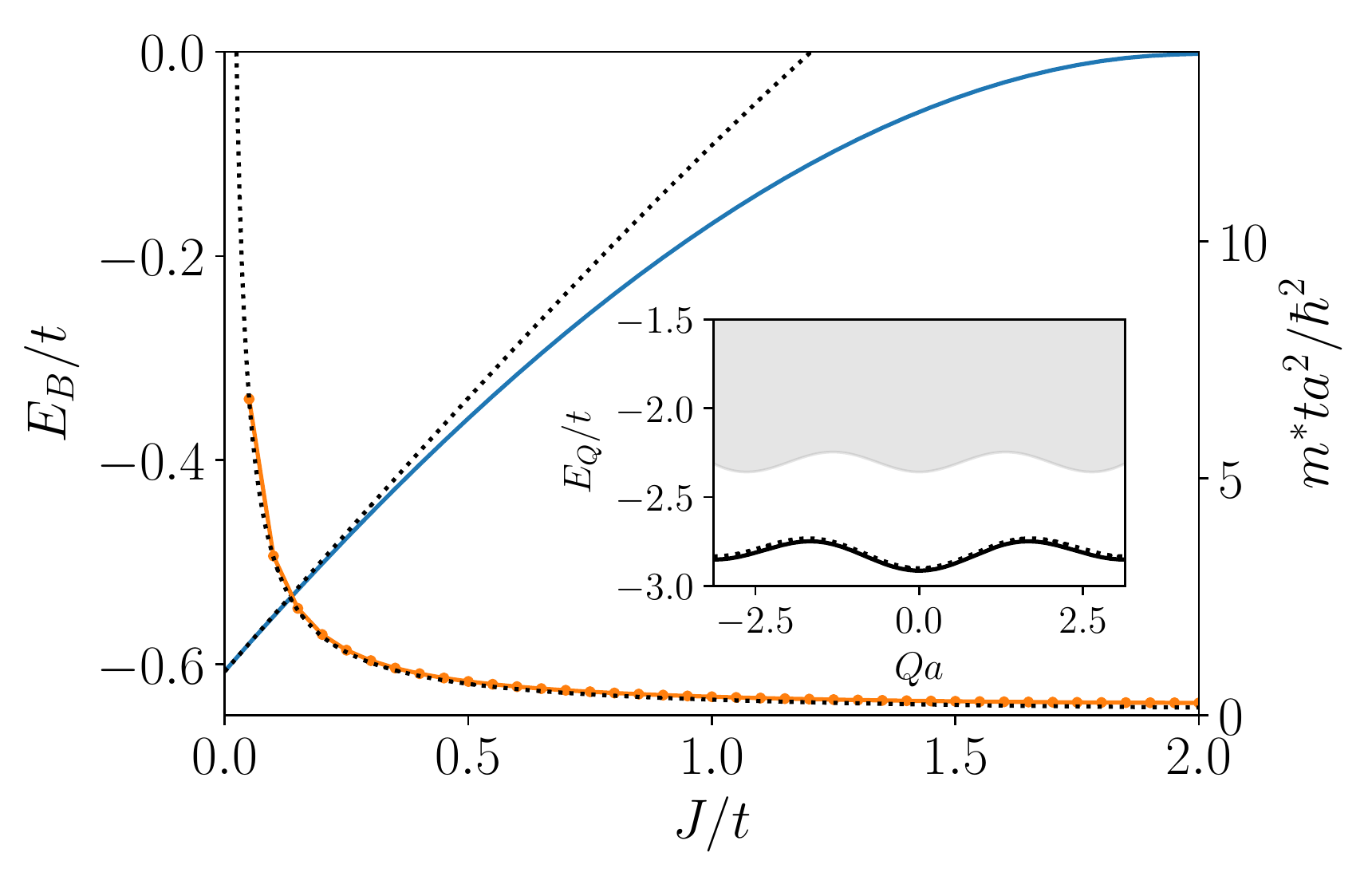} 
    \caption{Main panel: Binding energy $E_B$ (continuous line) and effective mass $m^*$ (dots) of the hole-magnon bound state in the fermionic zigzag ladder as a function of $J/t$. Inset: Band structure of the hole-magnon problem as a function of the total quasimomentum $Q$ for $J/t=0.1$. The grey area denotes the hole-magnon continuum and the continuous line the bound state band. Dotted lines correspond to the Born-Oppenheimer approximation, see main text.}
    \label{fig:EnergyMassBand}
\end{figure}

The minimum of the binding energy appears at $Q=0$ for any value of $J/t$. Therefore we define the binding energy at this quasi-momentum $E_B=E_{1H1M}-E_{1H}-E_{1M}$, see Fig.~\ref{fig:EnergyMassBand}. The binding energy is larger for small values of $J/t$ and it vanishes close to $J\approx 2t$ which can be easily interpreted as the point where the bare repulsion cancels the interexchange term. Since binding is stronger at $J/t=0$ we can neglect the magnon kinetic energy allowing us to perform a Born-Oppenheimer approximation. First, we solve Eq.~\eqref{Eq:EoM} for a static magnon and we obtain the hole energy $E^h$ and the wavefunction $\psi_Q^h(z)$. Then, we estimate the bound state energy by adding the kinetic energy of the magnon which is given by $E^m_Q= -J(\Gamma(e)\cos(Q)+\Gamma(2e)\cos(2Q))$,
where we define the overlap $\Gamma(e)=|\langle \psi^h(e)|\psi^h(0)\rangle|\approx |e|/3$ being $e=\pm a,\pm 2 a$. This gives us,
\begin{equation}
    E_Q = E^h - \frac{J}{3}\left(\cos(Qa)+2\cos(2Qa) \right).
    \label{Eq:BO}
\end{equation}
For small values of $J/t$ we observe a nice agreement between the Born-Oppenheimer approximation and the full solution of Eq.~\eqref{Eq:EoM}, see Fig.~\ref{fig:EnergyMassBand}. Moreover we can compute the effective mass of the hole-magnon bound state at the minimum of the band $(2m^*)^{-1}= |\frac{\partial^2E_Q}{\partial Q^2}|$. From the Born-Oppenheimer approximation we obtain $m^*=1/(3Ja^2)$ which diverges for $J/t\rightarrow 0$ indicating the immobile nature of the bound state, see Fig.~\ref{fig:EnergyMassBand}.

\subsection{Finite size scaling}
\begin{figure}[t]
    \centering
    \includegraphics[width=1\columnwidth]{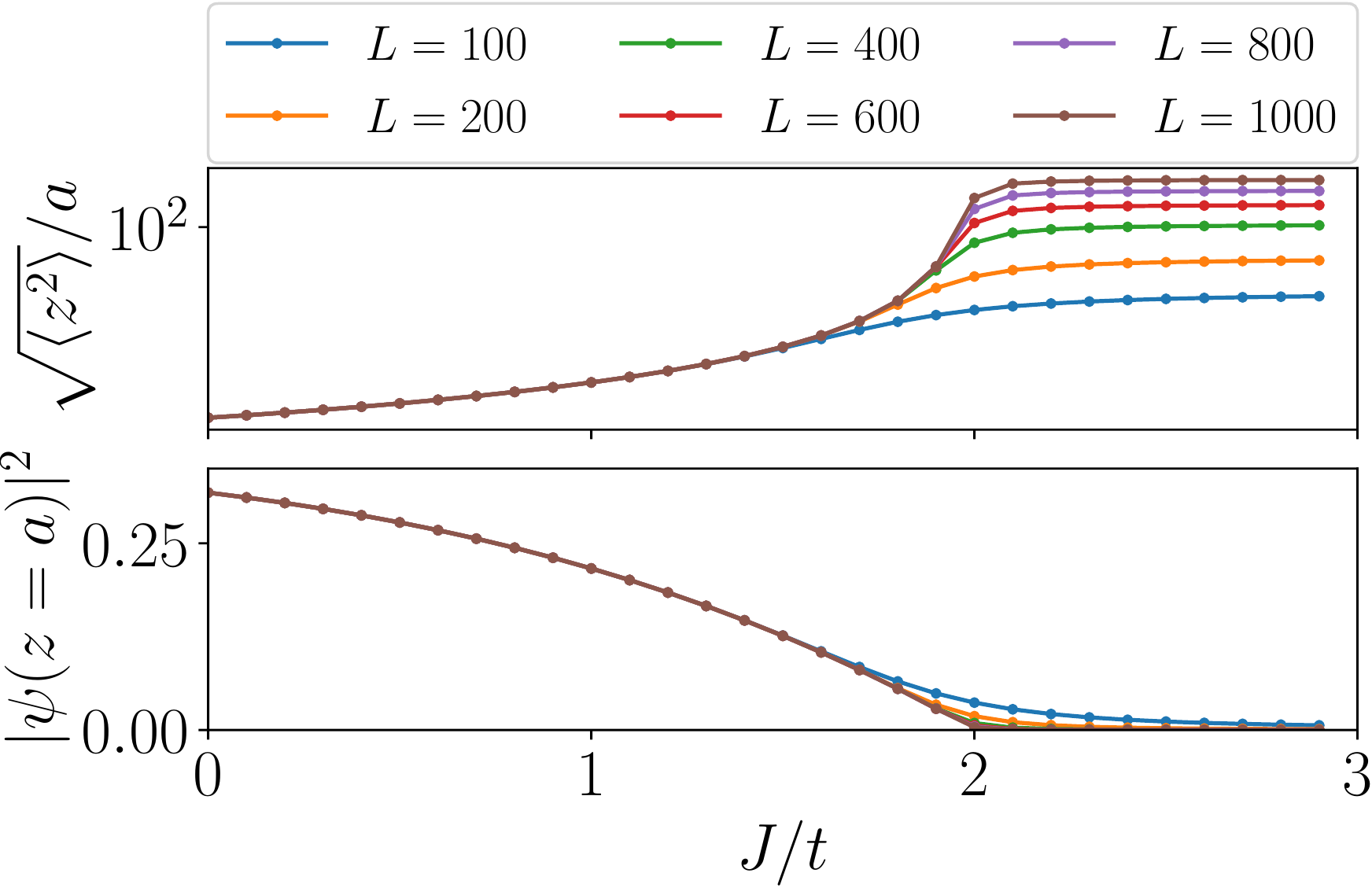} 
    \caption{Extension of the hole-magnon pair $\sqrt{\langle z^2 \rangle}/a$ (a) as a function of $J/t$ for different system sizes. Probability of finding a hole and a flipped spin at a distance equal to the lattice spacing $a$ (b) as a function of $J/t$ for different system sizes.}
    \label{fig:Scaling}
\end{figure}
In order to ensure the bound state nature of the hole-magnon pair for a range of $J/t$ we have performed a finite size scaling. We have obtained the binding energy, the extension of the pair and the probability of finding the hole and the magnon in nearest sites for large systems sizes and then we have extrapolated to the infinite size limit. The binding energy in the infinite size limit becomes negative for $J/t<2$ and the probability $a|\psi(z=a)|^2$ becomes zero for $J/t>2$ as shown in Fig.~\ref{fig:En_ext} and Fig.~\ref{fig:Scaling}. This signals a transition from bound to non-bound pair in the infinite size limit at $J/t=2$.  Moreover the size of the pair is independent of the system size for $J/t<2$ in the large size limit and it is strongly dependent on it for $J/t>2$. This also supports the appearance of a hole-magnon bound state at $J/t<2$ with a characteristic size independent of the system size.

\section{Antiferromagnetic spinbag model}
\label{AppendixB}
\begin{figure}[t!]
	\centering
		\includegraphics[width=\columnwidth]{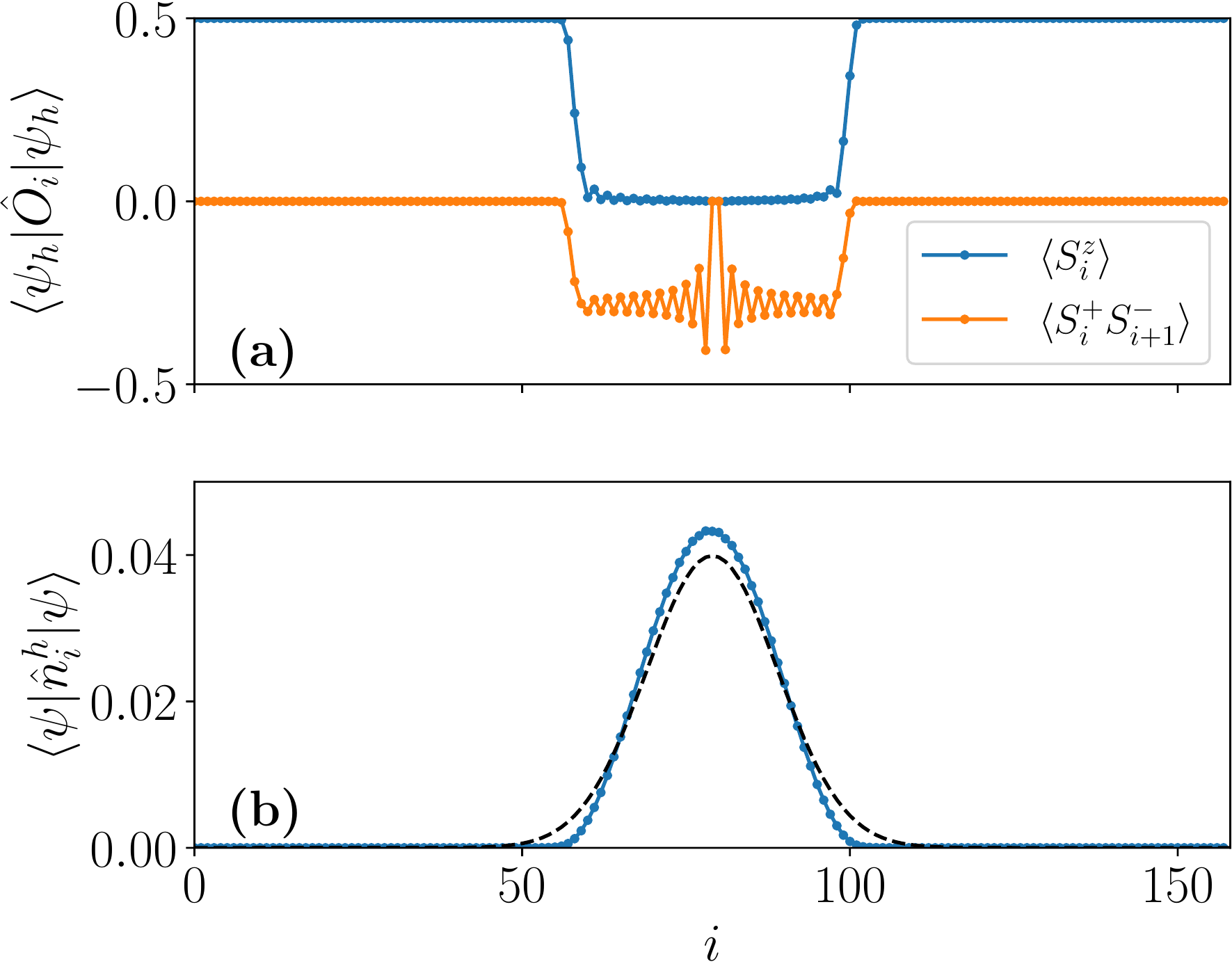}
	\caption{Magnetization and in-plane spin-spin correlations (a) computed over the projected wavefunction $|\psi_h\rangle$ for which we fix the hole position to be at the center of the lattice. Hole density (b) of the groundstate obtained from DMRG (blue dots) and from the variational calculation (dashed line). Both panels correspond to a computation of the fermionic zigzag ladder with a single hole and 20 magnons for $J/t=0$.} 
	\label{fig:AntSB}
\end{figure}
We now consider the situation of binding many magnons to a single hole thus creating a spin polaron or spinbag. An analytical treatment of this problem using the Hamiltonian~\eqref{Eq:tJHolMag} is very hard since the hole hopping disturbs the spin background. Therefore we change to a representation where the hole lives in the links of a new lattice with $2L-1$ sites i.e. the squeezed space~\cite{PhysRevB.41.2326,PhysRevB.48.16662,10.1080/13642810108208566,PhysRevB.70.075109,Hilker484,PhysRevB.97.125117}. In this way the hole hopping between the two legs of the ladder simply becomes $\left(t\sum_i \hat{h}^{\dagger}_{i+1}\hat{h}_{i}+\text{h.c.}\right)$ since it does not disturb the spin order in squeezed space. On the other hand, for the hole hopping between two sites of the same leg we have to take into account that two spins are exchanged in squeezed space. The operator doing such process is $\left( \hat{h}^{\dagger}_{i+2}\hat{h}_{i} +\hat{h}^{\dagger}_{i}\hat{h}_{i+2}  \right)\left(\frac{1}{2}+2\,\mathbf{S}_i \mathbf{S}_{i+1} \right)$. Concerning the spin interaction we have to take into account that two spins will not interact if a hole was occupying one of the two sites in the original lattice, giving the term $J\mathbf{S}_i \mathbf{S}_{i+1} (1-\hat{h}^{\dagger}_{i}\hat{h}_{i})$ and $J \mathbf{S}_i \mathbf{S}_{i+2} (1-\hat{h}^{\dagger}_{i}\hat{h}_{i})(1-\hat{h}^{\dagger}_{i+1}\hat{h}_{i+1})$. Moreover if a hole is present in the original lattice between two spins these two will be nearest neighbor in squeezed space and they will interact through the the superexchange coupling $J\mathbf{S}_i \mathbf{S}_{i+1} \hat{h}^{\dagger}_{i}\hat{h}_{i}$.
Therefore the Hamiltonian in squeezed space becomes,
\begin{eqnarray}
\hat{H}_{tJ}& =& t\sum_i \left( \hat{h}^{\dagger}_{i+1}\hat{h}_{i}+\text{h.c.} \right)+ J\sum_{i} \mathbf{S}_i \mathbf{S}_{i+1} \nonumber \\
&+&
t\sum_i \left( \hat{h}^{\dagger}_{i+2}\hat{h}_{i} +\hat{h}^{\dagger}_{i}\hat{h}_{i+2}  \right)\left(\frac{1}{2}+2\,\mathbf{S}_i \mathbf{S}_{i+1} \right)  \\
&+& J\sum_{i} \mathbf{S}_i \mathbf{S}_{i+2} (1-\hat{h}^{\dagger}_{i}\hat{h}_{i})(1-\hat{h}^{\dagger}_{i+1}\hat{h}_{i+1}). \nonumber
\label{Eq:Squeezed}
\end{eqnarray}
From this Hamiltonian we notice that the hole hopping in the same leg ($\hat{h}^{\dagger}_{i+2}\hat{h}_{i}$) provides a strong coupling of strength $t$ between the hole and the magnons. Notice that this coupling induces strong antiferromagnetic correlations between the two legs even though the original Ising interaction is not present for $J/t=0$. These antiferomagnetic correlations between the two legs will appear when a hole is present and the direct hopping dominates over the superexchange interaction $t/J\gg 1$. The original superexchange interaction presents magnetic frustration and at the semiclassical level it favours the formation of a spiral order of angle $\theta= \arccos(-1/4)$. Since we work at a fixed total mangetization (i.e. a fixed number of magnons) we expect that the system can be understood as consisting of two regions: Region (I) is adjacent to the hole and has spins in the XY plane with antiferromagnetic correlations between the legs. Region (II) extends outside of region (I), has nearly full $S^z$ polarization and spiral winding of the small XY components of the spins, see Fig.~\ref{fig:AntiferroSB1}. We name this object antiferromagnetic spinbag or antiferromagnetic spin polaron~\cite{NAGAEV199239}  since it is analogous to the ferromagnetic one~\cite{auerbach1994interacting} but its origin resides in the frustration of the system. 

To obtain qualitative expressions for the size and energy of the antiferromagnetic spinbag we consider a semiclassical approximation of the spin part of the wavefunction $\mathbf{S}_i \mathbf{S}_{i+1} = \Tilde{S}^2 \left(\sin(\varphi)^2 \cos(\theta) + \cos(\varphi)^2\right)$ with $\Tilde{S}^2=S(S+1)=3/4$ 
The angles $\theta$ and $\varphi$ determine the spin orientation in the XY plane and the angle respect to the $z$ axis, respectively (see top panel of Fig.~\ref{fig:AntSB}). Therefore we can relate the total number of magnons $N_m$ and the size of the antiferromagnetic spinbag $2R$ using the angle $\varphi_{II}$,
\begin{equation}
\cos(\varphi_{II})\left(L-R \right)=L-N_m.
\label{Eq:Mag}
\end{equation} 
By considering the limit $J/t\ll 1$ we assume a frozen magnetic order and solve the equation of motion for the hole with a fixed spin background along the lines of the Born-Oppenheimer approximation. The hole equation of motion in each region is given by,
\begin{eqnarray}
(I):\,  E_h h_i &=& t \left( h_{i+1} + h_{i-1}\right) - \frac{3J}{4} \left(h_i+h_{i+1} \right) \nonumber\\
&-& t \left( h_{i-2} + h_{i+2} \right), \\
(II):\,  E_h h_i &=& t \left( h_{i+1} + h_{i-1}\right) \nonumber \\ &+& \frac{3J}{32}\left(15\cos(\varphi_{II})^2-7 \right) \left(h_i+h_{i+1} \right) \\
&+& \frac{t}{8} \left( 15\cos(\varphi_{II})^2+1 \right) \left( h_{i-2} + h_{i+2} \right) \nonumber.
\label{Eq:Holeq}
\end{eqnarray}
These equations show that the hole energy is reduced when it is localized inside the antiferromagnetic region (I). Thus the antiferromagnetic background acts as an effective square well potential to the hole.
In order to obtain an analytical expression for the hole energy $E_h$ we solve the hole equation of motion by taking the continuum limit and proposing a gaussian wavefunction localized in the antiferromagnetic region $h_i=\exp\{-x_i^2/(R^2)\}/\sqrt{R\sqrt{\pi/2}}$, where we employ the size of the bag $2R$. Moreover Eq.~\eqref{Eq:Mag} establishes the relation between the size of the bag $R$ and the angle $\varphi_{II}$ when the total number of magnons is specified. Taking the limit $L\ll R \sim N_m\ll 1$ we obtain a hole energy that scales with the number of magnons as, 
\begin{equation}
E_h\propto 4.28 \frac{t}{N_m^2} -0.04 \frac{J}{N_m}.
\end{equation}

We now consider the magnetic energy cost of creating an antiferromagnetic spinbag. 
We compare the magnetic energies with the same number of magnons when there is an antiferromagnetic spinbag and when there is not. In the first case the antiferromagnetic region has zero magnetic energy and the spiral region has an energy $E_{M1}$ which is determined by the angle $\varphi_{II}$ (see Eq.~\eqref{Eq:Mag}) and the size of the spinbag $R$. In the second case the full ladder presents a spiral order with energy $E_{M2}$ and an angle satisfying $\cos(\varphi)L=L-N_m$. The difference between these two magnetic energies in the limit $L\gg R\gg 1$ is,
\begin{equation}
E_{M1}-E_{M2} \propto \frac{51}{8}JN_m.
\end{equation}
The magnetic energy increases when creating an antiferromagnetic spinbag because the angle $\varphi$ of the spiral order is reduced. The total energy of the antiferromagnetic spinbag is given by,
\begin{equation}
E_{ASB}=4.28 \frac{t}{N_m^2}-0.04 \frac{J}{N_m}+\frac{51}{8}JN_m,
\label{Eq:AFSB_energy}
\end{equation}
where we discard the constant terms with respect to the number of magnons. From Eq.~\eqref{Eq:AFSB_energy} we observe that increasing the size of the antiferromagnetic spinbag reduces the hole kinetic energy but increases the magnetic one. Therefore there is an optimal number of magnons forming the antiferromagnetic spinbag $N_m^*$. Considering the situation of a small antiferromagnetic region compared with the spiral region $L\gg R\sim N_m \gg 1$ we can estimate the number of magnons,
\begin{equation}
N_m^* \propto 1.10\, (t/J)^{1/3},
\label{Eq:MagN}
\end{equation}
considering the condition $(J/t)\ll 1$.

The semiclassical spinbag theory predicts that under doping the fermionic zigzag ladder with a single hole and magnons, a certain number of magnons, given by Eq.~\eqref{Eq:MagN}, accumulate around the hole with an antiferromagnetic ordering and the remaining ones are pushed away from this region forming a spiral order with a net magnetization pointing in the $z$ axis, see panel (a) of Fig.~\ref{fig:AntSB}. In order to benchmark our theory we have performed DMRG simulations of the fermionic zigzag ladder with a single hole and different number of magnons. We observe that for small values of $J/t$, the hole density accumulates in a certain region of the lattice, see panel (b) of Fig.~\ref{fig:AntSB}. This can be attributed to the large effective mass of the antiferromagnetic spinbag and the fact that we are working with open boundary conditions thus explicitly breaking the translational invariance of the system. By computing the projected wavefunction $\hat{P}^h_L|\psi\rangle =|\psi_h\rangle$ which has a hole fixed at the center of the lattice we obtain the spin ordering surrounding the hole~\cite{PhysRevB.55.6504}. Around the hole the magnetization suddenly decays indicating that the spins tilt to the XY plane. Moreover the in-plane spin-spin correlations show an antiferromagnetic order inside this region. Outside this region we recover a perfect ferromagnetic spin order. This confirms the formation of an antiferromagnetic spinbag for small values of $J/t$. Notice that the simulation is performed at $J/t=0$. Therefore all the magnons are contained in the antiferromagnetic spinbag and the spiral order simply becomes a ferromagnetic order since $\varphi=0$.

\end{document}